\documentstyle[12pt,aaspp4,flushrt]{article}
\newbox\grsign \setbox\grsign=\hbox{$>$} \newdimen\grdimen \grdimen=\ht\grsign
\newbox\simlessbox \newbox\simgreatbox
\setbox\simgreatbox=\hbox{\raise.5ex\hbox{$>$}\llap
     {\lower.5ex\hbox{$\sim$}}}\ht1=\grdimen\dp1=0pt
\setbox\simlessbox=\hbox{\raise.5ex\hbox{$<$}\llap
     {\lower.5ex\hbox{$\sim$}}}\ht2=\grdimen\dp2=0pt
\def\simgreat{\mathrel{\copy\simgreatbox}}
\def\simless{\mathrel{\copy\simlessbox}}

\def\vol#1  {{{#1}{\rm,}\ }}

\def\aj{{AJ}}  %Astronomical Journal%
\def\apj{{ApJ}} %Astrophysical Journal%
\def\apjs{{ApJS}} %Astrophysical Journal Supplements%
\def\pasp{{PASP}}  %Publications of the Astronomical%
\def\mnras{{MNRAS}} %Monthly Notices of the Royal%
\def\aa{{A\&A}}     %Astronomy & Astrophysics%
\def\etal{{\it et al.}\ }

\begin{document}
\title{A Study of Nine High-Redshift Clusters of Galaxies:\\
 II. Photometry, Spectra, and Ages of Clusters 0023+0423 and 1604+4304}

\author{Marc Postman}
\affil{Space Telescope Science Institute\altaffilmark{1},
3700 San Martin Drive, Baltimore, MD 21218}
\affil{Electronic mail: postman@stsci.edu}
 
\author{Lori M. Lubin\altaffilmark{2,3}}
\affil{Observatories of the Carnegie Institution of Washington, 813 Santa
Barbara St., Pasadena, CA 91101}
\affil{Electronic mail: lml@astro.caltech.edu}
 
\author{J. B. Oke}
\affil{Palomar Observatory, California Institute of Technology, Pasadena,
CA 91125}
\affil{and}
\affil{National Research Council Canada, Herzberg Institute of Astrophysics,
Dominion Astrophysical Observatory, 5071 W. Saanich Road, Victoria,
BC V8X 4M6}
\affil{Electronic mail: oke@dao.nrc.ca}
 
\vskip 1.0 cm
\centerline{Accepted for publication in the {\it Astronomical Journal}}
 
\altaffiltext{1}{Space Telescope Science Institute is operated by the
Association of Universities for Research in Astronomy, Inc.,
under contract to the National Aeronautics and Space Administration.}
 
\altaffiltext{2}{Hubble Fellow}
 
\altaffiltext{3}{Present Address : Palomar Observatory, California
Institute of Technology, Mail Stop 105-24, Pasadena, CA 91125}
 
\vfill
\eject

\begin{abstract}

We present an extensive photometric and spectroscopic study of two
high-redshift clusters of galaxies based on data obtained from the
Keck 10m telescopes and the Hubble Space Telescope.  The clusters
CL0023+0423 ($z=0.84$) and CL1604+4304 ($z=0.90$) are part of a
multi-wavelength program to study nine candidate clusters at $z
\simgreat 0.6$ (Oke, Postman \& Lubin 1998).  Based on these
observations, we study in detail both the field and cluster
populations. From the confirmed cluster members, we find that
CL0023+0423 actually consists of two components separated by
$\sim$2900 km s$^{-1}$. A kinematic analysis indicates that the two
components are a poor cluster with $\sim$3 $\times 10^{14}$ M$_\odot$
and a less massive group with $\sim$10$^{13}$ M$_\odot$.  CL1604+4304
is a centrally concentrated, rich cluster at $z = 0.8967$ with a
velocity dispersion of 1226 km s$^{-1}$ and a mass of $\sim$3 $\times
10^{15}$ M$_\odot$.

A large percentage of the cluster members show high levels of star
formation activity.  Approximately 57\% and 50\% of the galaxies are
active in CL0023+0423 and CL1604+4304, respectively.  These numbers
are significantly larger than those found in intermediate-redshift
clusters (Balogh \etal 1997). We also observe many old, red
galaxies. Found mainly in CL1604+4304, they have spectra consistent
with passive stellar evolution,
typical of the populations of early-type galaxies in low and
intermediate-redshift clusters. We have calculated their ages by
comparing their spectral energy distributions to standard Bruzual \&
Charlot (1995) evolutionary models. We find that their colors are
consistent with models having an exponentially decreasing star formation 
rate with a time constant of 0.6 Gyr.  We also observe a
significant luminosity brightening in our brightest cluster galaxies.
Compared to brightest cluster galaxies at $z \sim 0.1$, we find a
luminosity increase of $\sim 1$ mag in the rest $M_B$ and $\sim 0.8$
mag in the rest $M_V$.

In the field, we find that $\sim 76\%$ of the galaxies with $z > 0.4$
show emission line activity. These numbers are consistent with
previous studies (e.g.\ Hammer \etal 1997).  We find that an
exponentially decaying star formation rate is required to produce the
observed amount of star formation for the majority of the galaxies in
our sample. A time constant of $\tau = 0.6$ Gyr appears to be optimal.
We also detect several interesting galaxies at $z > 1$.  Two of these
galaxies are extremely luminous with strong MgII$\lambda$2800
absorption and FeII resonance line absorption. These lines are so
strong that we conclude that they must be generated within the
atmospheres of a large population of young, hot stars.

\end{abstract}
\eject
\section{Introduction}

The study of cluster galaxies at moderate look-back times of $50-70$\%
of the cosmic age ($0.7 \simless z \simless 1.5$) can provide
important constraints on the nature and duration of the processes
which have yielded the current epoch distribution of galaxy
properties. This intermediate redshift range is likely to be a period
where clusters are undergoing (or have recently completed)
virialization and where many galaxies are only 1 or 2 Gyr past the
peak in the cosmic star formation history (Madau, Pozzetti \&
Dickinson 1998). From a practical point of view, it is a regime which
has only recently become readily accessible, both photometrically and
spectroscopically, from space and ground-based optical and near-IR
telescopes.  In the redshift range from $z = 0.3$ to $0.6$ there have
been numerous photometric and/or spectroscopic studies of both
optically and X-ray selected clusters (e.g.\ \cite{koo81};
\cite{cou83}; \cite{ell85}; \cite{cou85}; \cite{cou87}; \cite{cou91};
\cite{fab91}; \cite{hen92}; \cite{dres92}; \cite{oke96}). Galaxies in
these systems can be studied easily down to 5\% of the characteristic
luminosity and contamination by interlopers is not a significant
problem.

A growing number of clusters with $z \leq 0.5$ have been extremely
well-studied.  With spectra of hundreds of cluster members, detailed
modeling of the infall patterns and intracluster chemical composition
gradients is now available (\cite{abr96}; \cite{elling97}).  The above
studies indicate that the cores of rich clusters are typically
dominated by a population of luminous early-type, red galaxies which
produce a remarkably narrow ridge (or ``red locus'') in the
color-magnitude (CM) relation. At $z \simless 0.4$ both the mean color
and the CM relation are consistent with those of present-day ellipticals
(e.g.\ Arag\'on-Salamanca \etal 1991; Dressler \& Gunn 1992; Stanford,
Eisenhardt \& Dickinson 1994,1997).  The bulk of these galaxies have
spectra which show no obvious signs of current or recent star
formation. However, there is a non-negligible fraction which show
post-starburst spectra with strong Balmer lines in absorption (e.g.\
Gunn \& Dressler 1988; Dressler \& Gunn 1992; Poggianti 1997). In
conjunction with this population is a fraction of blue cluster members
which is increasing with redshift, a phenomenon known as the
Butcher-Oemler effect (Butcher \& Oemler 1984). Most of these galaxies
appear as normal spirals or have peculiar morphologies (Dressler \etal
1994; Couch \etal 1994; Oemler, Dressler \& Butcher 1997). A large
fraction of this blue, spiral population exhibits exceptionally strong
Balmer lines and/or [OII] emission which indicates that a significant
fraction of the cluster members have recently undergone or are
currently undergoing a high level of star formation activity (e.g.\
Lavery \& Henry 1988; Gunn \& Dressler 1998; Lavery, Pierce \& McClure
1992; Dressler \etal 1994; Poggianti 1997).

Complimenting the ground-based work are several Hubble Space Telescope
(HST) programs to quantify the morphology of galaxies in these
clusters (e.g.\ \cite{cou94}; \cite{dres94}; \cite{sma97};
\cite{oem97}; \cite{ell97}). HST enables morphological classifications
which can be made on scales of $\sim 1~{\rm kpc}$, thus providing a
direct comparison to ground-based classifications of nearby galaxies.
In addition to relating the photometric and spectral characteristics
of a galaxy to its morphology, the HST studies can
be used to examine the overall morphological distribution in these
clusters. This work may indicate that the morphological composition of
clusters is evolving with redshift (Dressler \etal 1997; Oemler,
Dressler \& Butcher 1997), though these results are not yet certain
(e.g.\ Stanford, Eisenhardt \& Dickinson 1997; Lubin \etal
1998). These evolutionary changes are also apparently reflected in the
evolution of the morphology--density relation (\cite{dres80};
\cite{pg84}). This relation in intermediate-redshift clusters which
are centrally-concentrated and compact is qualitatively similar to
that in the local universe; however, unlike present-day clusters, the
relation is non-existent in the loose, open clusters (Dressler \etal
1997).

Studies of clusters of galaxies with redshifts greater than $z = 0.6$
are substantially more difficult because (1) the galaxies are
approaching the sensitivity limits of optical spectrographs on 4m
class telescopes, (2) the interloper contamination becomes
substantial, and (3) the well understood rest wavelength region
redward of 4000\AA\ moves into the near infrared.  Despite these
difficulties, several studies of high-redshift clusters have been
made. These studies indicate that the Butcher-Oemler effect continues
to strengthen up to $z = 0.9$ (Arag\'on-Salamanca \etal 1993; Rakos \&
Schombert 1995; Lubin 1996).  In addition, the red envelope of the
early-type cluster population moves bluewards with redshift. At $z
\sim 0.9$ there are few galaxies with colors as red as present-day
ellipticals (Arag\'on-Salamanca \etal 1993; Rakos \& Schombert 1995;
Oke, Gunn \& Hoessel 1996; Lubin 1996; Ellis \etal 1997; Stanford,
Eisenhardt \& Dickinson 1995,1997). This color evolution is consistent
with passive evolution of an old stellar population formed at an early
cosmic age.  The amount of color evolution is similar from cluster to
cluster at a given redshift and is independent of the cluster richness
or X-ray luminosity. These results indicate that the history of
early-type galaxies may be insensitive to environment; that is, these
galaxies appear to be coeval with a common star formation history
(Bower \etal 1992a,b; Arag\'on-Salamanca \etal 1993; Dickinson 1995;
Stanford, Eisenhardt \& Dickinson 1995,1997; Ellis \etal 1997).

There have also been many studies of high-redshift field galaxies.
Hamilton (1985) obtained spectra of 33 very red, field galaxies in the
redshift range of $z = 0.2$ to $z = 0.8$.  He found that the 4000 \AA\
break changed by less than 7\% over the observed $z$ range.  Songaila
\etal (1994) obtained $BVK$ photometry and spectra of a nearly
complete sample of 298 galaxies.  The redshifts are nearly all at $z <
1.0$. They find no $K$--band luminosity evolution. From measurements of
emission-line strengths and the 4000 \AA\ break, they infer that
galaxies are undergoing significantly more star formation at $z=1$
than at the present epoch.  The Canada-France-Redshift-Survey (CFRS)
group has used CFHT to carry out a very extensive spectroscopic survey
out to redshifts of 1.3 (\cite{CFRS}, \cite{ham97}).  They find that
the fraction of galaxies with significant emission lines (EW of [OII]
$> 15$ \AA) increases from about 13\% locally to over 50\% at $z
>0.5$.  The fraction of luminous, quiescent galaxies (no significent
[OII] emission) decreases with redshift from 53\% at $z=0.3$ to 23\%
for $z>0.5$.  They also find evidence that the metal abundance is
lower in emission-line galaxies at high redshifts than locally.  In
addition, Cohen \etal (1996a,b) have obtained spectra of high redshift
galaxies in the fields of both the Hubble Medium Deep Survey and the
Hubble Deep Field survey. They find that the redshifts are highly
clumped; the velocity dispersion in these clumps are similar to those
found in local groups of galaxies.  Further supporting this
observational evidence of strong velocity structure at high redshift,
Koo \etal (1996) carried out photometry and spectroscopy of 35
galaxies with redshifts of 0.3 to 1.6.  They found that half of the
redshifts in their sample are actually in two structures at $z=0.81$
and $z=1.0$.

Because studies of both field and cluster galaxies indicate that the
high-redshift universe is a place of substantial evolution, we have
undertaken an extensive program to study nine candidate clusters of
galaxies at $z \simgreat 0.6$ (Oke, Postman \& Lubin 1998; hereafter
Paper I). With the commissioning of the Keck 10 meter telescopes, this
detailed survey is now possible. The first paper in the series
describes the sample selection, data acquisition, and data reduction
procedures of the survey (Paper I). In this paper, we present our
analysis and interpretation of the spectra and photometry for the
first two clusters to be completed, CL0023+0423 ($z = 0.84$) and
CL1604+4304 ($z = 0.90$). In the third paper of this series (Lubin
\etal 1998; hereafter Paper III), the HST observations and the
resulting morphological composition of these two clusters are
described.

\section{Keck LRIS Observations}

Broad-band $BVRI$ and low-resolution spectroscopic data were obtained
for CL0023+0423 and CL1604+4304 using the Low Resolution Imaging
Spectrograph (LRIS; Oke \etal 1995) at the W.M. Keck Observatory.  The
details of these observations are presented in Paper I.  We present
here only a brief summary of the relevant information.  The LRIS
imaging for CL0023+0423 was obtained under photometric conditions.
Spectra were taken using six different masks.  The weather was
photometric for four of the slit masks but marginal for the other two.
Consequently, a second observation for one of the masks was obtained.
In the case of CL1604+4304 the weather was photometric for all the
broad-band imaging and for the six slit mask observations.  Figures
\ref{fig-cl00im} and \ref{fig-cl16im} show composite $BVR$ images for
these two clusters.

For the spectrophotometric observations a list was made of all objects
in a $600 \times 2000$ pixel area ($2.15 \times 7.16$ arcminutes) down
to a Johnson-Cousins R magnitude of 23.3.  The very few objects
brighter than $R = 18.3$ were excluded from the list, as well as
those objects which have $y$ coordinates too close to the two
selected set stars to allow spectra to be obtained.  This produced a
sample of 167 objects for CL0023+0423 and 168 for CL1604+4304. A
summary of the spectroscopic observations is provided here in Table
1. The numbered rows give in order (1) the number of objects in the
sampled region of sky, (2) the number of these objects which did not
have a slit positioned on them and, hence, no spectrum was obtained,
(3) the number of objects which have spectra but for which no redshift
was determined, (4) the number of stars and very low $z$ galaxies ($z \le
0.01$), (5) the number of quasars found, (6) the remaining number of
objects which have redshifts, and (7) the number of these which have
emission lines.

\subsection{Spectroscopy}

\subsubsection{Redshifts}

The spectra cover the wavelength range of $\sim$4500 \AA\ to 9500 \AA\
(see Paper I).  The redshift determination is described fully in Paper
I.  The full lists of candidates for which spectra were attempted are
given in Table 2 for the CL0023+0423 field and Table 3 for the
CL1604+4304 field.  Spectra were obtained for approximately 80\% of
the candidate objects; redshifts were determined for 90\% of those.
At redshifts above $z \sim 0.4 - 0.5$, the $[{\rm OII}]\lambda 3727$
line is observed. In the CL0023+0423 and CL1604+4304 fields, the
fraction of objects with spectra that have emission lines is 89\% and
79\%, respectively.  Assuming that the objects for which redshifts
were not obtained do not have emission lines, the fraction of faint
galaxies with emission lines is more like 78\% and 70\%, respectively,
for the two cluster fields.

The resulting redshifts and qualities of the redshifts are given in
columns 6 and 7 of Tables 2 and 3.  The number 9.0000 in the tables
means that a spectrum was obtained but no redshift could be derived.
Stars and galaxies with redshifts less than 0.01 are listed as $z =
0.0000$. The quality of a redshift is specified by a number from 1 to 4 which
roughly corresponds to the number of features identified.  A quality
of 4 means that the redshift is certain.  Quality 3 indicates that the
redshift is almost certainly correct.  Quality 2 means the redshift is
probably correct while quality 1, which corresponds to only 1 emission
line being seen , means the redshift is possible.  In the cases
where a single emission line can only be
identified with [OII] the quality is set to 2.
The distributions of spectroscopic targets on the sky and in
redshift for the two cluster fields are shown in Figures
\ref{fig-cl00targ} and \ref{fig-cl16targ}.  The results are summarized
in Table 4 which lists the number of objects and the mean redshift for
the significant structures in the two fields.

Sample spectra are shown in Figure~\ref{fig-samplespectra}
where the flux, represented by the AB magnitude, is plotted against
the observed wavelength. The locations of the more prominent emission
and absorption lines in each spectrum are marked.  The relative values
of AB are also plotted for a typical night sky spectrum.

\subsubsection{Equivalent Width Measurements}

The features which have equivalent widths measured are the $[{\rm
OII}]\lambda3727$ emission line, the absorption feature centered at
3835 \AA\ (which includes CN, metal lines, Balmer H9), Balmer H8, the
CaII H and K lines, H$\epsilon$, H$\delta$, the G-band, H$\gamma$,
H$\beta$, and [OIII] $\lambda 5007$.  These spectral features are
prominent and, except for the last two, are located in a spectral
range where the S/N is relatively good.  To measure the equivalent
width two ``continuum'' bands are defined on either side of the
feature, and a continuum level is derived by linearly interpolating
between the two bands.  The equivalent width is then an integration of
the continuum subtracted signal over a band centered on the feature.
The continuum and line bands used for each feature are defined in
Table 5.  The observed spectra are shifted to zero redshift before
carrying out the calculations.  The equivalent widths are in Angstroms
and are positive if the line is in absorption and negative if the line
is in emission.  The [OII] and [OIII] lines should be negative except
for noise variations.  Equivalent widths of the Balmer lines can be
negative or positive.  The rest-frame equivalent widths of $[{\rm
OII}]\lambda3727$, H$\beta$, and [OIII] $\lambda 5007$ are listed in
columns 10, 11, and 12, respectively, of Tables 2 and 3.  Errors in
the equivalent widths of [OII], derived from the actual measurements
in objects with no emission line, are about 4 \AA\ ; any equivalent
widths below this value are usually null detections.  Errors for
H$\beta$ and [OIII] are about 6 or 7 \AA\ since they tend to lie redwards
of 8000\AA\ where the spectra are noisy, and the sky subtraction is
difficult. The rest frame equivalent widths of [OII] are plotted
against $z$ for the two cluster fields in Figures \ref{fig-cl00eqwz}
and \ref{fig-cl16eqwz}.

If we define active star formation by the presence of an [OII] line
with an equivalent width of $\simgreat 15.0$ \AA\ (as used by Hammer
\etal 1997), 76\% of the field (non-cluster) sample with redshifts of
$z > 0.$4 are active. Splitting this sample into the redshift bins of
Hammer \etal (1997), we find that 79\% of the field galaxies with $0.4
\le z < 0.9$ are active, while 62\% with $z \ge 0.9$ are active. In
their field sample, Hammer \etal (1997) find 65\% and 90\% for the
same redshift ranges.  Since both samples are small, there is no
significant difference.  Of the CL0023+0423 and CL1604+4304 cluster
members, 57\% and 50\% are active galaxies, respectively.  These
fractions are much larger, however, than the $\simless 15\%$ in clusters 
between $z=0.2$ and $z=0.55$ (\cite{bal97}).

\subsubsection{The 4000 \AA\ Break and the Balmer Jump}
 
The traditional estimator for the 4000 \AA\ break amplitude is defined
as the ratio of the integrated flux per unit wavelength in the rest
band 4050--4250 \AA\ to that in the rest band 3750--3950 \AA\ (Bruzual
1983).  One problem with this measurement is that it is substantially
influenced by the overall color of the galaxy since the two
measurements are separated by 300 \AA.  It also has some sensitivity
to reddening.  This traditional measurement works fairly well for
early-type galaxies where (a) the spectrum is dominated by late-type
stars, (b) the spectral energy distribution $f_\lambda$ is nearly
constant so that the jump has only a small color term embedded in it,
and (c) the jump is produced by metal line absorption between 3750 and
3980 \AA.

For younger objects, such as those that we are dealing with in this
paper, the Balmer jump can be much more important than the 4000 \AA\
break. Hammer \etal (1997) introduced a Balmer jump index; however, it
is simply a flux ratio in two bands and does not remove the overall
color change. We, therefore, introduce a new definition of the break
amplitude which is based on the flux between 3400 \AA\ and 4280 \AA.
Two continuum bands, one from 3400--3700 \AA\ and the other from
4050--4280 \AA, are defined.  The flux in each band is fitted by a
first order equation.  In practice, the two slopes that are generated
are quite similar although they are not identical because of the
intrinsic character of the spectrum and noise.  The average of the two
slopes is adopted for both bands and used to extrapolate the spectrum
in each band to a rest wavelength of 3850 \AA.  The ratio of the two
fits evaluated at 3850 \AA\ defines the jump.  Since we have
extrapolated with the same slope, the measured jump is just the
vertical separation between the two linear fits.  (The wavelength for
the fit is arbitrary since, by definition, the vertical separation is
the same anywhere in the 3700--4000 \AA\ range.)  We will refer to
this parameter as $J$ which is simply the intensity ratio expressed in
magnitudes.  The advantages of our $J$ estimator is that it can
measure either the 4000 \AA\ break or the Balmer jump.  In addition,
it is insensitive to redenning and to the slope of the energy
distribution throughout the spectral range from 3400--4280 \AA.

We use the analytic fit in the 3400--3700 \AA\ region to extrapolate
the flux between 3700 and 4000 \AA.  The ratio of the observed flux
(in 50 \AA\ segments) to this extrapolated flux is noted.  If the jump
is the Balmer jump, the ratios will increase rapidly between 3750 and
3875 \AA\ and then stay relatively constant.  If the jump is the 4000
\AA\ break, the ratio stays close to unity from 3700 to 3950 \AA\ and
then quickly increases.  The wavelength at which the intensity ratio
increases from unity to the value corresponding to $J$ indicates
whether it is the Balmer or 4000 \AA\ jump or somewhere between the
two.  The measured jump ($J$) and the transition wavelength
($\lambda_J$), where measurable, are listed in column 13 of Tables 2
and 3.  In addition, we have also calculated the traditional 4000 \AA\
break D(4000) as defined above.  These values are listed in column 14
of Tables 2 and 3.

\subsection{Photometry}

\subsubsection{Broad Band Colors}

The photometric survey was conducted in four broad band filters,
$BVRI$, which match the Cousins system well. The response curves of
these filters are shown in Figure 1 of Paper I. The Keck observations
have been calibrated to the standard Cousins-Bessell-Landolt (Cape)
system through exposures of a number of Landolt standard star fields
(Landolt 1992).  The FOCAS package (Valdes 1982) was used to detect,
classify, and obtain aperture and isophotal magnitudes for all objects
in the co-added $BVRI$ images.  For each galaxy in the field, we have
derived magnitudes in a circular aperture with a
radius of $3^{''}$. This corresponds to a physical radius of
$\{14.60~14.92\}~h^{-1}~{\rm kpc}$ at $z = \{0.84~ 0.90\}$, the
redshifts of CL0023+0423 and CL1604+4304, respectively.  The limiting
magnitudes are $B = 25.1$, $V = 24.1$, $R = 23.5$, and $I = 21.7$ for
a 5-$\sigma$ detection in our standard aperture (for more details, see
Sects.\ 3.1 and 4.1 of Paper I). From these magnitudes, we have
generated the corresponding AB values of ABB, ABV, ABR, and ABI using
Equations 2--5 in Paper I. The corresponding AB values for all
galaxies which have spectra are given in Tables 2 and 3 for the
CL0023+0423 and CL1604+4304 fields, respectively.

\subsubsection{Absolute Luminosities}

One method for determining the absolute luminosities of distant
galaxies is to use observed broad-band magnitudes and k-corrections.
In our case this is very difficult as we are dealing with young
objects at high redshifts where there are few observations from which
to calculate the appropriate k-corrections.  Instead, we make use of
the fact that we have the observed AB magnitudes (see Sect.\ 2.2.1)
and a best-fit evolutionary model to the energy distribution (for
details on the evolutionary models and the fitting procedure, see
Sect.\ 4) for each galaxy.  We can relate apparent and absolute
magnitude using the formalism of Equations 6, 9, and 10 in Gunn \& Oke
(1975). This relation becomes :

\begin{eqnarray}
M_{AB_{\nu(1+z)}} = m_{AB_\nu} - 2.5~{\rm log}~\left[ \frac{9.00 \times
10^{20} {\pounds}^{2}_{q}(z)(1+z)} {H_o^2}\right] 
\end{eqnarray}

\noindent where ${\pounds_q(z)}$ is given in Equation 9 of Gunn \& Oke
(1975).

We can easily measure $m_{AB_\nu}$, the value of AB at the redshifted
wavelength of the $B$ filter, for example. For a given $H_o$ and $q_o$, we
can use this value to calculate the absolute magnitude at the rest $B$
wavelength from Equation 1 above.  We have chosen to use a rest $B$
wavelength to eliminate or minimize any extrapolation. The absolute
magnitude in this band is hereafter referred to as $M_{ABB}$.  For
redshifts of $z < 0.92$ the redshifted $B$ filter position is within
the observed wavelength range, and an interpolation of the best-fit
evolutionary model can be made.  Above $z = 0.92$ the rest-frame $B$
filter wavelength is above the observed $I$ band, and an extrapolation
is necessary.  This extrapolation is done by using the best-fit
evolutionary model to the four observed AB values (see Sect.\ 4) to
extrapolate to the appropriate frequency.  The uncertainty in the
resulting absolute AB can be estimated from the uncertainty in the fit
of the observations to the model.

We have done the calculations for $h = 1.0$ where $h = H_o/100~{\rm
km~s^{-1}~Mpc^{-1}}$ and for $q_o = 0.1$ (e.g.\ Carlberg \etal 1996).
The results are listed in Tables 2 and 3.  For other values of $h$,
simply add $5~{\rm log}~h$ to numbers in the tables.

\section{Cluster Kinematics}

One of the primary goals of the survey is an analysis of the kinematic
properties of the clusters. We have acquired redshifts for $15-35$
cluster members in each system, enabling an accurate estimate of the
cluster velocity dispersion to be made and, potentially, the cluster
mass.  Cluster velocity dispersions are calculated by first defining a
broad redshift range, typically $\Delta z = \pm 0.06$, in which to
conduct the calculations.  This range is manually chosen to be
centered on the approximate redshift of the cluster.  We then compute
the bi-weight mean and dispersion of the velocity distribution
(\cite{beers}) and identify the galaxy with the largest deviation from
the mean. Velocity offsets from the mean are taken to be $\Delta v =
c(z - \overline{z})/(1 + \overline{z})$ which corrects for
cosmological and relativistic effects. In the case of bi-weight
statistics, $\overline{z}$ is the median of the distribution.  If the
galaxy with the largest velocity deviation differs from the bi-weight
median by either more than 3$\sigma$ or by more than 3500 km s$^{-1}$,
it is excluded and the computations are redone. The procedure
continues until no further galaxies satisfy the above criteria.  The
3500 km s$^{-1}$ limit is based on extensive data available for low
$z$ clusters. For example, 95\% of the galaxies within the central
3$h^{-1}$ Mpc region of the Coma cluster and with $cz \le 12,000$ km
s$^{-1}$ lie within $\pm 3500$ km s$^{-1}$ of the mean Coma
redshift. This clipping procedure is conservative and does not impose
a Gaussian distribution on the final redshift distribution: for
CL0023+0423 we find that the resulting redshift distribution is
inconsistent with a Gaussian at more than the 97\% confidence level.

Figure~\ref{fig-velhist} shows histograms of the velocity offsets
relative to the mean cluster redshifts for the two clusters. In the
case of CL1604+4304, the clipping procedure concludes after rejecting
the 4 most deviant redshifts, all of which differ from the mean
redshift (when the outlier is included in the computation) by more
than 3500 km s$^{-1}$. We then get a mean redshift of $\overline{z} =
0.8967$ and a dispersion of 1226 km s$^{-1}$ (corrected for cosmological
effects) based on 22 galaxies.
All dispersions quoted here have also been corrected for redshift
measurement errors which are typically about 100 km s$^{-1}$ at $z
\sim 0.9$, and the uncertainty in the dispersion is computed following
the prescription of Danese \etal (1980). The Danese \etal prescription
assumes the errors in velocity dispersions can be modeled
as a $\chi^2$ distribution and that a galaxy's velocity deviation from the
mean cluster redshift is independent of the galaxy's mass ({\it i.e.}, the
cluster is virialized).  The angular and redshift
distributions of the galaxies in CL1604+4304 show no obvious
substructure (see Figure~\ref{fig-cl16targ}).  In the case of
CL0023+0423, the velocity dispersion is 1497 km s$^{-1}$ after the
galaxies with velocity deviations of 3500 km s$^{-1}$ or greater are
excluded (and the clipping process concludes because subsequent
deviations are less than $3\sigma$).  However, the galaxy distribution
for CL0023+0423 shows clear bimodal structure in redshift space with
peaks at $z = 0.8274$ (7 galaxies) and $z = 0.8453$ (17 galaxies),
corresponding to a cosmologically-corrected velocity difference of
$2922 \pm 216$ km s$^{-1}$ (see Figure~\ref{fig-velhist}).  These
peaks are separated on the sky as well (see
Figure~\ref{fig-cl00targ}).  If we identify all CL0023+0423 galaxies
with negative velocity offsets as belonging to separate group, we then
find that the $z = 0.8274$ system has a dispersion of 158 km s$^{-1}$
and the $z = 0.8453$ system has a dispersion of 415 km s$^{-1}$. Table
6 gives the mean redshifts and dispersions for the clusters along with
their errors.

The CL0023+0423 system highlights one of the inherent difficulties of
studying cluster kinematics at high redshift -- one requires at least
15 cluster members and, ideally, more before an accurate estimate of
the velocity dispersion can be obtained.  Figure~\ref{fig-veldisp}
shows how the velocity dispersion for each cluster changes as the
extreme outliers are sequentially excluded.  The specific dispersion
values discussed above are highlighted by arrow marks. The 95\%
confidence limits on our dispersions always overlap the adjacent value
had we stopped the clipping process one step earlier or later.

\subsection{Mass Estimation}

A number of estimators are available for determining masses of
virialized systems given a set of galaxy positions and redshifts. The
pairwise mass estimator is defined as

\begin{eqnarray}
M_{PW} = {{3\pi}\over{G}}\sigma_{1}^{2}R_h
\end{eqnarray}

\noindent where $\sigma_1$ is the observed radial (1D) velocity
dispersion and $R_h$ is the mean harmonic radius, defined as

\begin{eqnarray}
R_h^{-1} = \left( {{N(N-1)} \over 2} \right)^{-1} \sum_{i<j} {1 \over {|r_i - r_j|}}
\end{eqnarray}

\noindent where $r_i$ and $r_j$ are the coordinates of the $i$th and
$j$th galaxies and $N$ is the total number of galaxies.  The mean
harmonic radius tends to overweight close pairs but has the advantage
that one does not need to specify a cluster center.  Two estimators
which tend to give somewhat more robust and accurate mass estimates
are the projected mass estimator (\cite{bt81}; \cite{htb85}) and the
ringwise mass estimator (\cite{carl1996}).  The projected mass is
defined as

\begin{eqnarray}
M_{PM} = {10.2 \over {G(N-1.5)}} \sum_i (\Delta v_i)^2 R_{i}
\end{eqnarray}

\noindent where $\Delta v_i$ is the velocity offset of the $i$th
galaxy from the mean cluster redshift and $R_{i}$ is its projected
distance from the cluster center.  The ringwise mass is defined as

\begin{eqnarray}
M_{RW} = {{3\pi}\over {2G}} \sigma_1^2 R_{rw}
\end{eqnarray} 

\noindent where $R_{rw}$ is

\begin{eqnarray}
R_{rw}^{-1} = N^{-2} \sum_{i<j} c(k)/({\pi \over 2}[R_i + R_j])
\end{eqnarray}

\noindent where $R_i$ and $R_j$ are the projected distances from the
cluster center of the $i$th and $j$th galaxies, $k = \sqrt{4 R_i R_j /
(R_i + R_j)^2}$, and $c(k)$ is the complete elliptic integral of the
first kind (\cite{press}). The reader is referred to Carlberg \etal
(1996) for further details.

For completeness, we compute mass estimates using all three methods
and present the results in Table 6, along with the corresponding
$1\sigma$ errors.  In all cases, the pairwise mass estimate is the
smallest value owing to the high weight given to close pairs (which
tends to decrease the value of the harmonic radius and hence the
mass). The projected and ringwise mass estimates are often in better
agreement although the latter tends to produce the highest mass
estimates.

For each system we generate mass estimates within the central
250$h^{-1}$ kpc and 500$h^{-1}$ kpc (so long as there is a sufficient
number of galaxies within these bins) in addition to a mass estimate
based on all the data shown in Figure~\ref{fig-velhist}.  One obvious
concern in evaluating the mass estimates in Table 6 is whether or not
the clusters are indeed virialized. Recently, Small \etal (1998) have
shown that most virial mass estimators do a reasonable job even when a
bound system is not in equilibrium -- estimates remain unbiased with a
scatter of $\sim25$\% when $\delta\rho/\rho > 5$. In the worst case, a
virial mass estimate of a marginally bound (and, thus, {\it
un}virialized) system can be too high by not more than a factor of
2. The velocity histogram for CL1604+4304 is consistent with a
Gaussian distribution (the probability that the observed redshift is
drawn from a Gaussian distribution is 27.7\%).  For CL0023+0423, the
velocity histograms are not well modeled by a Gaussian (probabilities
$\simless 2.2$\%).
 
The velocity dispersion and mass of CL1604+4304 are comparable with
those in current epoch richness class 2 clusters. In the CL0023+0423
system, the high velocity component at $z=0.8453$ is significantly
less massive and would be better compared with current epoch richness
class 0 clusters, while the lower velocity component at $z=0.8274$ has
a dispersion and mass which are comparable to local groups of galaxies
(Zabludoff \& Mulchaey 1998).

CL1604+4304 has also been detected in X-rays by the ROSAT PSPC
(\cite{cast94}). Its X-ray luminosity is $L_x(0.1-2.4~{\rm keV}) =
2.70 \pm 0.58 \times 10^{43}~h^{-2}$ erg s$^{-1}$.  Because of the
large scatter in the local optical-X-ray relations, CL1604+4304 is
consistent with the $L_x-\sigma$ relation of nearby clusters
(\cite{mush97}). However, its X-ray luminosity is low for its velocity
dispersion and, therefore, the estimated cluster mass.  A similar
trend is observed in the Couch \etal (1991) optically-selected cluster
sample at intermediate redshift. \cite{bower97} have examined several
clusters from this sample in the range $0.38 \le z \le 0.51$ and find
that they also have higher velocity dispersions for a given X-ray
luminosity. This may indicate that the galaxies and the gas are not in
thermal equilibrium or that clusters at these earlier epochs are still
experiencing significant infall.

\subsection{Mass-to-Light Ratios}

We use our BVR imaging to generate mass-to-light ratios. The
luminosity in a given metric radius is

\begin{eqnarray}
L = f(M_{Lim},\phi(M))\left[ \left( \sum_i 10^{0.4(m_z - m_i + A + K(z))} \right) - L_{bg} \right]
\end{eqnarray}

\noindent where the sum is over all galaxies within the prescribed
radius, $m_i$ is the galaxy's apparent magnitude, $A$ is a correction
for galactic extinction, $K(z)$ is a k-correction, $m_z$ is a
zeropoint adjustment to make the units absolute solar luminosities
(and, thus, depends on the luminosity distance), $L_{bg}$ is a
background subtraction, and $f(M_{Lim},\phi(M))$ is a correction for
the unsampled faint end of the cluster luminosity function. We define
$f(M_{Lim},\phi(M))$ to be

\begin{eqnarray}
f(M_{Lim},\phi(M)) = 1 + \left( {{\int_{M_{Lim}}^{M_{Ref}} \phi(M) dM} \over
{\int_{M_{Bright}}^{M_{Lim}} \phi(M) dM}} \right)
\end{eqnarray}

\noindent where we have assumed a Schechter form for $\phi(M)$ with
$\alpha = -1.1$, $M_{Lim}$ is the absolute magnitude corresponding to
the survey flux limit at the redshift of the cluster, $M_{Ref} =
-10.2$, and $M_{Bright} = -26.5$. The background subtraction is based
on deep galaxy counts performed by Gardner \etal (1996), Metcalfe
\etal (1995) [for $B$-band]; Gardner \etal (1996), Smail \etal (1995)
[for $V$ band]; and Smail \etal (1995) [for $R$-band].  Our own $BVR$
data on two non-cluster fields agrees well with the counts from these
publications.  We use the SED from the tau0.6 model (see \S4.1) with a
color age of 2 Gyr, the approximate mean color age of the cluster
members (see Figures~\ref{fig-cl00agez} and \ref{fig-cl16agez}), to
compute the appropriate k-corrections for each passband.

Tables 7 and 8 show the M/L ratios for the two cluster fields as
functions of metric radius and passband. The uncertainties listed
reflect the errors in the estimates of both the mass and the
luminosity.  For CL0023+0423, we compute results based on the centroid
of the $z = 0.8453$ component only. We use galaxies brighter than 25 mag
in $B$, brighter than 24.5 mag in $V$, 
and brighter than 23.5 mag in $R$ for the cluster luminosity
computations. As indicated above, however, all luminosities
are corrected to reflect an integration to a common fiducial
absolute luminosity given by $M_{Ref}$.  The results in 
Tables 7 and 8 are based on $M^*_B = -19.4,\ M^*_V = -20.1,\ M^*_R = -21.0$,
which are values appropriate for $z = 0$ clusters. If we assume that
$M^*$ evolves as $M^*(z) \approx M^*(0) - z$ (\cite{CFRS}), the M/L
ratios increase by about 40\%, 20\%, and 14\% in $B$, $V$, and $R$,
respectively.  The increase is a result of $f(M_{Lim},\phi(M))$
decreasing as $M^*$ decreases (becomes brighter).

The similarity of the M/L ratios in these two clusters to those in low
$z$ clusters of similar richness suggests that cluster M/L ratios, at
least in the central 500$h^{-1}$ kpc, have not changed much since $z
\sim 0.8-0.9$.  The uncertainties are large however and are dominated
by uncertainties in the derived luminosity ({\it e.g.}, the
k-correction is a strong function of a galaxy's age and morphology which
makes it difficult to derive a precise luminosity without redshifts for
many more cluster members than we currently have available here).

\section{Comparison of Observations with Galaxy Evolutionary Models}

In this section we compare the photometric and spectroscopic
observations with synthetic models of galaxy evolution.  The
comparison of the models with the photometric observations yield what
we refer to as ``color ages'', while the comparison with the observed
spectral properties yield ``spectrum ages''.  Both of these ages
measure the time since the last period of major star formation.  We
have examined several different families of models. These families
differ in their star formation scenarios and, therefore, give best-fit
``ages'' to the observations which are different.  Consequently, the
ages derived from this model fitting are not physical values but only
parameters which characterize the data.  To convert these model ages
into real ages, we must choose, firstly, the family of models which
fits best the observations and, secondly, the appropriate metal
abundance. In this section, we have attempted to do this.  Thirdly, we
must estimate the amount of time between creation of the galaxy and
the first period of star formation.  This process is discussed in
Sects.\ 5 \& 6.

\subsection{Galaxy Evolutionary Models} 

In order to establish the star formation history of the field and
cluster galaxies, we compare the observed data to an appropriate set
of spectrophotometric evolution models. We choose to use the Bruzual
\& Charlot (1995) family of stellar evolutionary models.  These models
are constructed using stars with solar metal abundances.  This choice
has the advantage that there exists a large database of spectra
representing stars over the whole Hertzsprung-Russell diagram.  The
spectral energy distributions (SEDs) of the models fit very well the
energy distributions measured for nearby present-age elliptical
galaxies (\cite{bruz1993}), as well as galaxies in clusters with
redshifts near 0.5 (Oke, Gunn \& Hoessel 1996).  In a practical sense
the models are ideal since they include an absolute spectral energy
distribution with a spectral resolution and range which is similar to
that of our observations.  The spectra can be converted directly to AB
magnitudes, and broad-band AB values can be calculated (see Paper I).
We have chosen models with a Salpeter luminosity function (Salpeter
1955) and a maximum stellar mass of $125~{\rm M_\odot}$.  Comparisons
with models constructed with a Scalo luminosity function (Scalo 1986)
show no significant difference at the level of accuracy that we can
achieve.

We also need to choose the model with the most appropriate star
formation history.  The simplest model is to assume that there is a
large, initial burst of star formation after which the galaxy fades in
accordance with passive stellar evolution models.  These are called
ssp models by Bruzual \& Charlot.  The next simplest models are those
where star formation begins at $t=0.0$ and decreases exponentially
with a fixed time constant.  We have considered three such families of
models with time constants of 0.3, 0.6, and 1.0 Gyr; these models are
hereafter referred to as the tau0.3, tau0.6, and tau1.0 models,
respectively.

In Figure~\ref{fig-ABlognu} we have plotted the broad-band AB values,
that is ABB, ABV, ABR, and ABI, as a function of log$\nu$ where $\nu$
is the frequency in Hz for a series of tau0.6 models of various ages
ranging from 0 to 8 Gyr for a redshift of 0.897 (the redshift of the
cluster in CL1604+4304).  Over the age range from 0.05 to 6 Gyr, the
models predict rapid and dramatic color evolution.  However, there is
very little color sensitivity with age beyond 6 Gyr as demonstrated by
the similarity between the 6 and 8 Gyr curves.  The broken curve in
Figure~\ref{fig-ABlognu} shows the energy distribution at 10 Gyr.  It
is virtually the same as the 8 Gyr curve.  When the observed spectra
have energy distributions in this range, we can only determine a
minimum age. (The exact minimum age depends somewhat on the quality of
the observations.)  A similar family of curves can be plotted for the
other models and for other redshifts.  At a given redshift the
predicted SEDs are very similar (provided that $z \ge 0.5$) except
that the age at any color is somewhat different.

The Bruzual \& Charlot models are also capable of generating detailed
spectra.  A sample of tau0.6 model spectra for different ages is
shown in Figure~\ref{fig-modspec}.  Here, the relative AB magnitudes
are plotted against wavelength in \AA.  Within the age range shown in
Figure~\ref{fig-modspec}, the most significant change with time, apart
from the overall changes in color discussed above, is the variation in
the strengths of the metal lines relative to the strengths of the
Balmer lines. 

\subsection{Broad-Band Energy distributions}

For each galaxy we have the four observed values ABB, ABV, ABR, and
ABI.  In addition, for any particular family of models (e.g.\ the
tau0.6 models) we have the corresponding values of AB for each model
age.  We can now characterize the observations by finding the model
with the age that best fits the observations.  Since the relationship
between AB and log $\nu$ for the models is not quite linear (see
Figure~\ref{fig-ABlognu}), the fitting is done using a maximum
likelihood technique. The broad-band energy distribution of each
galaxy is then characterized by a single parameter which we will refer
to as the color age.  Since the uncertainty in the fitted color age is
directly related to the uncertainty in the fitted linear slope, we use
this relation to derive uncertainties in the color ages.  The
$1\sigma$ uncertainties are typically $0.1-0.4$ Gyr.

The above procedure has been carried out for several different
families of models, including ssp, tau0.3, and tau0.6 models.  As
mentioned above, the resulting color ages are different for different
models.  Therefore, the color age should not be interpreted as a real
age since some initial epoch.  Rather it is a representative parameter
which describes the energy distribution of a particular galaxy.
Although the general progression of colors and spectral features with
age are qualitatively similar for the ssp and various tau models,
there are quantitative differences between them. Specifically, the
relative strengths of spectral features are not coupled to the colors
independently of the family of models.  In an attempt to determine
which model is more suited to our data, we have used the observational
data for the confirmed cluster members. Firstly, we compare their
observed spectral energy distributions (that is, the four AB values)
with each model SED at the appropriate redshift in order to determine
which model fits best.  Using $\chi^2$ as a measure of the fit, we
find that the tau0.3, tau0.6, and tau1.0 models provide equally good
fits and are, in general, slightly better than the ssp models.  The
differences are small enough that we conclude that the color fits
cannot be used to discriminate between models.

For reasons given in the next section, we have adopted the color ages
provided by the tau0.6 family of models.  The color age and the
estimated error for each galaxy energy distribution, as derived using
the tau0.6 model, are listed in columns 8 and 9 of Tables 2 and 3.
Occasionally, the age is based on data from only three (or, in rare
cases, two) passbands.  As evident from Figure~\ref{fig-ABlognu} and
the derived ages listed in Tables 2 and 3, most of the fits occur in a
domain where the models are changing rapidly with age. In a few cases
the energy distributions imply ages which are sufficiently long that
the energy distribution is insensitive to the model age.  In these
cases only a minimum model age can be derived (see Sect.\ 4.1).

\subsection{Emission Line Strengths}

In galaxies where star formation is ongoing or has occurred until very
recently, the observed emission lines are primarily generated within
HII regions.  In a radiation limited case one has the classical
Stromgren sphere.  Every photon below the Lyman limit is absorbed by
the ISM and eventually produces a Balmer line photon which can be
observed in the visible spectrum.  This argument was used by Zanstra
(1931) to infer the flux beyond the Lyman limit of a star by measuring
the Balmer line strength.  This method has also been used by Kennicutt
(1983) to calculate H$\alpha$ equivalent widths for studies of spiral
galaxies.  The excess energy of the original photon above the binding
energy of HI is converted into kinetic energy in the surrounding
gaseous nebula and, in equilibrium, reappears primarily in the
forbidden lines of [OII], [OIII], etc.  One complication which can
arise is that absorption of visual and UV photons by dust and the
subsequent re-radiation of this energy at IR wavelengths can upset the
simple equilibrium of photons.

If dust absorption is negligible, one can readily calculate the
equivalent width of a Balmer emission line and estimate the equivalent
widths of the $[{\rm OII}] + [{\rm OIII}]$ emission lines, provided
that the UV flux below the Lyman limit and the spectral energy
distribution of the radiation in the visual is known. The Bruzual \&
Charlot models provide the necessary stellar flux information both
below the Lyman limit and in the visible regime.  In doing this
calculation, we have represented the Lyman continuum flux by the flux
at the Lyman limit multiplied by an appropriate bandwidth. This
bandwidth was determined by integrating the model photon flux from the
Lyman limit down to 250 \AA\ and dividing by the flux at the Lyman
limit.  Our computations were done in detail only for the tau0.6 model
with an age of 1 Gyr.  This is adequate since the UV energy
distributions relative to the flux at the Lyman limit are very similar
for tau0.6 models of different ages.  Assuming that 66\% of the Balmer
photons are H$\alpha$ and 17\% are H$\beta$, the equivalent widths in
\AA\ for $\rm H\alpha$ and $\rm H\beta$ are found to be:

\begin{eqnarray}
  EW(\hbox{\rm H}\alpha) &=& 2483 \frac{f_{\nu}(912)} {f_{\nu}(6563)} \\
  EW(\hbox{\rm H}\beta) &=& 472 \frac{f_{\nu}(912)} {f_{\nu}(4861)}  
 \end{eqnarray}

The calculation of the intensities and equivalent widths of the [OII]
and [OIII] lines are much more complex.  We have, therefore, used the
models of McCall, Rybski \& Shields (1985) to provide the ratios of
intensities of these lines to that of H$\beta$.  They interpret the
variations in line strengths from one HII region to another to be due
to oxygen abundance differences.  In our case we are looking at a
entire galaxy and presumably a large number of HII regions with
different oxygen abundances.  Consequently, we adopt a mean value for
the oxygen abundance; their observations suggest that an abundance of
1.3 times the solar abundance is representative.  Their calculated
model for this case gives the desired line-intensity ratios (see Table
9).  With the equivalent width of H$\beta$ as calculated from Equation
9 and the known energy distribution of the galaxy model, the
equivalent widths of [OII] and [OIII]$\lambda 5007$ can readily be
calculated.  The results for a series of tau0.6 models of different
ages are given in Table 9.

Searle (1971) and Baldwin \etal (1981) have also measured emission
line ratios for many extragalactic HII regions.  The model that we
have chosen represents a mean of these data quite well.  We have
equivalent widths of [OIII]$\lambda5007$ and H$\beta$ for some of our
objects (see Tables 2 and 3).  Although they usually have very large
errors because they lie in the wavelength range 8000\AA\ -- 1$\mu$
where the spectra are noisy, they do cluster around the values selected in 
Table 9.

Since the strength of the [OII] emission line is an indicator of the
star formation rate, its observed strength provides a guide as to
whether an ssp or a tau model is a more appropriate model for making
age estimates.  If we fit an ssp model to the observed colors, we
derive minimum possible ages.  The strength of the rest $[{\rm OII}]$
equivalent width versus the ssp color age is shown in
Figure~\ref{fig-eqwcolssp} for the $z = 0.8274$ and $z = 0.8453$
objects in the CL0023+0423 field and the $z = 0.8290$ and $z = 0.8967$
objects in the CL1604+4304 field.  The calculated emission-line
equivalent widths indicate that for ssp models the ionizing flux
decreases so rapidly with time that an emission line should not be
visible after a model age of 0.02 Gyr. The fact that we do see
emission lines in objects which have ssp model ages of at least one
Gyr indicates that star formation must have continued for at least
that long.  This implies that strict ssp models are not valid for a
majority of the galaxies; that is, these models do not produce enough
star formation as a function of age.

Instead, consider a tau model which includes an exponentially
decreasing star formation rate.  Figure~\ref{fig-eqwcoltau0.6} is a
plot similar to that in Figure~\ref{fig-eqwcolssp}, except that the
color ages are derived from tau0.6 models.  Table 9 lists the
calculated equivalent width of the [OII] line for various galaxy ages
in a tau0.6 model. These values are plotted as the solid line in
Figure~\ref{fig-eqwcoltau0.6}.  This curve is essentially a negative
exponential with a time constant of 0.6 Gyr since the strength of
the [OII] line is controlled by the nearly instantaneous star formation
rate.  Within a factor of 2, the calculated curve represents very well
the observed [OII] emission line strengths.  The two points which are
far to the right of the curve correspond to galaxies with poor spectra
and uncertain redshifts.  A similar plot with tau0.3 or tau1.0 models
would show that the calculated curve fits the observations equally
well. Therefore, we conclude that an exponentially decaying SFR is
required to reproduce the amount of [OII] emission observed in our
high-redshift sample.  Such a star formation rate is most readily made
with a tau model although models with a series of bursts which
decrease in intensity or frequency with time would also work.

\subsection{Analysis of the 4000\AA\ Break and the Balmer Jump}

We examine the relation between the break measure $J$ (defined in
Sect.\ 2.1.3) and the color age as determined from the evolutionary
models. Since the observed jumps are noisy for individual objects, a
better comparison of the observational data with the models can be
made by grouping galaxies together and averaging their spectra.  The
groups are selected using proximity in redshift and in color age, now
crudely defined as young ($\leq$2 Gyr), medium (2-3 Gyr), and old
($\geq$3 Gyr).  The jump ($J$) and the wavelength where the jump
($\lambda_J$) is centered of each average spectrum is calculated.  The
value of D(4000) is also determined.  In
Figure~\ref{fig-sspjump}, we plot the resulting $J$ values (top
panel), the wavelengths where the jump is centered $\lambda_J$ (middle
panel), and the conventional 4000 \AA\ break D(4000) [bottom panel]
versus the logarithm of the color age as derived from fittings to the
ssp models.  Figure \ref{fig-taujump} shows a similar plot using the
color ages derived from fitting to the tau0.6 models. We list the
relevant values of this analysis for the tau0.6 models in Table
10. The group and the number of members used in each average spectrum
are listed in columns 1 and 2. The mean color ages, the resulting
jumps and the central wavelengths, and D(4000) are listed in columns
3, 6, and 7, respectively.

We can use the synthetic spectral energy distributions of the Bruzual
\& Charlot models described above to derive the theoretical relation
between the break measure $J$ and the color age. The resulting
relation for the ssp family of models is shown as the solid curve in
the top panel of Figure~\ref{fig-sspjump}.  Between 0.1 and 0.3 Gyr
the jump $J$ is the Balmer jump and corresponds to the dominant stars
in the galaxy changing from B- to A-type stars.  From 0.3 to 1.0 Gyr
the dominant stars are later than A0, and the Balmer jump declines. From
1.0 to 4.0 Gyr the jump is the traditional 4000 \AA\ break
dominated by the CaII H and K lines and other strong metal lines.  The
wavelength where the jump occurs for the same ssp models is shown by
the solid line in the middle panel of Figure~\ref{fig-sspjump}. At
left it is at 3875 \AA\ and is due to the Balmer jump.  At right it is
at 4000 \AA\ and is the traditional 4000 \AA\ break.  In the bottom
panel, the solid curve represents the theoretical relation between
D(4000) and color age for the ssp models.  It should be noted that
between 1.0 and 10.0 Gyr our value of $J$ is nearly constant, while
the traditional 4000 \AA\ jump D(4000) is still increasing ({\it e.g.},
see Figure 13 of Bruzual \& Charlot 1993).  At lower ages D(4000)
decreases rapidly as metal line absorption decreases, while $J$
actually increases because the Balmer Jump is large.
 
We have generated these relations for the tau0.6 family of models as
well. The solid curves in Figure~\ref{fig-taujump} show these
relations.  As expected, the results are similar, but the transitions
occur at later color ages.  The interpretation of the curves is
analogous to that for the ssp models given above.  A comparison of
Figures~\ref{fig-sspjump} and \ref{fig-taujump} shows that the 
$J$ data fit the tau0.6 models better than the ssp models.  However, for
galaxies with smaller color ages, the measured jump is smaller than
predicted by the models. This trend suggests that a tau model with a
decay time constant even longer than 0.6 Gyr may be more appropriate.
Indeed, the fit using a tau1.0 family of models does bring the
observed points closer to the analytic curve.  To summarize, the jump
as measured by $J$ indicates that tau models are in general more
appropriate than ssp models for galaxies with young color ages.  For
the oldest galaxies, either family of models is satisfactory.

The Bruzual \& Charlot models assume solar metal abundances; however,
we need to determine whether such models are appropriate for our data.
We can do this by comparing the values of D(4000) predicted by various
Bruzual \& Charlot models to the observed values for the oldest
galaxies in our sample.  These galaxies are defined in lines 1, 4, 7
and 10 of Table 10.  (We omit those galaxies in line 15 as their
redshifts are $z <0.5$.)  Averaging the data by the total number of
galaxies in each line, we find that the observed--to--model ratio of
D(4000) is $0.92 \pm 0.05$ for the ssp models and $0.91 \pm 0.06$ for
the tau0.6 models.  Here, we have assumed that the color ages which
are listed as lower limits are actually 1 Gyr older; however, this
changes the ratio by only 1--2\%.  The equivalent solar metal
abundance models of Worthey (1994) give values of D(4000) which are
about $7\%$ smaller than those of Bruzual \& Charlot.  Consequently,
the observed--to--model ratios for the ssp and tau0.6 models of
Worthey would be 0.98 and 0.97, respectively. We, therefore, conclude
that our observations are consistent with solar abundances.
  
Worthey (1994) finds that D(4000) changes by about 12\% when the
abundance changes by 0.25 in the logarithm relative to solar.  At the
same time the overall colors of Worthey's models also change by
substantial amounts.  Unfortunately, we are unable to make a detailed
comparison to the models as the rest wavelength range defined by our
colors is much shorter than that of Worthey; however,
it is evident that the change in color with metal abundance
approximately compensates for the change in D(4000).  In other words,
an ssp model with a metal abundance of -0.25 and an age of 1.5 Gyr
gives the same color and D(4000) as a solar abundance model with an
age of 3.0 Gyr.  Since it is not possible to measure the metallicity
directly with our data, we cannot separate age and metallicity.
In other observations, Hammer \etal (1997) find a
significant population of field galaxies at $z = 0.7$ with small
D(4000) indices which suggest lower than solar metal abundances or
younger ages.

\subsection{Ages Derived from Spectral Absorption Features}

In addition to the color ages, a spectroscopic age estimate is
obtained by comparing the equivalent widths of the seven selected
spectral absorption features $\lambda$3835, H8, K, H, H$\delta$,
G-band, and H$\gamma$ with those for the tau models. The best-fit
spectroscopic-based age is then just that of the model with the
minimum $\chi^2$ value.  The resulting ages, however, are quite
uncertain because the individual equivalent widths are themselves very
uncertain. Therefore, we have decided not to estimate spectrum ages
for individual galaxies, but rather to use the average spectra for the
groups already defined in Table 10.  For each averaged spectrum, an
age is obtained from the maximum likelihood fit.  In addition, we have
also estimated a spectrum age by visually comparing the averaged data
with the model spectra.  The best fit was judged by eye, and the
corresponding model age was noted.  The visual estimate of the
spectrum age and the best-fit spectrum age (along with the $\chi^2$
value) are listed in columns 4 and 5 of Table 10.  A comparison of the
color ages and spectrum ages shows reasonable agreement although the
scatter is large. In these plots of color age versus spectrum age the
scatter is lower for the visually determined spectrum ages than for
the $\chi^2$ fitted ages, probably because the eye allows for
inconsistencies in the apparent line strengths. As discussed in the
previous section, we are unable to distinguish between spectrum age
and metallicity for the reddest objects.

\subsection{The Choice of Model}

In Sect.\ 4, we have discussed the broad-band energy distributions,
the emission line equivalent widths, the Balmer or 4000 \AA\ jump $J$,
the 4000 \AA\ jump D(4000), and the absorption features in the galaxy
spectra in terms of their consistency with the solar abundance Bruzual
\& Charlot families of models.  Only the [OII] equivalent width and
possibly the jump $J$ can serve to discriminate among the families of
models.  These spectral features, as observed, clearly rule out ssp
models, but present good overall fits for the tau models.  The color
and spectrum ages also depend strongly on the assumed metallicity.
Because there is no way to measure this with our data, we have chosen
to calculate ages based on tau0.6 models with solar abundances (see
Tables 2 and 3).
  
It should again be emphasized that the color and spectrum ages should
not be interpreted as real ages since some initial epoch.  The ages
derived depend very much on the family of models and the assumed
metallicity.  That is, derived ages for tau1.0 models are older than
those for tau0.6 models which are, in turn, older than the ssp
models. The color age principally describes the overall color of the
galaxy. To emphasize this, we note that the ages of galaxies within a
single cluster show a large spread even though they are probably
coeval (see Sect.\ 5). The cluster member ages could be made identical
by using a different star-formation-rate decay time or a different
metallicity for each individual galaxy.  Since the calculated color is
actually determined in the rest violet and ultraviolet regimes which
is dominated by the more recently formed blue main sequence stars, the
color age is really a measure of how long ago star formation was
important.  The [OII] equivalent width, on the other hand, represents
the present star formation rate relative to this past star formation
history.

\section{Cluster Data}

In this section, we examine the relationship between the estimated
ages and other galaxy properties for the cluster and group members in
the two fields. We use the ages derived from the colors, as they are
considerably more precise than those derived from spectral features
(see Sect.\ 4).

As noted in Table 6, the CL0023+0423 field contains two poor clusters
or groups of galaxies at $z = 0.8274$ and $z =0.8453$ with 7 and 17
known members, respectively.  In CL1604+4304 there is a cluster with
$z = 0.8967$ with 22 known members.  There is also a system of 8
galaxies with $z = 0.8290$ which is most likely a sheet of galaxies
given its distribution on the sky (see Figure~\ref{fig-cl16targ}).
The rest [OII] equivalent widths are plotted against the tau0.6 model
color ages for these galaxies in Figure~\ref{fig-eqwcoltau0.6}.  The
figure shows that the [OII] equivalent widths are small for almost
every galaxy with a color age greater than 2.5 Gyr.  As expected, this
implies that the reddest galaxies do not currently have significant
star formation.  We now plot in Figures~\ref{fig-colabs0023} and
\ref{fig-colabs1604} the tau0.6 color age versus the absolute
magnitude $M_{ABB}$ (see Sect.\ 2.2.2) for the confirmed group/cluster
members in CL0023+0423 and CL1604+4304, respectively.  The reddest
galaxies not only have very weak or absent [OII] emission, but they
are the most luminous galaxies in the cluster.  A similar result was
found for field galaxies (Songaila \etal 1994).  An obvious
interpretation is that very luminous galaxies (and presumably very
massive ones) begin star formation quickly and exhaust their gaseous
component with an exponentially decreasing time constant of about 0.6
Gyr.  Less massive galaxies convert gas to stars more slowly with an
exponentially decreasing time constant which can be longer than 0.6
Gyr. 

This correlation between age and luminosity is obvious in both groups
of the CL0023+0423 system (Figure~\ref{fig-colabs0023}). In the high
velocity component at $z = 0.8453$, there are few luminous galaxies;
correspondingly, there are only 2 (out of 17) galaxies which are red
and old. In the low velocity component at $z = 0.8274$, there are
relatively more luminous galaxies and also relatively more (4 out of
7) cool, red objects.  The location of these old, red galaxies in the
cluster is also of some interest.  In CL0023+0423 the two old galaxies
of the $z = 0.8453$ system are close together in the densest part of
this group.  The four red objects in the $z = 0.8274$ system are also
clumped together near the center of the group (see
Figure~\ref{fig-cl00targ}).  In the cluster of CL1604+4304 which
contains the most old, red members, these galaxies are strung out in a
line in the inner half diameter of the cluster (see
Figure~\ref{fig-cl16targ}).

The only objects which can potentially yield model ages which are
actual epochs are the reddest galaxies in the clusters.  These
galaxies have tau0.6 color ages of $\simgreat 3.8$ Gyr in
Figures~\ref{fig-colabs0023} and \ref{fig-colabs1604}.  These galaxies
also make up the ``red locus'' observed in the color-magnitude
diagrams of low and intermediate redshift clusters (e.g.\ Butcher \&
Oemler 1984; Arag\'on-Salamanca et al.\ 1991,1993; Stanford,
Eisenhardt \& Dickinson 1995,1997).  HST observations indicate that
these galaxies are typically early-type (elliptical or S0)
galaxies. Similarly, we find that the red group/cluster members in our
two fields are also classified as early spiral or early-type galaxies
(see Paper III).  These galaxies are fit well by either a single
initial burst ssp model or by a tau model, such as tau0.6, where the
star formation decay rate is rapid enough so that essentially no star
formation has occurred within the last Gyr. Choosing those galaxies
with accurate photometry and tau0.6 color ages of $> 3.8$ Gyr, we find
that there is one such object in the $z=0.8274$ group in the
CL0023+0423 field, while there are five in the $z=0.8967$ cluster and
one in the $z=0.8290$ structure in the CL1604+4304 field.  These
galaxies are listed in the footnote to Table 11 along with their
galaxy classification where available (see Paper III).

The reddest of these galaxies have energy distributions for which the
sensitivity to age is rapidly decreasing.  All seven objects have been
averaged to produce mean values of AB.  Since all the galaxies have
fairly similar measured apparent magnitudes, each galaxy has been
given a weight of unity.  The mean values of AB are listed in Table
11.  The mean ABs have been fitted in the usual manner with solar
metallicity ssp, tau0.6, and tau1.0 models. The ABs for the best
fitting models are listed.  The resulting ages and their estimated
uncertainties (assuming solar metallicities) are also given in Table
11 along with the model values of AB.  In all cases the fits are by no
means perfect since the observations and the models have distinctly
different curvatures.  As expected the ssp ages are the shortest.  The
tau0.6 and tau1.0 model ages are longer and show that the derived age
is very sensitive to the particular model.  In a standard cosmology of
$H_o = 65$ and $q_o = 0.1$, the age of the universe is 12.8 Gyr, and
the age at $z = 0.90$ is 5.8 Gyr.  The ssp models, which we have
already seen are not appropriate (see Sect.\ 4), give a time for the
formation of the first stars at $2.5 \pm 0.5$ Gyr after creation or $z
\sim 3$.  The tau0.6 models, which appear to fit the observed energy
distributions and spectral features (see Sect.\ 4), give $0.8 \pm 0.5$
Gyr ($z \sim 5$) which allows approximately 1 Gyr for the first stars
to form.  The tau1.0 models give ages which are much too long for this
cosmology. If we had used a metallicity of $-0.25$, the ssp 
ages would have been about a factor of two smaller; the tau0.6 and tau1.0 
ages would have been lower by less than a factor of two because tau models
mix in young hot stars in which metallicity effects are smaller.

Most of the member galaxies with absent or weak [OII] emission have
high color ages; however, there are a number of galaxies which appear
to be young although they have very little [OII] emission.  Their blue
color and lack of [OII] emission indicate recent but no current star
formation. Since star formation most likely occurs in bursts which
exponentially decrease in intensity or frequency with time, these
galaxies are just cluster members where the most recent burst occurred
at least 0.1 Gyr ago.

\section{Brightest Cluster Galaxies}

The brightest cluster galaxy (BCG) in the CL0023+0423 system is in the
low-velocity $z=0.8274$ group. It is Keck \#2055 and has $M_{ABB} = -
21.63 + 5~{\rm log}~h$.  This galaxy is classified as an Sa(pec) and
is shown in Figure 15 of Paper III.  This group has an Abell richness
class (RC) of 0 at best but is actually most similar in nature to a
small group of galaxies (Zabludoff \& Mulchaey 1998).  The brightest
cluster galaxy in the CL1604+4304 cluster at $z = 0.8967$ is Keck
\#2855. It has $M_{ABB} = - 21.91 + 5~{\rm log}~h$ and is classified
as an Sa (see Figure 16 in Paper III).  This cluster is quite rich,
corresponding to an Abell richness class of about 2.  Since not all of
the galaxies in the fields were spectroscopically observed, we have
checked to see if any other galaxy near the cluster (or group) centers
which had not been observed could be the brightest cluster members.
No candidates were found in either field.  From (1) the color age of
the galaxy, (2) the relation between color age and $B-V$ for the
best-fit tau0.6 model, and (3) the conversion between ABB and $B$
given in Equation 3 of Paper I, we can derive the absolute rest
luminosities of $M_B$ and $M_V$.  These values for the two BCGs are
listed in columns 4 and 5 of Table 12.

We compare the absolute luminosities of our BCGs to those in the
nearby universe by examining the results of Schneider, Gunn \& Hoessel
(1983).  For this comparison, we will compare the BCG in CL0023+0423
to a present-day ${\rm RC} = 0$ BCG and the BCG in CL1604+42304 to a
present-day ${\rm RC} = 2$ BCG.  Schneider, Gunn \& Hoessel (1983)
have measured and analyzed the brightest cluster galaxies in 84 Abell
clusters.  Using their definition of the reduced absolute magnitude
(RAM) and converting their results to our default standard cosmology
of $H_0=100$ and $q_0=0.1$, we find that the brightest cluster galaxy
in a richness class 0 and 2 cluster at low redshift has a luminosity
of $M_g = -21.13$ and $-21.41$, respectively (see column 7 of Table
12).  We can convert the absolute luminosity $M_g$ to $M_B$ and $M_V$
using the conversion between the two passbands of $g$ and $B$, and the
typical color of $B-V = 0.97$ for the brightest cluster galaxies which
are given in Schneider, Gunn \& Hoessel (1983).  The resulting values
of $M_B$ and $M_V$ for a BCG in a low-redshift, richness class 0 and 2
cluster are given in columns 8 and 9, respectively, of Table 12.
Based on the data presented in Table 12, we find that the brightening
between $z = 0.10$ (the mean redshift of the Schneider, Gunn \&
Hoessel sample) and $z \sim 0.90$ is $\Delta M_B = 1.02$ and $\Delta
M_V = 0.85$.  These results are identical for both the group in
CL0023+0423 and the cluster in CL1604+4304; however, the accuracy of
the magnitude differences is at best $0.2-0.3$ mag.  This brightening
is independent of the value of $H_0$ chosen.  However, it does depend
on $q_0$; that is, increasing $q_0$ from 0.1 to 0.5 would decrease the
brightening by 0.37 mag.

We can use our best-fit tau0.6 models of these galaxies to calculate
the expected brightening between $z=0.10$ and $z=0.90$.  The results
depend on both $q_0$ and $H_0$.  We also assume that there is an
interval of 1 Gyr between the time of creation and the time when star
formation begins.  For our standard values of $H_0= 100~{\rm
km~s^{-1}~Mpc^{-1}}$ and $q_0=0.1$, the brightening is 0.9 and 0.7 mag
for $M_B$ and $M_V$, respectively.  For $H_0 = 65$ and $q_0 = 0.1$ the
predicted brightening is 0.8 and 0.7 mag, respectively.  Given $H_0 =
65$, increasing $q_0$ to 0.5 increases the predicted brightening to
1.1 and 0.9 mag for $M_B$ and $M_V$, respectively.  Since the standard
deviations in the brightening of $M_B$ and $M_V$ are $\sim$0.2 mag, we
find a formal value of $q_0= 0.25 \pm 0.25$ for $H_0=65$ provided the
metallicity is solar.  Similarly, Arag\'on-Salamanca \etal (1993) have
looked for luminosity evolution in the brightest cluster galaxies
using the infrared $K$ band.  They have examined clusters at redshifts
of $z \simless 0.90$, including CL1604+4304, but find no brightening.
Our best-fit tau0.6 models predict that in $K$ the luminosity
evolution should be about 60\% of the luminosity evolution observed in
$B$.

\section{Galaxies at $z > 0.9$}

\subsection{Luminosities}

We have examined the highest redshift galaxies in our sample for those
which may be of particular interest because of their spectral nature
or luminosity. At these high redshifts, the luminosity brightening
depends very strongly on $H_0$.  Assuming a time of 1 Gyr for the
first stars to form and $q_0 = 0.1$, the tau0.6 models predict a
luminosity brightening in the brightest cluster galaxies of 1.6 mag
for $H_0 = 100$ and 0.5 mag for $H_0 = 65$ between $z=0.90$ and
$z=1.50$.  For a reasonable Hubble constant $H_0 = 65$, almost all of
the galaxies that we observed in this redshift range are no brighter
than the predicted brightest cluster galaxies (see Sect.\ 6).
However, there are a few interesting exceptions.  At a redshift of $z
= 1.0908$ Keck \#3152 in the CL0023+0423 field has an absolute
luminosity of $M_{ABB} = -22.96$, nearly 1 mag brighter than a BCG at
this redshift.  This object, however, is a quasar or AGN.  In
addition, Keck \#1023 ($z = 1.2316$) in the CL1604+4304 field is about
0.2 mag brighter than expected; however, otherwise appears normal.  Of
special interest, Keck \#3560 ($z = 1.5029$) in the CL0023+0423 field
and Keck \#2858 ($z = 1.3757$) in the CL1604+4304 field are 0.4 to 0.6
mag brighter than the expected luminosity of a BCG at these redshifts.
These two objects also distinguish themselves by showing strong UV
absorption lines. We discuss the implications of the ultraviolet lines
of these two galaxies below.
 
\subsection{Ultraviolet Absorption Lines}
  
The MgII doublet at 2800 \AA\ becomes visible in the models at an age
of 0.5 Gyr and then strengthens rapidly (see Figure
\ref{fig-modspec}).  Spectra taken with the Keck 10m telescope of the
galaxy 53W091 at $z = 1.55$ show this feature clearly (Dunlop \etal
1996).  In addition, Cowie \etal (1995) observed these lines in a Keck
spectrum of a $z = 1.614$ galaxy. The MgII doublet is often seen in
nearby star-forming and Seyfert 2 galaxies (\cite{sto95};
\cite{kin96}). It is visible in all the spectra shown in Figure
\ref{fig-samplespectra}.  In the older and redder objects this doublet
is accompanied by numerous other absorption lines, the most prominent
of which are 2881 \AA\ due mainly to SiI and 2852 \AA\ due mainly to
MgI. Some of these lines can also be seen in the spectra of
Figure~\ref{fig-samplespectra}.  These features are strong in stars
such as the sun and $\alpha$ Canis Minoris (Morton \etal 1977).  At
still shorter wavelengths, one finds the FeII resonance absorption
lines at 2586.2 \AA\ and 2600.2 \AA.  Keck spectra of a $z = 1.614$
galaxy obtained by Cowie \etal (1995) and a $z = 1.6036$ galaxy
obtained by Koo \etal (1996) show these lines.  These lines are often
seen in nearby star-forming and Seyfert 2 galaxies with equivalent
widths of $1-8$ \AA\ for the combined $\lambda$2586, 2600 pair
(\cite{sto95}; \cite{kin96}). In our spectra they can be seen only
when the redshift is well above 1.0, and the spectra have high
signal-to-noise ratios. We detect these features in four objects in
CL0023+0423 and three in CL1604+4304.
 
As shown in Figures~\ref{fig-obj3560} and \ref{fig-obj2858}, two
galaxies, Keck \#3560 in CL0023+0423 at $z = 1.5029$ and Keck \#2858
in CL1604+4304 at $z = 1.3757$, not only have the two FeII resonance
lines and the MgII 2800 \AA\ feature, but they also have all the other
resonance lines of FeII in multiplets UV1 to UV5 (Fuhr \etal 1988).
Three of these lines were identified by Cowie \etal (1995) in the
$z=1.614$ galaxy.  All of these resonance lines are seen in HST
spectra of high-redshift quasar including H821+643 and are caused by
interstellar absorption in the Galaxy (Bahcall \etal 1992; Savage
\etal 1993).  Because in our case the lines are being created in
distant galaxies, the question arises whether these lines originate in
the atmospheres of the stars in the galaxy or whether they are of
interstellar origin.  The strongest argument for their interstellar
origin is that only the true resonance lines are seen.  Lines from
slightly excited levels, such as $\lambda$2631 from multiplet UV1, are
not seen.  On the other hand, the strength of these features argues
against an interstellar origin.  The rest equivalent widths of the
strongest interstellar FeII lines in the Galaxy are 1 to 2 \AA\
(Bahcall \etal 1992; Savage \etal 1993).  In Keck \#3560
(CL0023+04323) the rest equivalent widths of $\lambda2585$ and
$\lambda2601$ are 3.9 and 3.0 \AA, respectively, while in the rather
noisy spectrum of Keck \#2858 (CL1604+4304) the combined pair has a
total equivalent width of 13 \AA.  In our distant galaxies we are
looking at a whole galaxy, and the range in velocity along the
line-of-sight can be somewhat larger than in our own Galaxy.  There
appears, however, to be no way to generate sufficiently high
velocities along the line-of-sight to produce interstellar lines with
the large rest equivalent widths that we observe in Keck \#3560 and
\#2858. In addition, these FeII lines are commonly seen in stars which
should be present in these high-redshift galaxies.  The strengths that
we observe are comparable with the line strengths in star-forming and
Seyfert 2 galaxies (\cite{sto95}; \cite{kin96}).  These lines are also
strong in A-type stars such as Vega (Rogerson 1989) and Sirius
(Rogerson 1987) and in early F-type stars such as $\alpha$ Aquilae
(Morton \etal 1977).  
See also IUE spectra of B, A, and F type stars (Wu {\it et al.} 1991).
We, therefore, conclude that the UV FeII and
MgII lines must in fact be generated in the atmospheres of the young,
hot stars in the galaxies.

\section{Conclusions}

In this paper, we have presented the results from a deep photometric
and spectroscopic survey of two fields which are centered on the
candidate, high-redshift clusters of CL0023+0423 at $z = 0.84$ and
CL1604+4304 at $z = 0.90$.  Based on these observations, we conclude
the following :

\newcounter{discnt}
 
\begin{list}
{\arabic{discnt}.}  {\usecounter{discnt}}

\item We have spectroscopically observed magnitude-limited (R $\le
23.3$) samples of two fields with candidate clusters of galaxies at
$z > 0.8$.  Both candidates are confirmed as real cluster-like density
enhancements, but only $\sim20$\% of the observed galaxies are
actually members of the clusters. The large foreground/background
contamination is a common hazard when working in this redshift range
and implies that one requires a minimum of $\sim 100-150$ redshifts
per 15 square arcminutes if accurate kinematic parameters are desired.

\item We find that the main cluster in the CL0023+0423 field consists
of two components at nearly identical redshifts of $z = 0.8274$ and
$0.8453$ (corresponding to a radial velocity difference of $\sim
2900~{\rm km~s^{-1}}$).  Their calculated velocity dispersions and
corresponding masses indicate that these two components are more
similar to groups or poor clusters of galaxies. In addition, the
morphological analysis of Paper III indicates that these systems have
a morphological composition which more closely resembles that of the
field.  The fraction of spiral galaxies is 66\% or more.  Though it is
possible that these two groups are simply a chance projection on the
sky, they may also be in the process of merging. The dynamics of the
merging scenario is presented in another paper of this series (Lubin,
Postman \& Oke 1998).

\item We have confirmed that there is a centrally-concentrated, rich
cluster at $z=0.8967$ in the CL1604+4304 field. The velocity
dispersion and implied mass are consistent with an Abell richness
class 2 cluster.  In addition, the morphological composition of
CL1604+4304 is characteristic of a normal, present-day rich
cluster. Early-type galaxies comprise 76\% of the galaxies in the
central $\sim$0.5 $h^{-1}~{\rm Mpc}$.  In this population, the ratio
of S0 to ellipticals is $\sim$1.7, consistent with galaxy populations
found in local clusters (Paper III).

\item The mass-to-light ratios in both clusters are similar to those
seen locally but the uncertainties are high due to the sensitivity of
the derived luminosity to the amount of color and luminosity
evolution.  Our k-corrected $R$-band M/L values are, none the less,
significantly less (by $3 - 4\sigma$) than the zero redshift $R$-band
M/L of $\sim1300$ expected for an $\Omega = 1$ universe.

\item Defining active star formation as a rest equivalent width of
[OII] $\lambda3727$ which is greater than 15 \AA, 79\% of the field
(non-cluster) galaxies with $0.4 \le z < 0.9$ are active, while 62\%
with $z \ge 0.9$ are active.  The fraction of field galaxies which
have active star formation appears to be relatively independent of
redshift.  Of cluster members, 57\% and 50\% are active in the
CL0023+0423 and CL1604+4304, respectively. These numbers are
significantly higher than those of intermediate-redshift clusters
(Balogh \etal 1997). Even though the cluster masses are very different
in our two cases, the fractions of active galaxies are approximately
the same; this suggests that star formation activity may not be a
strong function of the cluster mass.

\item The traditional estimator D(4000) of the 4000 \AA\ break has
been supplemented by a measurement $J$ which is color independent,
measures either the Balmer jump or the 4000 \AA\ break, and is able to
distinguish between the two.  For blue objects which have a Balmer
jump, the measured jump is smaller than that predicted by standard
Bruzual \& Charlot models.  For the old, red galaxies our measurements
of D(4000) are consistent with the measured colors of the galaxies.
Because changes in D(4000) are accompanied by compensating changes in
color when the metallicity is varied, it is not possible to derive the
metallicity from our data.

\item We have compared our photometric and spectroscopic observations
with various families of Bruzual \& Charlot evolutionary models.  The
equivalent widths of the [OII]$\lambda3727$ line show that models with
a single burst of star formation followed by passive evolution (ssp
models) are in most cases not satisfactory.  However, models with an
exponentially decreasing star formation rate after time zero (tau
models) do predict the correct equivalent widths of [OII]; a time
constant of 0.6 Gyr appears to be optimum.  Our parameter $J$ also
indicates that an exponentially decaying star formation model is
better than the assumption of a single, monolithic burst.

\item Fitting the observed spectra to the tau0.6 synthetic models
indicate that the red galaxies, found mainly in the CL1604+4304
cluster at $z=0.8967$, have ages of $\sim 5$ Gyr for an assumed solar
metallicity.  For a cosmology of $H_0=65$ and $q_0=0.1$, this derived
age allows nearly 1 Gyr between the time of creation and the first
onset of star formation.  Of the other Bruzual \& Charlot models that
we have examined, the tau1.0 models, which are otherwise acceptable,
predict ages for these early-type galaxies which appear to be too old
for this cosmology. The ages derived using ssp models, which we have
shown are not generally appropriate for other observational reasons,
give ages which are only $\sim 3$ Gyr. The morphological analysis
presented in Paper III reveals that these old galaxies belong to the
Hubble class of spheroid or early spiral galaxies. The derived ages
are consistent with previous studies of high-redshift, early-type
cluster members (Dickinson 1995; Stanford, Eisenhardt \& Dickinson
1997).

\item Comparing the brightest cluster galaxies in our clusters with
those in Abell clusters of similar richness at redshifts of $z \sim
0.1$, we find a luminosity brightening of 1.02 mag for rest $M_B$ and
0.85 mag for rest $M_V$.  For the same cosmological parameters above,
the tau0.6 models predict a brightening of 0.8 and 0.7 mag for $M_B$
and $M_V$, respectively.  Increasing $q_0$ from 0.1 to 0.5 increases
the predicted brightening to 1.1 and 0.9 mag, respectively.  At a
given $q_0$, increasing $H_0$ also slightly increases the predicted
brightening.  Unlike the vast majority of their low-redshift
counterparts, these two BCGs are disk systems (early-type spirals).

\item Most of the field galaxies at $z > 0.9$ have absolute
luminosities which are not brighter than the predicted luminosity of a
brightest cluster galaxy at the same redshift; however, there are two
interesting exceptions. One galaxy at $z=1.5029$ in the CL0023+0423
field and one galaxy at $z=1.3757$ in the CL1604+4304 field are
overluminous by nearly 1 mag compared with what is expected for
brightest cluster galaxies at these redshifts. These galaxies have
strong MgII$\lambda2800$ absorption and FeII resonance line
absorption.  Although one argument suggests that in other galaxies
these lines are mainly interstellar, they are too strong in these two
galaxies for this to be the case.  Therefore, they presumably come
from late B-type stars in the galaxies.  

\end{list}

\vskip 0.5cm
\acknowledgments

The authors would like to thank the referee for numerous valuable
comments.  We also thank Dr. J.E. Gunn and Dr. D. Schneider for a
number of enlightening discussions.  C.D. Fassnacht is thanked for the
use of his superb spectral plotting program. Support for this research
was provided, in part, by NASA through grant number GO-06000.01-94A
from the Space Telescope Science Institute, which is operated by the
Association of Universities for Research in Astronomy, Inc., under
NASA contract NAS5-26555. LML graciously acknowledges support from a
Carnegie Fellowship.
 
Observational material for this project was obtained at the W.M. Keck
Observatory which is operated as a scientific partnership between the
California Institute of Technology, the University of California, and
the National Aeronautics and Space Administration.  It was made
possible by the generous financial support of the W. M. Keck
Foundation.

\newpage
\pagestyle{empty}
\renewcommand{\baselinestretch}{1}
\baselineskip 14pt

\begin{deluxetable}{lrr}
\tablewidth{0pt}
\tablenum{1}
\tablecaption{ Information on Spectra}
\tablehead{
\colhead{~} & 
\colhead{CL0023+0423} & 
\colhead{CL1604+4304}}
\startdata
(1) Number in sample & 167 & 168 \\
(2) Number not observed & 39 & 31 \\
(3) Number observed, no result & 14 & 13 \\
(4) Number of z=0.000 cases & 6 & 20 \\
(5) QSOs & 2 & 1 \\
(6) Number with significant redshifts & 106 & 103 \\
(7) Number with emission lines & 94 & 81 \\
\enddata
\end{deluxetable}

\newpage

\textheight=9.0in
\textwidth=7.0in
\voffset -0.85in
\hoffset -0.90in

\begin{deluxetable}{cccccccccrrrrrcc}
\scriptsize
\tablewidth{0pt}
\tablenum{2}
\tablecaption{Objects with Spectra in the CL0023+0423 Field}
\tablehead{
\colhead{n} & \colhead{ABB} & \colhead{ABV} & \colhead{ABR} & \colhead{ABI} & 
\colhead{z} & \colhead{Qual} & \colhead{colage} & \colhead{err} & 
\colhead{EW[OII]} & \colhead{EW($H\beta$)} & \colhead{EW[OIII]} & 
\colhead{Jump($\lambda_J$)} & \colhead{D(4000)} & \colhead{$M_{ABB}$}} 
\startdata

   30&  23.91&  23.33&  22.64&  21.44& 0.8470&  4& 1.8& 0.2&  -7.2&   2.5&  15.3& 0.36(3900)& 
1.10 & -20.41\\
  40&  23.82&  23.09&  22.66&  22.16& 0.3307&  4& 1.0& 0.3&  \nodata&  -0.6& -10.8&\nodata&  \nodata& -17.14\\
  53&  \nodata&  \nodata&  22.74&  22.12& 0.8435&  4& 0.4& 0.2&  -6.1& -20.3&  16.7& 0.99(3875) & 1.46 & -19.91\\
  96&  24.74&  23.64&  22.88&  21.37& 0.6230&  4& 3.4& 0.4&  -2.7&   0.2&   4.5&\nodata& 
 1.58 & -19.15\\
 134&  24.06&  23.96&  23.81&  21.95& 0.8027&  4& 1.1& 0.2& -53.7&  \nodata&  \nodata& 0.39(\phantom{0}---\phantom{0})&  \nodata& -19.35\\
 145&  24.04&  23.70&  23.77&  22.18& 1.0730&  2& 1.4& 0.3& -26.4&  \nodata&  \nodata&-0.16(\phantom{0}---\phantom{0})&  1.31 & -20.70\\
 157&  26.16&  24.71&  26.43&  22.32& 0.8447&  1& 3.6& 2.0& -46.7&  -5.2&  12.9& 0.80(3850) 
&1.21 & -19.52\\
 194&  24.10&  24.05&  23.49&  22.53& 9.0000&\nodata&\nodata&\nodata&  \nodata&  \nodata&  \nodata&\nodata&  \nodata&  \nodata\\
 217&  24.58&  23.73&  23.02&  21.85& 0.8514&  1& 2.2& 0.3& -10.9&  -8.2&  17.4&-0.28(\phantom{0}---\phantom{0})& 
 1.44 & -20.06\\
 264&  23.15&  22.52&  21.93&  21.36& 0.7946&  4& 1.4& 0.1& -40.5&  38.1&  33.9& 0.22(\phantom{0}---\phantom{0})& 
 1.14 & -20.62\\
 394&  23.73&  23.45&  22.78&  22.94& 0.7947&  4& 1.1& 0.2& -29.5&  30.9&  22.0&\nodata&  \nodata& -19.71\\
 451&  24.18&  23.43&  22.81&  21.95& 9.0000&\nodata&\nodata&\nodata&  \nodata&  \nodata&  \nodata&\nodata&  \nodata&  \nodata\\
 468&  23.90&  23.25&  22.70&  21.68& 0.7718&  4& 1.7& 0.2& -21.5& -42.9& -18.6&\nodata&
  1.66 & -19.90\\
 509&  \nodata&  \nodata&  22.47&  21.57& 0.8458&  4& 0.8& 0.3& -13.5&   2.7&   5.4& 0.18(\phantom{0}---\phantom{0})& 
 1.09 & -20.30\\
 557&  24.18&  23.49&  23.47&  23.46& 0.3642&  4& 0.4& 0.3&  \nodata&  -6.9& -26.4&\nodata&
  1.38 & -16.78\\
 569&  24.25&  24.13&  23.46&  22.22& 0.7180&  1& 1.1& 0.2&-111.9& -79.9&  38.7&\nodata&  \nodata& -18.79\\
 588&  25.17&  24.41&  23.26&  21.40& 9.0000&\nodata&\nodata&\nodata&  \nodata&  \nodata&  \nodata&\nodata&  \nodata&  \nodata\\
 605&  24.70&  24.28&  23.96&  22.32& 9.0000&\nodata&\nodata&\nodata&  \nodata&  \nodata&  \nodata&\nodata&  \nodata&  \nodata\\
 644&  23.79&  23.07&  22.86&  22.31& 0.0260&  1& 3.5& 1.6&  \nodata&  \nodata&  \nodata&\nodata&  \nodata& -10.76\\
 658&  24.11&  23.54&  22.85&  22.11& 0.9150&  2& 1.8& 0.3& -18.7&   7.3& \ 10.8& 0.74(3950)
& 1.18 & -20.45\\
 701&  20.68&  19.91&  19.57&  19.03& 0.2388&  4& 2.0& 0.2&  \nodata&  -7.0&  -4.9&\nodata&  \nodata& -19.42\\
 702&  24.01&  23.53&  23.02&  22.09& 0.6266&  4& 1.0& 0.2& -30.7&   5.9& -18.2& 0.83(3800)
&\nodata & -18.61\\
 792&  24.82&  24.35&  23.38&  21.72& 0.6432&  4& 2.6& 0.4& -14.7&  -9.7&  -7.3& 0.53(3950)
& 1.41 & -18.81\\
 805&  23.97&  23.82&  23.29&  \nodata& 1.0886&  3& 1.6& 0.3&  -3.6&  \nodata&  \nodata& 0.49(\phantom{0}---\phantom{0})&  1.14 & -20.83\\
 825&  22.84&  22.35&  22.11&  21.83& 0.4436&  4& 0.3& 0.1&  22.2& -16.8& -81.4&\nodata&  \nodata& -18.47\\
 926&  23.65&  23.23&  22.72&  21.90& 0.6900&  4& 1.0& 0.2& -31.4&  -8.0&   3.8& 0.63(3775)
&  1.39 & -19.24\\
 946&  25.86&  25.31&  25.12&  22.99& 0.7213&  2& 2.0& 1.0& -19.3& -22.9&  -4.6&\nodata
&  1.61 & -17.91\\
 980&  22.53&  23.74&  23.46&  22.08& 1.4870&  1& 0.3&\nodata&  \nodata&  \nodata&  \nodata&\nodata&  \nodata& -21.25\\
 984&  24.52&  22.85&  \nodata&  \nodata& 0.0000&\nodata&\nodata&\nodata&  \nodata&  \nodata&  \nodata&\nodata&  \nodata&  \nodata\\
1052&  24.38&  22.53&  20.16&  19.82& 0.7445&  4&  $\geq$5.0&\nodata&  -0.9&   2.0&  25.8&
 0.54(4000) & 1.79 & -21.84\\
1099&  23.46&  22.81&  22.63&  22.01& 0.2700&  4& 0.6& 0.3&  \nodata& -13.0& -51.1&\nodata&  \nodata& -16.77\\
1104&  22.82&  22.47&  21.82&  20.75& 0.4010&  4& 1.8& 0.2&  \nodata& -22.1& -86.6&\nodata&  \nodata& -18.68\\
1124&  \nodata&  21.65&  21.14&  19.88& 0.4005&  4&  $\geq$6.0&\nodata&  \nodata& -10.9& -89.0&\nodata&  \nodata& -19.33\\
1230&  25.70&  24.39&  23.50&  21.42& 0.9330&  4&  $\geq$4.0&\nodata&   0.8&   6.7&  -8.2
& 0.51(4000) & 1.74 &  -20.61\\
1253&  23.99&  23.53&  22.94&  21.75& 0.8458&  4& 1.6& 0.2& -22.3&  -5.0&   3.6&
 0.64(3850) & 0.86 & -20.03\\
1266&  23.30&  22.83&  22.51&  20.78& 9.0000&\nodata&\nodata&\nodata&  \nodata&  \nodata&  \nodata&\nodata&  \nodata&  \nodata\\
1302&  24.02&  23.19&  23.39&  22.17& 9.0000&\nodata&\nodata&\nodata&  \nodata&  \nodata&  \nodata&\nodata&  \nodata&  \nodata\\
1330&  21.99&  21.80&  21.74&  21.22& 0.2541&  1& 0.1& 0.1&  \nodata&   0.8&  -0.4&\nodata&  \nodata& -17.64\\
1336&  22.93&  22.35&  21.83&  21.23& 0.5776&  4& 1.0& 0.1& -24.8&  -0.7&   2.9& 0.47(4000)
&  1.19 & -19.50\\
1374&  24.28&  23.91&  24.38&  22.51& 0.8442&  1& 0.9& 0.3& -60.2& -14.2&  15.5& 0.58(3850)
&  1.24 & -19.16\\
1375&  23.06&  22.68&  22.36&  21.65& 0.6284&  4& 0.6& 0.1& -43.2& -17.3& -37.8& 0.80(3875)
&  1.09 & -19.22\\
1392&  23.49&  23.09&  22.84&  22.15& 0.9142&  1& 1.0& 0.1&-125.3&  -1.8&  11.5&\nodata&  \nodata& -20.33\\
1412&  23.40&  22.95&  \nodata&  \nodata& 0.9137&  2& 1.6& 0.2& -44.6&   2.6& \nodata.9
& 0.98(3875) &  1.25 & -20.94\\
1453&  23.37&  22.89&  22.45&  21.82& 0.7725&  3& 1.1& 0.1& -40.3& -39.7&  -8.9& 0.48(3800)
&  1.14 & -20.01\\
1470&  24.61&  23.99&  23.52&  22.17& 0.4416&  4& 2.0& 0.5& -14.9&   1.9& -29.3& 0.36(\phantom{0}---\phantom{0})&  \nodata& -17.37\\
1478&  24.02&  23.77&  24.09&  22.46& 0.7732&  1& 0.6& 0.1& -79.3&  11.1&  -0.8& \nodata&  \nodata& -18.76\\
1511&  23.90&  23.88&  23.50&  22.48& 1.3351&  2& 1.8& 0.3& -35.8&  \nodata&  \nodata&\nodata&  \nodata& -21.62\\
1526&  22.66&  21.97&  21.51&  21.07& 0.4083&  4& 1.0& 0.1&  \nodata&  -6.3& -14.3&\nodata&  \nodata& -18.87\\
1534&  23.75&  23.34&  23.08&  22.05& 9.0000&\nodata&\nodata&\nodata&  \nodata&  \nodata&  \nodata&\nodata&  \nodata&  \nodata\\
1552&  21.62&  20.94&  20.67&  20.37& 0.1852&  4& 1.4& 0.1&  \nodata&  -1.7&  -8.7&\nodata&  \nodata& -17.63\\
1566&  24.18&  23.75&  23.55&  22.17& 0.8470&  4& 1.1& 0.3& -53.9&  -7.0& -29.2& 0.85(3800)
& 1.51 & -19.49\\
1572&  20.73&  20.06&  19.81&  19.46& 0.1851&  4& 1.4& 0.1&  \nodata&  -1.6&  -4.5&\nodata&  \nodata& -18.53\\
1628&  23.31&  22.55&  22.35&  21.84& 0.5875&  1& 0.9& 0.1&   1.6& -27.7&   8.9&-0.19(\phantom{0}---\phantom{0})&  \nodata& -19.15\\
1659&  23.65&  23.17&  22.96&  22.25& 0.5708&  4& 0.6& 0.2& -72.5& -19.0& -83.7&\nodata
& 1.18 & -18.41\\
1662&  23.28&  22.73&  22.16&  21.01& 9.0000&\nodata&\nodata&\nodata&  \nodata&  \nodata&  \nodata&\nodata&  \nodata&  \nodata\\
1689&  23.47&  23.36&  23.21&  22.04& 1.4679&  2& 2.0& 0.2& -21.6&  \nodata&  \nodata&\nodata&  \nodata& -22.49\\
1699&  24.60&  23.26&  22.37&  20.91& 0.8435&  4& 3.5& 0.2&  -2.8&   5.4&  -2.7& 0.33(3975)
&  1.53 & -20.93\\
1749&  23.51&  22.95&  22.92&  22.25& 0.0811&  1& 1.6& 0.4&  \nodata&   1.8&  -0.5&\nodata&  \nodata& -13.62\\
1832&  24.05&  23.68&  \nodata&  \nodata& 9.0000&\nodata&\nodata&\nodata&  \nodata&  \nodata&  \nodata&\nodata&  \nodata&  \nodata\\
1833&  23.81&  23.04&  22.39&  20.91& 9.0000&\nodata&\nodata&\nodata&  \nodata&  \nodata&  \nodata&\nodata&  \nodata&  \nodata\\
1846&  24.34&  24.53&  23.19&  21.08& 0.8465&  2& 2.7& 0.2&   2.2&   1.1&  20.4& 0.46(3950)
&  1.34 & -20.50\\
1912&  24.56&  24.22&  23.12&  21.76& 0.8444&  3& 2.2& 0.4& -38.8& -19.4&  25.2& 0.46(3975)
& 1.14 & -19.95\\
\tablebreak
1921&  24.31&  23.71&  23.01&  21.93& 0.8456&  4& 1.8& 0.3& -30.4&   3.8&   4.8& 0.83(3825)
&  1.03 & -19.92\\
1922&  23.07&  22.90&  22.94&  21.95& 1.3341&  2& 1.3& 0.2& -12.6&  \nodata&  \nodata&\nodata&  \nodata& -21.93\\
1949&  24.54&  23.74&  \nodata&  \nodata& 9.0000&\nodata&\nodata&\nodata&  \nodata&  \nodata&  \nodata&\nodata&  \nodata&  \nodata\\
1968&  23.59&  23.54&  23.11&  22.48& 9.0000&\nodata&\nodata&\nodata&  \nodata&  \nodata&  \nodata&\nodata&  \nodata&  \nodata\\
1971&  22.96&  22.61&  22.19&  21.45& 1.1074&  4& 1.6& 0.1& -18.9&  \nodata&  \nodata
& 0.99(3850)&  1.28 & -22.07\\
2003&  22.93&  22.40&  22.47&  21.71& 0.2480&  4& 0.3& 0.2&  \nodata&  -3.5& -22.2&\nodata&  \nodata& -16.88\\
2035&  24.22&  23.88&  22.44&  20.60& 0.8274&  4& 3.5& 0.2&   0.6&   3.3&  -4.1& 0.69(4000)
&  1.84 & -20.95\\
2055&  24.11&  22.66&  21.71&  20.06& 0.8266&  4& 3.9& 0.2&  -4.0&   6.7&   3.1& 0.64(3975)
& 1.56 & -21.63\\
2075&  23.25&  22.97&  22.96&  21.72& 0.8447&  4& 0.8& 0.1& -82.7& -12.4& -98.6
& 0.08(\phantom{0}---\phantom{0}) & 0.94 & -20.02\\
2092&  24.25&  23.78&  22.78&  22.11& 0.5162&  3& 1.5& 0.4&  \nodata& -27.3&  -0.1&\nodata
&  1.04 & -18.12\\
2103&  24.09&  23.53&  23.23&  22.05& 0.3524&  4& 1.1& 0.3&  \nodata& -23.4& -87.5&\nodata&  \nodata& -17.05\\
2108&  23.56&  23.19&  23.71&  21.97& 0.9375&  2& 0.8& 0.1&  -9.0& -10.6&  -2.9& \nodata 
& \nodata & -20.19\\
2110&  23.43&  23.07&  23.06&  21.91& 1.0858&  4& 1.1& 0.1& -33.7&  \nodata&  \nodata
& 0.54(3950) & 1.29 & -21.00\\
2121&  23.56&  22.85&  22.24&  21.09& 0.8281&  4& 1.9& 0.2&  -1.4&  35.6&   2.6& 0.84(3900)
& 1.05 & -20.67\\
2144&  23.69&  23.32&  22.84&  22.45& 0.8444&  2& 1.1& 0.2& -28.1& \nodata&  23.1& 0.59(3875)
& 1.26 & -20.01\\
2166&  23.93&  22.94&  22.20&  20.94& 0.8288&  3& 2.6& 0.2&  -1.1&   2.4&  -4.8& 0.43(4000)
& 1.38 & -20.85\\
2182&  22.01&  21.37&  \nodata&  \nodata& 0.0000&  4&\nodata&\nodata&  \nodata&  \nodata&  \nodata&\nodata&  \nodata&  \nodata\\
2212&  23.37&  22.66&  22.09&  21.29& 0.3297&  4& 1.9& 0.2&  \nodata&  -5.3& -22.9&\nodata&  \nodata& -17.71\\
2213&  24.04&  23.76&  23.27&  22.64& 0.8380&  2& 1.0& 0.2& -44.8&   4.4&   1.6& 0.48(3825)
& 1.09 & -19.52\\
2232&  \nodata&  23.02&  22.29&  21.29& 0.9931&  1& 1.8& 0.1&  -8.7&  \nodata&  \nodata
& 0.27(3850)& 1.10 &  -21.41\\
2271&  23.86&  23.12&  \nodata&  \nodata& 0.3528&  4& 0.6&\nodata&  \nodata&  -3.9& -42.2&\nodata&  \nodata& -17.13\\
2336&  24.96&  \nodata&  25.23&  22.46& 0.8251&  1& 1.7& 0.5& -13.1&  -7.1& -33.3& \nodata&  \nodata& -19.29\\
2348&  22.49&  21.63&  21.24&  20.86& 0.8474&  4& 1.3& 0.1& -33.2& -18.8&  22.0& 0.45(3775)
& 1.38 & -21.49\\
2400&  23.66&  23.36&  23.23&  22.02& 1.1079&  2& 1.3& 0.2& -25.8&  \nodata&  \nodata
& 0.16(\phantom{0}---\phantom{0})&  0.90 & -20.92\\
2415&  24.79&  23.75&  23.43&  21.68& 0.8451&  4& 2.4& 0.3& -34.6&   2.6&  16.4
&-0.20(\phantom{0}---\phantom{0})&  1.29 & -19.97\\
2433&  22.83&  22.27&  22.05&  21.77& 0.2388&  4& 0.6& 0.2&  \nodata&  -6.7& -28.7&\nodata&  \nodata& -17.01\\
2477&  24.06&  23.31&  23.09&  22.10& 0.1034&  2& 2.8& 1.0&  \nodata&   1.6&  23.1&\nodata&  \nodata& -13.77\\
2548&  23.99&  23.27&  23.01&  22.14& 0.2208&  3& 1.6& 0.4&  \nodata&   3.5&   2.8&\nodata&  \nodata& -15.79\\
2563&  23.01&  22.29&  22.40&  21.41& 9.0000&\nodata&\nodata&\nodata&  \nodata&  \nodata&  \nodata&\nodata&  \nodata&  \nodata\\
2598&  23.52&  23.13&  22.22&  22.20& 9.0000&\nodata&\nodata&\nodata&  \nodata&  \nodata&  \nodata&\nodata&  \nodata&  \nodata\\
2615&  23.63&  23.30&  22.61&  22.17& 0.6124&  4& 0.9& 0.2& -45.5&  -3.3&  -9.5&\nodata
&  1.92 & -18.83\\
2616&  24.00&  23.54&  22.99&  22.03& 0.7193&  4& 1.2& 0.2& -31.3& -29.5&   1.3& 0.89(3875)
&  0.88 & -19.14\\
2640&  24.63&  24.07&  23.53&  22.19& 0.9131&  4& 1.9& 0.3& -12.5&  -6.8& -19.0& 0.94(4000)
&  1.34 & -19.95\\
2664&  23.69&  23.43&  22.84&  22.07& 0.7208&  4& 0.9& 0.1& -67.6&-108.2&-102.7& \nodata
& \nodata & -19.20\\
2680&  23.84&  23.29&  22.83&  21.60& 1.0235&  3& 1.9& 0.2& -15.2&  \nodata&  \nodata
& 0.33(3850) &  1.51 & -21.12\\
2733&  24.02&  23.62&  23.34&  22.39& 0.4072&  4& 0.5& 0.2& -44.7& -21.9& -66.3&\nodata
&  1.13 & -17.15\\
2764&  24.06&  23.46&  23.14&  22.41& 0.7210&  1& 1.1& 0.2& -31.4&   6.9& -19.4&\nodata
&  1.17  & -19.10\\
2773&  23.35&  22.92&  22.93&  22.14& 0.9404&  2& 0.7& 0.1& -85.4&  15.3&  -1.9
& 0.23(\phantom{0}---\phantom{0}) &  0.98 & -20.39\\
2792&  23.20&  22.32&  22.03&  21.76& 0.2983&  4& 1.0& 0.3&  \nodata&   4.4&  -7.9&\nodata&  \nodata& -17.54\\
2798&  24.59&  23.81&  23.46&  22.05& 0.9752&  1& 2.0& 0.7& -33.5&  \nodata&  \nodata& 0.38(4000)& \nodata& -20.49\\
2832&  23.81&  23.54&  23.02&  22.15& 1.3645&  4& 1.4& 0.2& -14.9&  \nodata&  \nodata&\nodata&  \nodata& -21.01\\
2872&  23.82&  23.66&  23.60&  22.50& 2.5950&  4&\nodata&\nodata&  \nodata&  \nodata&  \nodata&\nodata&  \nodata&  \nodata\\
2924&  25.93&  24.46&  23.30&  21.17& 0.0000&\nodata&\nodata&\nodata&  \nodata&  \nodata&  \nodata&\nodata&  \nodata&  \nodata\\
2986&  23.45&  23.21&  23.07&  22.12& 1.2436&  4& 1.5& 0.2& -47.5&  \nodata&  \nodata&\nodata&  \nodata& -21.59\\
3045&  23.85&  23.04&  22.82&  22.28& 0.2971&  1& 0.8& 0.4&  \nodata& -10.5&   1.1&\nodata&  \nodata& -16.80\\
3075&  24.35&  23.42&  22.86&  22.24& 0.3306&  4& 2.4& 0.7&  \nodata&  -5.3& -13.2&\nodata&  \nodata& -16.87\\
3079&  24.64&  23.65&  23.20&  22.01& 0.0000&  4&\nodata&\nodata&  \nodata&  \nodata&  \nodata&\nodata&  \nodata&  \nodata\\
3083&  23.92&  22.97&  22.56&  22.21& 0.3356&  4& 1.4& 0.5&  \nodata&   2.7& -10.6&\nodata&  \nodata& -17.27\\
3108&  24.21&  23.36&  23.24&  22.13& 0.3683&  4& 1.2& 0.5& -20.3&  -9.7& -39.3&\nodata
&  1.43 & -17.14\\
3119&  23.91&  23.72&  23.20&  22.77& 0.4989&  1& 0.3& 0.2& -22.8& -12.1& \ 10.2& 0.85(3875)
&  0.88 & -17.54\\
3128&  23.69&  23.03&  \nodata&  \nodata& 0.3779&  4& 0.6&\nodata& -54.4& -64.3&-277.2&\nodata&  \nodata& -17.46\\
3152&  23.16&  22.56&  \nodata&  \nodata& 1.0908&  1& 3.0&\nodata&  \nodata&  \nodata&  \nodata&\nodata&  \nodata& -22.96\\
3175&  \nodata&  22.78&  22.38&  21.55& 0.8280&  4& 0.9& 0.2& -28.6&   0.3& -17.4& 0.86(3825)
& 1.03 & -20.30\\
3265&  21.71&  20.35&  19.62&  18.17& 0.0000&\nodata&\nodata&\nodata&  \nodata&  \nodata&  \nodata&\nodata&  \nodata&  \nodata\\
3371&  24.30&  24.08&  24.30&  22.13& 0.7472&  4& 1.2& 0.3& -34.8&  -1.5&  -6.0& 0.33(3825)
& 1.25 & -18.93\\
3416&  24.10&  23.78&  23.19&  22.17& 0.7734&  2& 1.1& 0.2& -71.8&   9.1& -59.2& 0.68(3850)
& 1.12 & -19.23\\
3435&  23.11&  22.77&  \nodata&  \nodata& 0.4953&  4& 0.5&\nodata&  -4.0&  -0.9&   2.7&\nodata&  \nodata& -18.50\\
3436&  23.91&  23.66&  23.49&  22.49& 0.8906&  2& 0.8& 0.2& -40.2&  19.0&  -4.3& 0.05(\phantom{0}---\phantom{0})&  0.98 & -19.59\\
3465&  25.31&  23.93&  \nodata&  \nodata& 0.0000&  4&\nodata&\nodata&  \nodata&  \nodata&  \nodata&\nodata&  \nodata&  \nodata\\
3508&  22.14&  21.44&  21.15&  21.10& 0.1471&  4& 1.4& 0.2&  \nodata&   1.3&  -7.3&\nodata&  \nodata& -16.56\\
3560&  23.19&  22.96&  22.79&  22.12& 1.5029&  3& 2.0& 0.2& -65.9&  \nodata&  \nodata&\nodata&  \nodata& -22.88\\
3611&  23.36&  22.91&  22.54&  21.40& 0.8224&  2& 1.3& 0.1& -20.0&  \nodata&  -1.6& 0.46(4000)&  1.37 & -20.37\\
3640&  24.35&  24.04&  23.35&  22.33& 0.3690&  4& 1.1& 0.4& -53.0&  -0.2&  -4.8&\nodata&  \nodata& -16.87\\
3671&  24.15&  23.76&  23.36&  22.39& 0.5248&  4& 0.7& 0.2& -36.4&  -7.7& -33.5&\nodata
&  0.99 & -17.78\\
3703&  22.41&  21.49&  20.92&  20.53& 0.3604&  4& 1.9& 0.2&  \nodata&  -5.5&  -9.5&\nodata&  \nodata& -19.05\\

\enddata
\end{deluxetable}

\newpage
 
\begin{deluxetable}{cccccccccrrrrcc}
\scriptsize
\tablewidth{0pt}
\tablenum{3}
\tablecaption{Objects with Spectra in the CL1604+4304 Field}
\tablehead{
\colhead{n} & \colhead{ABB} & \colhead{ABV} & \colhead{ABR} & \colhead{ABI} & 
\colhead{z} & \colhead{Qual} & \colhead{colage} & \colhead{err} & 
\colhead{EW[OII]} & \colhead{EW($H\beta$)} & \colhead{EW[OIII]} &
\colhead{Jump($\lambda_J$)} & \colhead{D(4000)} & \colhead{$M_{ABB}$}} 
\startdata

  98&  \nodata&  \nodata&  26.82&  22.44& 0.3459&  1&\nodata&\nodata&  \nodata&  -4.9&  -0.7&\nodata&  \nodata&  \nodata\\
 142&  23.56&  23.24&  22.48&  21.15& 0.8216&  2& 1.6& 0.1& -39.0&   6.1&   1.4&\nodata
&  0.93 & -20.39\\
 144&  22.64&  22.26&  21.75&  20.93& 0.6825&  1& 1.0& 0.1& -20.4&  -0.9& -14.6& 0.63(3850)
& 1.18 & -20.18\\
 205&  24.01&  24.02&  22.74&  21.60& 0.4258&  4& 2.2& 0.2&  \nodata&  -4.0& -26.4&\nodata&  \nodata& -17.87\\
 293&  23.48&  23.44&  22.85&  21.49& 0.9748&  1& 1.4& 0.2& -57.9&  11.6&  \nodata&\nodata&  \nodata& -20.98\\
 320&  23.44&  23.22&  22.71&  21.40& 0.9738&  2& 1.4& 0.2& -77.1&  12.8&  \nodata&\nodata&  \nodata& -21.05\\
 336&  23.64&  22.62&  21.96&  20.80& 0.3018&  4& 5.0& 1.0&  \nodata&  -6.5&  -4.0&\nodata&  \nodata& -17.51\\
 343&  24.11&  23.81&  22.41&  21.16& 0.8074&  4& 2.4& 0.2&  -7.8&  -8.2&  -3.6&\nodata
&  2.13 & -20.44\\
 381&  23.14&  22.84&  22.30&  21.23& 0.4965&  4& 0.9& 0.1& -60.2&  -8.4& -20.3&\nodata
&  1.43 & -18.72\\
 407&  22.91&  22.53&  21.67&  20.74& 0.0000&  4&\nodata&\nodata&  \nodata&  \nodata&  \nodata&\nodata&  \nodata&  \nodata\\
 413&   2.18&  24.92&  23.42&  21.74& 9.0000&\nodata&\nodata&\nodata&  \nodata&  \nodata&  \nodata&\nodata&  \nodata&  \nodata\\
 416&  23.17&  23.00&  22.44&  21.40& 0.8938&  3& 1.1& 0.1& -73.8&  \nodata&  \nodata
& 0.41(3800) & 1.03 & -20.62\\
 450&  \nodata&  25.90&  23.93&  21.13& 0.0000&  4&\nodata&\nodata&  \nodata&  \nodata&  \nodata&\nodata&  \nodata&  \nodata\\
 519&  23.30&  22.59&  22.37&  21.52& 0.1893&  4& 1.9& 0.3&  \nodata&  -0.8& -17.1&\nodata&  \nodata& -16.07\\
 528&  23.08&  22.98&  22.42&  21.27& 0.8720&  4& 1.1& 0.1& -71.3&   0.4&   3.8& 0.03(\phantom{0}---\phantom{0})&  \nodata& -20.75\\
 539&  24.39&  24.30&  24.62&  21.84& 0.2980&  4& 3.5& 0.6&  \nodata&  -7.9& -15.3&\nodata&  \nodata& -16.55\\
 774&  23.90&  22.96&  22.83&  21.70& 0.3839&  2& 1.5& 0.3&  \nodata&  -1.1&  -2.2&\nodata&  \nodata& -17.65\\
 796&  25.42&  24.46&  23.22&  21.00& 1.0905&  1& $\geq$4.0&\nodata&  -7.4&  \nodata&  \nodata& 0.08(\phantom{0}---\phantom{0}) &  1.19 & -22.04\\
 807&  23.12&  22.84&  22.70&  21.71& 0.0000&  2&\nodata&\nodata&  \nodata&  \nodata&  \nodata&\nodata&  \nodata&  \nodata\\
 821&  23.65&  22.80&  22.33&  21.31& 0.0000&  3&\nodata&\nodata&  \nodata&  \nodata&  \nodata&\nodata&  \nodata&  \nodata\\
 895&  23.70&  22.40&  21.61&  20.57& 0.3891&  4& 6.0& 1.0&  -9.6&   4.4&  -3.2&\nodata&  \nodata& -18.59\\
 978&  25.20&  24.29&  23.74&  22.38& 9.0000&\nodata&\nodata&\nodata&  \nodata&  \nodata&  \nodata&\nodata&  \nodata&  \nodata\\
1023&  24.20&  23.74&  22.93&  21.33& 1.2316&  2& 3.0& 0.2& -56.3&  \nodata&  \nodata&\nodata&  \nodata& -22.40\\
1028&  22.94&  22.52&  22.08&  20.98& 0.8288&  4& 1.3& 0.1&  -4.9&  -8.1&  -2.6& 0.30(4000)
& 1.58 & -20.76\\
1037&  23.63&  22.95&  22.59&  21.40& 0.3142&  4& 1.9& 0.2&  \nodata&  -8.3& -34.5&\nodata&  \nodata& -17.34\\
1065&  22.67&  22.22&  21.79&  21.05& 0.4954&  4& 0.8& 0.1& -59.8& -14.2& -63.3&\nodata&  \nodata& -19.16\\
1111&  23.15&  22.34&  22.02&  21.35& 0.2733&  4& 1.4& 0.2&  \nodata&  -1.3&  -9.8&\nodata&  \nodata& -17.32\\
1165&  23.19&  22.00&  21.28&  20.19& 0.0000&\nodata&\nodata&\nodata&  \nodata&  \nodata&  \nodata&\nodata&  \nodata&  \nodata\\
1275&  20.64&  20.00&  19.66&  18.76& 0.0000&  4&\nodata&\nodata&  \nodata&  \nodata&  \nodata&\nodata&  \nodata&  \nodata\\
1365&  24.45&  24.04&  23.03&  20.82& 0.8280&  4& 3.5& 0.3& -13.2&  -2.5&   0.7
& 0.02(\phantom{0}---\phantom{0}) &  1.94 & -20.70\\
1487&  24.34&  23.98&  22.89&  21.20& 0.8329&  2& 2.6& 0.2&  -3.8&   1.9&   3.4& 0.55(4000)
& 1.46 & -20.27\\
1492&  25.02&  23.85&  22.99&  21.25& 0.0000&  4&\nodata&\nodata&  \nodata&  \nodata&  \nodata&\nodata&  \nodata&  \nodata\\
1510&  23.68&  23.36&  23.36&  21.87& 0.9742&  3& 1.1& 0.2& -37.3&  \nodata&  \nodata
& 0.28(3850) & 1.47 & -20.48\\
1543&  22.44&  22.01&  21.46&  20.81& 0.4998&  4& 1.0& 0.1& -52.2&   0.8&  -6.7
&\nodata&  1.29 & -19.51\\
1574&  25.26&  24.98&  24.60&  22.33& 9.0000&\nodata&\nodata&\nodata&  \nodata&  \nodata&  \nodata&\nodata&  \nodata&  \nodata\\
1618&  23.27&  22.79&  22.26&  20.99& 0.8798&  4& 1.7& 0.1& -29.0&  \nodata&  \nodata
&-0.01(\phantom{0}---\phantom{0}) &  1.18 & -21.03\\
1649&  24.00&  23.71&  23.01&  21.95& 0.5479&  4& 1.1& 0.2&-131.9&   3.2& -99.6&-0.24(\phantom{0}---\phantom{0})&  \nodata& -18.28\\
1653&  23.35&  22.88&  22.25&  21.11& 0.8804&  2& 1.7& 0.1& -14.2&  -3.9&  -9.8&\nodata &  \nodata& -20.98\\
1693&  24.36&  23.48&  22.83&  21.65& 0.0000&  3&\nodata&\nodata&  \nodata&  \nodata&  \nodata&\nodata&  \nodata&  \nodata\\
1771&  24.14&  23.91&  23.16&  21.46& 9.0000&\nodata&\nodata&\nodata&  \nodata&  \nodata&  \nodata&\nodata&  \nodata&  \nodata\\
1787&  23.47&  23.47&  22.34&  20.63& 0.8707&  2& 2.2& 0.1& -88.2& \nodata.1& -15.7&\nodata
&  \nodata& -21.12\\
1832&  24.21&  23.98&  23.53&  21.66& 0.7787&  1& 1.3& 0.2& -92.2& \nodata.9& -20.6& 0.76(4000)& \nodata& -19.51\\
1869&  23.88&  23.93&  23.94&  22.17& 1.0966&  4& 1.1& 0.2& -48.2&  \nodata&  \nodata
& \nodata& \nodata & -20.53\\
1875&  23.59&  23.09&  22.78&  21.83& 0.4909&  2& 0.7& 0.2&   0.0&   0.5&  -5.4&\nodata&  \nodata& -18.25\\
1915&  22.42&  21.56&  20.87&  19.85& 0.0000&  4&\nodata&\nodata&  \nodata&  \nodata&  \nodata&\nodata&  \nodata&  \nodata\\
1954&  23.88&  23.23&  23.15&  21.88& 0.2405&  4& 1.3& 0.3&  \nodata&  -4.7& -15.6&\nodata&  \nodata& -16.15\\
1969&  23.36&  22.91&  22.09&  20.62& 0.8984&  4& 2.2& 0.1&  -8.0&  16.3&   6.7& 0.66(3950)
& 1.21 & -21.34\\
1978&  24.80&  24.04&  23.68&  21.19& 0.4964&  2& $\geq$7.0&\nodata& -91.0&  -3.6& 10.8&\nodata&  1.26 & -18.46\\
2006&  23.86&  23.31&  23.21&  21.24& 0.8294&  4& 1.8& 0.2&  -0.3&   3.6&  \nodata
& 1.09(3850) &  1.47 & -20.23\\
2041&  23.83&  23.44&  23.38&  21.37& 9.0000&\nodata&\nodata&\nodata&  \nodata&  \nodata&  \nodata&\nodata&  \nodata&  \nodata\\
2134&  24.18&  23.46&  23.44&  22.51& 0.4142&  3& 0.7& 0.3& -18.8&   4.8&  -6.0&\nodata
&  1.03 & -17.27\\
2159&  25.48&  23.97&  23.46&  21.07& 0.8968&  2& $\geq$4.0&\nodata& -16.5&  \nodata&  \nodata& 0.42(3825) & 1.06 & -21.02\\
2167&  24.09&  23.70&  23.47&  21.59& 0.8893&  1& 1.7& 0.2& -34.8&  21.1&  -3.1
& 0.26(\phantom{0}---\phantom{0}) &  1.43 & -20.23\\
2194&  25.14&  23.92&  23.32&  21.02& 1.0682&  2& $\geq$4.0&\nodata& -37.0&  \nodata&  \nodata& 0.39(3800) & 0.93 &  -21.81\\
2232&  25.33&  24.61&  23.92&  21.30& 0.8998&  4& $\geq$4.0&\nodata&   2.1&   9.4&  -6.0
& 0.25(3975) & 1.56 & -20.79\\
2307&  22.91&  22.74&  22.66&  21.41& 0.9005&  2& 0.7& 0.1& -37.8&  \nodata&  \nodata& \nodata&  \nodata& -20.48\\
2312&  24.44&  23.25&  22.56&  20.46& 0.0000&  4&\nodata&\nodata&  \nodata&  \nodata&  \nodata&\nodata&  \nodata&  \nodata\\
2315&  24.97&  24.12&  23.32&  21.77& 0.3796&  2& $\geq$6.0&\nodata&  \nodata&   6.9&  -0.4&\nodata&  \nodata& -17.00\\
2459&  24.39&  23.30&  22.39&  20.64& 0.8956&  4& 3.5& 0.2&   1.2&  -5.7&  -2.2& 0.64(3975)
&  1.72 & -21.32\\
2463&  21.28&  19.96&  19.49&  18.69& 0.2415&  4& 6.0& 0.5&  \nodata&   3.8&   1.6&\nodata&  \nodata& -19.30\\
2515&  23.91&  23.26&  23.23&  21.91& 0.4779&  2& 1.0& 0.2&  \nodata&   1.6&   0.1&\nodata&  \nodata& -18.04\\
2704&  23.66&  23.38&  22.56&  21.12& 0.9851&  2& 2.2& 0.1&  -4.4&  12.4&   0.0& 0.69(4000)
& 1.74 & -21.47\\
\tablebreak
 2793&  24.06&  23.42&  22.86&  22.36& 0.3278&  1& 1.1& 0.4&  \nodata&  -1.6&  -5.3&\nodata&  \nodata& -16.94\\
2837&  23.37&  23.63&  22.65&  21.28& 0.8852&  2& 1.3& 0.1& -31.9&   2.0& -79.6& 0.82(3975)
& 1.01 & -20.54\\
2851&  24.77&  23.72&  22.53&  20.14& 0.8962&  4& $\geq$5.0&\nodata&   0.5&  -3.3&  15.7
& 0.33(4000) & 1.69 & -21.88\\
2855&  24.02&  23.35&  22.09&  20.00& 0.8964&  4& 4.0& 0.2&  -2.4&   0.1&   7.6& 0.55(3950)
& 1.50 & -21.91\\
2858&  22.64&  22.49&  22.41&  21.18& 1.3757&  4& 1.7& 0.1& -75.8&  \nodata&  \nodata&\nodata&  \nodata& -22.90\\
2945&  24.20&  23.85&  23.17&  \nodata& 0.4430&  4& 0.6& 0.4& -79.4& -19.7& -73.3&\nodata&  \nodata& -17.26\\
2985&  24.51&  23.98&  23.12&  21.13& 0.8990&  4& 3.0& 0.2&   2.5&  -0.7&  -8.7
&-0.15(\phantom{0}---\phantom{0}) &  1.94 & -20.76\\
2999&  23.98&  23.43&  22.61&  21.58& 0.9076&  3& 2.0& 0.2& -10.1&   8.5& -64.9& 0.82(3800)
&  1.36 & -20.65\\
3091&  23.28&  22.57&  21.79&  20.52& 0.7413&  4& 2.2& 0.1& -22.9&  -1.5&   1.1
& 0.39(\phantom{0}---\phantom{0}) &  0.98 & -20.71\\
3101&  23.24&  22.45&  21.72&  20.37& 0.8650&  4& 2.4& 0.1&  -4.8&   4.3&   1.5& 0.39(3900)
& 1.14 &  -21.58\\
3107&  23.25&  22.59&  22.47&  21.65& 0.2392&  4& 0.8& 0.2&  \nodata&   3.4&  11.0&\nodata&  \nodata& -16.66\\
3201&  21.70&  20.78&  20.33&  19.56& 0.2746&  4& 2.8& 0.2&  \nodata&   0.3&  -1.4&\nodata&  \nodata& -18.99\\
3239&  24.27&  23.48&  22.90&  20.93& 0.0000&  4&\nodata&\nodata&  \nodata&  \nodata&  \nodata&\nodata&  \nodata&  \nodata\\
3268&  23.74&  22.24&  21.76&  19.89& 0.0000&  4&\nodata&\nodata&  \nodata&  \nodata&  \nodata&\nodata&  \nodata&  \nodata\\
3308&  23.76&  23.01&  23.04&  21.39& 0.4940&  1& 1.6& 0.2&  \nodata&  -1.0& -19.8& 0.62(3950) & \nodata & -18.55\\
3352&  25.18&  23.71&  22.62&  20.66& 0.8978&  4& $\geq$4.0&\nodata&   1.6&   5.8&  -3.4
& 0.45(4000) & 1.80 & -21.43\\
3405&  23.85&  23.52&  22.91&  21.36& 1.1269&  2& 2.4& 1.0& -14.5&  \nodata&  \nodata
& 0.45(3850) & 1.24 & -21.87\\
3413&  22.09&  21.25&  21.01&  20.62& 0.0000&  4&\nodata&\nodata&  \nodata&  \nodata&  \nodata&\nodata&  \nodata&  \nodata\\
3499&  23.82&  22.79&  23.04&  21.65& 0.4982&  4& 1.6& 0.2&  \nodata& -11.1& -31.7& 0.43(3800)&\nodata& -18.59\\
3542&  22.71&  22.59&  22.20&  21.01& 1.0854&  3& 1.4& 0.1& -69.9&  \nodata&  \nodata& \nodata&  \nodata& -21.94\\
3566&  22.73&  22.53&  22.18&  21.08& 0.9732&  4& 1.1& 0.1& -55.8&  \nodata&  \nodata&\nodata&  \nodata& -21.46\\
3594&  23.22&  22.82&  22.56&  21.05& 0.8298&  4& 1.3& 0.1& -44.1&  -5.2& -13.1& 0.28(3800)
& 1.22 & -20.43\\
3609&  21.37&  20.81&  20.59&  20.01& 0.1337&  4& 1.9& 0.1&  \nodata&  20.2& \nodata.6&\nodata&  \nodata& -17.00\\
3671&  23.40&  22.95&  22.63&  21.80& 0.9373&  3& 1.4& 0.1& -13.7& -50.8&  19.0
& 0.20(\phantom{0}---\phantom{0}) & 0.77 & -20.88\\
3716&  23.53&  23.09&  22.35&  21.35& 0.6088&  4& 1.3& 0.2& -42.9&   4.5& -19.8& 0.51(3850)
& 1.19 & -19.20\\
3719&  24.35&  23.52&  22.90&  20.98& 0.9032&  2& 3.0& 0.2&  -6.8&  \nodata&  \nodata
& 0.25(4000) & 1.98 & -20.96\\
3763&  22.45&  21.86&  21.60&  20.86& 0.3292&  4& 1.1& 0.1&  \nodata& -14.0& -59.4&\nodata&  \nodata& -18.45\\
3808&  23.75&  23.00&  22.67&  21.65& 0.6000&  1& 1.3& 0.2& -58.5&  -4.1&  40.0
&-0.25(\phantom{0}---\phantom{0}) &  1.11 & -18.97\\
3858&  24.36&  23.17&  22.57&  20.16& 0.0000&  4&\nodata&\nodata&  \nodata&  \nodata&  \nodata&\nodata&  \nodata&  \nodata\\
4014&  23.71&  23.04&  22.15&  20.49& 0.9038&  4& 2.8& 0.1&  -4.8&   5.7& -11.9& 0.38(4000)
& 2.01 & -21.45\\
4051&  23.28&  22.56&  22.07&  21.10& 0.3678&  2& 1.9& 0.2&  \nodata&  -4.3&  -5.7&\nodata&  \nodata& -18.09\\
4064&  25.44&  23.75&  22.56&  20.93& 0.8320&  2& 5.0& 0.4&  -0.4&   6.4&   8.1& 1.19(3800)
& 1.84 & -20.77\\
4091&  24.26&  23.52&  22.88&  21.38& 9.0000&\nodata&\nodata&\nodata&  \nodata&  \nodata&  \nodata&\nodata&  \nodata&  \nodata\\
4131&  24.62&  23.77&  23.32&  21.48& 0.9111&  3& 2.6& 0.2& -11.1&  \nodata&  \nodata
&-0.03(\phantom{0}---\phantom{0}) &  1.46 & -20.46\\
4210&  23.41&  23.13&  22.98&  21.83& 9.0000&\nodata&\nodata&\nodata&  \nodata&  \nodata&  \nodata&\nodata&  \nodata&  \nodata\\
4220&  23.58&  22.13&  21.36&  20.04& 0.0000&  4&\nodata&\nodata&  \nodata&  \nodata&  \nodata&\nodata&  \nodata&  \nodata\\
4238&  23.41&  23.15&  23.02&  21.83& 9.0000&\nodata&\nodata&\nodata&  \nodata&  \nodata&  \nodata&\nodata&  \nodata&  \nodata\\
4268&  23.81&  23.23&  22.92&  22.10& 0.4555&  4& 0.7& 0.3& -72.5& -21.1& -57.0&\nodata&  \nodata& -17.87\\
4319&  23.90&  22.82&  22.97&  21.89& 0.8938&  1& 1.7& 0.2&-114.5&  \nodata&  \nodata
& 0.48(3800) & \nodata & -20.70\\
4321&  23.93&  23.00&  22.13&  20.09& 0.0000&  4&\nodata&\nodata&  \nodata&  \nodata&  \nodata&\nodata&  \nodata&  \nodata\\
4329&  23.28&  23.03&  22.74&  21.60& 1.1999&  2& 1.7& 0.1& -92.0&  \nodata&  \nodata&\nodata&  \nodata& -21.88\\
4339&  23.35&  22.62&  22.41&  21.40& 0.2992&  4& 1.2& 0.2&  \nodata&  -8.5& -21.4&\nodata&  \nodata& -17.32\\
4376&  27.52&  24.38&  22.65&  20.81& 0.8296&  4& $\geq$4.0&\nodata&   3.3&  -0.4&   4.3
& 0.28(4000) & 1.92 & -20.85\\
4408&  22.04&  21.59&  21.40&  20.98& 0.0718&  4& 1.8& 0.1&  \nodata& -34.2&  -6.6&\nodata&  \nodata& -14.75\\
4576&  22.83&  21.59&  20.86&  19.39& 0.0000&  4&\nodata&\nodata&  \nodata&  \nodata&  \nodata&\nodata&  \nodata&  \nodata\\
4902&  23.72&  22.26&  21.18&  19.07& 0.0000&  4&\nodata&\nodata&  \nodata&  \nodata&  \nodata&\nodata&  \nodata&  \nodata\\
4979&  23.54&  22.80&  22.49&  21.68& 0.3335&  4& 1.1& 0.3&  \nodata&  -7.1& -28.5&\nodata&  \nodata& -18.49\\
5040&  23.83&  22.79&  22.04&  20.85& 0.3266&  3& 7.0& 1.0&  \nodata&  -0.5&  -1.3&\nodata&  \nodata& -17.66\\
5110&  24.10&  23.50&  22.86&  21.37& 0.6401&  4& 2.0& 0.2& -20.3&  -2.1&  -2.3& 0.85(3875)
& 1.41 & -19.18\\
5123&  23.57&  23.32&  22.85&  21.34& 0.9027&  4& 1.4& 0.1& -26.9&  21.9&  -4.3& 0.76(3800)
& 1.00 & -20.53\\
5229&  23.51&  22.78&  22.52&  21.60& 0.4564&  4& 1.0& 0.2& -45.5&  -0.4& -28.4&\nodata&  \nodata& -18.34\\
5244&  22.97&  22.63&  22.04&  21.05& 0.6007&  4& 1.0& 0.2& -23.3&   0.3&  -1.1&\nodata
&  1.39 & -19.46\\
5254&  23.53&  22.79&  22.54&  21.40& 0.2979&  4& 1.9& 0.2&  \nodata&-182.0&-168.7&\nodata&  \nodata& -17.21\\
5345&  25.00&  23.83&  23.20&  21.37& 0.0000&  2&\nodata&\nodata&  \nodata&  \nodata&  \nodata&\nodata&  \nodata&  \nodata\\
5393&  25.27&  23.82&  23.45&  21.08& 0.0000&  4&\nodata&\nodata&  \nodata&  \nodata&  \nodata&\nodata&  \nodata&  \nodata\\
5403&  24.36&  23.78&  23.34&  21.37& 9.0000&\nodata&\nodata&\nodata&  \nodata&  \nodata&  \nodata&\nodata&  \nodata&  \nodata\\
5435&  23.77&  23.46&  23.58&  21.68& 9.0000&\nodata&\nodata&\nodata&  \nodata&  \nodata&  \nodata&\nodata&  \nodata&  \nodata\\
5471&  23.54&  23.47&  23.56&  21.77& 0.6316&  1& 0.7& 0.1&   0.2& -39.6&  13.5
& 0.21(\phantom{0}---\phantom{0}) &  1.06 & -18.78\\
5541&  24.23&  23.77&  22.99&  21.08& 1.0360&  1& 3.0& 0.2&  -6.7&  \nodata&  \nodata
& \nodata& \nodata& -21.66\\
5567&  23.03&  22.92&  22.63&  21.41& 0.7745&  4& 0.7& 0.1& -87.3&-163.9&-198.4& 0.79(3850)& \nodata& -19.88\\
5593&  23.15&  23.04&  22.90&  21.74& 9.0000&\nodata&\nodata&\nodata&  \nodata&  \nodata&  \nodata&\nodata&  \nodata&  \nodata\\
5600&  23.91&  23.22&  22.72&  21.21& 0.4713&  4& 2.6& 0.2& -12.0&   0.9&  -1.7&\nodata&  \nodata& -18.43\\
5609&  24.04&  23.51&  22.89&  21.79& 0.6394&  4& 1.4& 0.2& -11.2&  -0.1&  -0.1& 0.54(\phantom{0}---\phantom{0})&  \nodata& -18.90\\
5638&  23.68&  23.12&  22.50&  20.92& 0.7804&  4& 2.0& 0.1& -32.2& -15.8&   2.4& 0.68(3825)
& 1.17 & -20.34\\
5720&  25.97&  24.12&  23.38&  20.96& 9.0000& \nodata&\nodata&\nodata&  \nodata&  \nodata&  \nodata&\nodata&  \nodata&  \nodata\\

\enddata
\end{deluxetable}

\clearpage

\begin{center}
Table 4. Mean Redshifts of Significant Structures
 
\begin{tabular}{lrcl} \hline \hline
Field Name & $N_{gal}$ & $\overline{z}$ & Comments \\ 
\hline
CL0023+0423 &  5 & 0.720 & Sheet?\\
            &  7 & 0.827 & Group \\
            & 17 & 0.845 & RC=0 cluster \\
CL1604+4304 &  7 & 0.497 & Group  \\
            &  8 & 0.829 & Sheet? \\
            & 22 & 0.897 & RC=2 cluster \\
\hline
\end{tabular}
\end{center}
\clearpage

\newpage

\renewcommand{\arraystretch}{0.8}
\begin{center}
Table 5. Spectrum Features Bands

\begin{tabular}{ccc}
\hline \hline
Continuum Band & Line Band & Identification  \\
\hline
3699-3711 &          &                       \\
          & 3712-3742 & [OII]                \\
3749-3761 &           &                      \\
3799-3821 &           &                      \\
          & 3819-3845 & CN, H9, etc.         \\
3849-3871 &           &                      \\
          & 3879-3899 & H8                   \\
3899-3921 &           &                      \\
          & 3923-3943 & CaII K               \\
          & 3959-3981 & CaII H, H$\epsilon$  \\
3985-4011 &           &                      \\
4065-4085 &           &                      \\
          & 4089-4111 & H$\delta$            \\
4115-4135 &           &                      \\
4239-4281 &           &                      \\
          & 4280-4320 & G-band               \\
          & 4329-4351 & H$\gamma$            \\
4355-4375 &           &                      \\
4789-4829 &           &                      \\
          & 4839-4879 & H$\beta$             \\
4899-4939 &           &                      \\
          & 4987-5027 & [OIII] 5007          \\
5029-5069 &           &                      \\
\hline
\end{tabular}
\end{center}

\newpage

\renewcommand{\arraystretch}{1.0}
\begin{center}
Table 6. Cluster Dynamical Parameters 

\begin{tabular}{cccccccccrrrccc} \hline \hline
{Cluster} &
{N$_z$} &
{$\overline z$} &
{$\sigma$} &
{M$_{\rm PW} \ (10^{14}$M$_\odot$)} &
{M$_{\rm PM} \ (10^{14}$M$_\odot$)} &
{M$_{\rm RW} \ (10^{14}$M$_\odot$)} &
{Radius (kpc)} \\
\hline
0023+0423A&  7 & 0.8274&158$^{+42}_{-33}$   & 0.10$^{+0.05}_{-0.04}$ & 0.36$\pm0.05$ & 0.33$\pm0.05$ & Unlimited \\
 & & & & & & \\
0023+0423B&  7 & 0.8438&418$^{+212}_{-93}$  & 1.49$^{+1.51}_{-0.67}$ & 1.47$\pm0.58$ & 1.99$\pm0.79$ & 250 \\
0023+0423B& 12 & 0.8448&381$^{+121}_{-68}$  & 1.68$^{+1.07}_{-0.61}$ & 1.70$\pm0.39$ & 2.93$\pm0.67$ & 500 \\
0023+0423B& 17 & 0.8453&415$^{+102}_{-63}$  & 2.60$^{+1.27}_{-0.79}$ & 4.17$\pm 0.68$ & 5.61$\pm0.91$& Unlimited \\
 & & & & & & \\
1604+4304 & 11 & 0.8964& 921$^{+303}_{-155}$ & 2.45$^{+1.68}_{-0.95}$ & 4.21$\pm0.76$ &  9.25$\pm1.67$ & 250 \\
1604+4304 & 19 & 0.8967&1300$^{+286}_{-173}$ & 6.86$^{+3.11}_{-1.97}$ & 27.3$\pm2.94$ &  31.3$\pm3.37$ & 500 \\
1604+4304 & 22 & 0.8967&1226$^{+245}_{-154}$ & 7.77$^{+3.19}_{-2.10}$ & 25.3$\pm2.33$ &  31.3$\pm2.88$ & Unlimited \\
\hline
\end{tabular}
\end{center}
\clearpage

\begin{center}
Table 7. M/L Results for CL0023+0423

\begin{tabular}{ccrrl} \hline \hline
     &Radius &   &    &   \\
Band & (kpc) &  $M_{PW}/L$ & $M_{PM}/L$ & $M_{RW}/L$ \\
\hline
   B & 250 &  75$^{+ 81}_{- 44}$ &   74$\pm40$& 100$\pm 55$ \\
     & 500 & 183$^{+122}_{- 75}$ &  183$\pm55$& 312$\pm 94$ \\
   V & 250 &  83$^{+ 88}_{- 45}$ &   82$\pm41$& 110$\pm 56$ \\
     & 500 &  67$^{+ 44}_{- 26}$ &   67$\pm19$& 113$\pm 32$ \\
   R & 250 & 172$^{+181}_{- 93}$ &  169$\pm84$& 229$\pm114$ \\
     & 500 &  98$^{+ 72}_{- 43}$ &   98$\pm26$& 166$\pm 45$ \\
\hline
\end{tabular}
\end{center}
\clearpage
 
\begin{center}
Table 8. M/L Results for CL1604+4304
 
\begin{tabular}{ccrll} \hline \hline
     &Radius &   &    &   \\
Band & (kpc) &  $M_{PW}/L$ & $M_{PM}/L$ & $M_{RW}/L$ \\
\hline
   B & 250 & 186$^{+147}_{-102}$ & 320$\pm138$& 702$\pm304$ \\
     & 500 &  49$^{+ 24}_{- 17}$ & 195$\pm 43$& 224$\pm 49$ \\
   V & 250 & 129$^{+ 96}_{- 63}$ & 222$\pm 77$& 487$\pm168$ \\
     & 500 &  52$^{+ 25}_{- 17}$ & 206$\pm 37$& 236$\pm 43$ \\
   R & 250 & 142$^{+105}_{- 68}$ & 244$\pm 83$& 536$\pm181$ \\
     & 500 & 123$^{+ 59}_{- 40}$ & 490$\pm 90$& 562$\pm103$ \\
\hline 
\end{tabular}
\end{center}
\clearpage

\begin{table}
\begin{center}
Table 9. Calculated Emission Line Equivalent widths and Relative Intensities

\begin{tabular}{rrrrrrrr} \hline \hline
Age(Gyr) &EW(H$\beta$)\tablenotemark{a} &EW(H$\alpha$)\tablenotemark{b} &EW[OII] &EW[OIII]  
&I[OII]/I[OIII]\tablenotemark{c} &I[OII]/I(H$\beta$)\tablenotemark{c} &
I[OIII]/I(H$\beta$)\tablenotemark{c} \\
\hline
0.3 & 31.2 & 155.0 & 126.0 & 55.6 & 1.44 & 3.23 & 2.24 \\
1.0 & 9.4 & 43.0 & 56.4 & 16.8 & \nodata & \nodata & \nodata \\
2.0 & 3.1 & 11.9 & 19.6 & 5.6 & \nodata & \nodata & \nodata \\
3.0 & 0.9 & 3.0 & 7.2 & 1.7 & \nodata & \nodata & \nodata\\
4.0 & 0.3 & 1.0 & 2.6 & 0.6 & \nodata & \nodata & \nodata \\
\hline
\tablenotetext{a}{Calculated from equation 10}
\tablenotetext{b}{Calculated from equation 9}
\tablenotetext{c}{From McCall, Rybski, and Shields (1985)}
\end{tabular}
\end{center}
\end{table}

\clearpage

\begin{center}
Table 10. Cluster Age Estimates

\begin{tabular}{lrrrrrc} \hline \hline
Group & Number & Colage & Spage(eye) & Spage($\chi^2$) & Jump($\lambda_J$) & D(4000) \\
\hline
CL0023,z=0.8453,0.8267, old & 3 & 3.7 & 2.4 & 2.8(0.28) & 0.45(4000) & 1.45 \\
CL0023,z=0.8453,0.8267, med & 9 & 2.3 & 2.0 & 2.3(0.40) & 0.40(4000) & 1.29 \\
CL0023,z=0.8453,0.8267, young & 9 & 1.1 & 1.3 & 2.7(0.76) & 0.53(3800) & 1.13 \\
CL0023, z=0.9140,0.937, old & 1 & $\geq$4.0 & 3.0 & 6.0(0.97) & 0.52(4000) & 1.74 \\
CL0023, z=0.914,0.937, young & 4 & 1.2 & 1.0 & young  & 0.68(3800) & 1.16 \\
CL0023, z=0.7200,0.7730, young & 7 & 1.3 & 1.8 & 2.4(0.44) & 0.53(\phantom{0}---\phantom{0})
& 1.10 \\
CL0023, z=0.6-0.8, old & 2 & 4.7 & 3.0 & 3.6(0.45) & 0.49(4000) & 1.79 \\
CL0023, z=0.6-0.8, med & 1 & 2.6 & 1.6 & 2.9(0.84)  & 0.54(4000) & 1.41 \\
CL0023, z=0.6-0.8, young & 4 & 1.0 & 1.3 & 1.6(1.15)  & 0.60(\phantom{0}---\phantom{0})
& 1.20  \\
CL1604, z=0,8967, old   & 5 & $\geq$4.0 & 3.0 & 3.4(0.40) & 0.47(4000) & 1.66 \\
CL1604, z=0,8967, med & 6 & 2.6 & 2.6 & 2.6(0.23) & 0.43(4000) & 1.61 \\
CL1604, z=0,8967, young & 7 & 1.4 & 1.0 & 2.5(1.07) & 0.43(3800) & 1.02 \\
CL1604, z=0.8290, med & 3 & 2.8 & 2.0 & 3.0(0.71) & 0.02(\phantom{0}---\phantom{0}) 
& 1.82 \\
CL1604, z=0.8290, young & 4 & 1.5 & 1.0 & 2.2(0.79) & 0.51(3850) & 1.92 \\
CL1604, z=0.4959, old & 2 & $\geq$7.0 & 3.0 & 5.0(0.57) & & 2.05 \\
CL1604, z=0.4959, young & 8 & 1.1 & 1.4 & 2.2(0.88) & & 1.32 \\
CL1604, z=0.7782, young  & 3 & 1.3 & 1.6: & 3.6(1.52) & 0.74(3800) & 1.04 \\
CL1604, z=0.9762, young  & 2 & 1.4 & 1.6 & 2.5(0.57) & 0.53(3850) & 1.53 \\
\hline
\end{tabular}
\end{center}
\clearpage

\begin{table}
\begin{center}
Table 11. Least Active Galaxies

\begin{tabular}{lrrrrr} \hline \hline
Obs/Model & ABB & ABV & ABR & ABI & Age(Gyr) \\
\hline
observed\tablenotemark{a}  & 24.90 & 23.67 & 22.70 & 20.59 & \nodata \\
ssp    & 25.28 & 23.46 & 22.36 & 20.79 & 3.2$\pm0.5$ \\
tau0.6 & 25.40 & 23.48 & 22.31 & 20.72 & 5.0$\pm0.5$ \\ 
tau1.0 & 25.25 & 23.60 & 22.39 & 20.72 & 8.0$\pm0.5$ \\
\hline
\tablenotetext{a}{Keck \#2055[Sa(pec)] in the CL0023+0423 field;  Keck \#2159, 
2232(E+E), 2851, 2855(Sa), and 3352(E) in the CL1604+4304 field.} 
\end{tabular}
\end{center}
\end{table}

\clearpage

\begin{table}
\begin{center}
Table 12. Brightest Cluster Galaxies

\begin{tabular}{lcccc|cccc|cc} \hline \hline
Cluster & Keck \# & \multicolumn{3}{c|}{Observed BCG}&\multicolumn{4}{c|}{Equivalent Low $z$ BCG\tablenotemark{a}}&\multicolumn{2}{c}{Comparison} \\
&&$M_{ABB}$ &$M_B$ &$M_V$ & RC &$M_g$ &$M_B$&$M_V$& $\Delta M_B$ & $\Delta M_V$ \\
\hline 
CL0023+0423 & 2055\tablenotemark{b}& -21.63 & -21.47 & -22.27 & 0 & -21.13 & -20.45 & -21.42 & 1.02 & 0.85 \\
CL1604+4304 & 2855\tablenotemark{c}& -21.91 & -21.75 & -22.55 & 2 & -21.41 & -20.73 & -21.70 & 1.02 & 0.85 \\
\hline

\tablenotetext{a}{The absolute magnitudes of brightest cluster
galaxies in clusters at low redshift (Schneider, Gunn \& Hoessel 1983)
with similar richness classes (RC) to CL0023+0423 (richness class 0)
and CL1604+4304 (richness class 2).}  

\tablenotetext{b}{$z = 0.8266$; RA = $00^{\rm h}$ $23^{\rm m}$
$54.5^{\rm s}$, Dec = $04^{\rm o}$ $23'$ ${8}\farcs{9}$.}  

\tablenotetext{c}{$z = 0.8964$; RA = $16^{\rm h}$ $04^{\rm m}$
$25.0^{\rm s}$, Dec = $43^{\rm o}$ $04'$ ${51}\farcs{5}$.}  

\end{tabular}
\end{center}
\end{table}

\clearpage

\textheight=9.5in
\textwidth=7.0in
\hoffset -0.2in
\voffset -0.5in

% plot specification

\def\eps@scaling{.95}
\def\epsscale#1{\gdef\eps@scaling{#1}}

\def\plottwo#1#2{\centering \leavevmode
    \epsfxsize=.45\columnwidth \epsfbox{#1} \hfil
    \epsfxsize=.45\columnwidth \epsfbox{#2}}

\begin{figure}
\begin{center}
PLATE \#1 HERE
\end{center}
\caption{A composite BVR image of the central 3.6 arcminutes
of the CL0023+0423 field. Confirmed cluster members are marked by open squares.
North is at top, west is to the right.}
\label{fig-cl00im}
\end{figure}
 
\clearpage
 
\begin{figure}
\begin{center}
PLATE \#2 HERE
\end{center}
\caption{A composite BVR image of the central 3.6 arcminutes
of the CL1604+4304 field. Confirmed cluster members are marked by open squares.
North is at top, west is to the right.}
\label{fig-cl16im}
\end{figure}

\clearpage

\begin{figure}
\epsscale{1.0}
\plotone{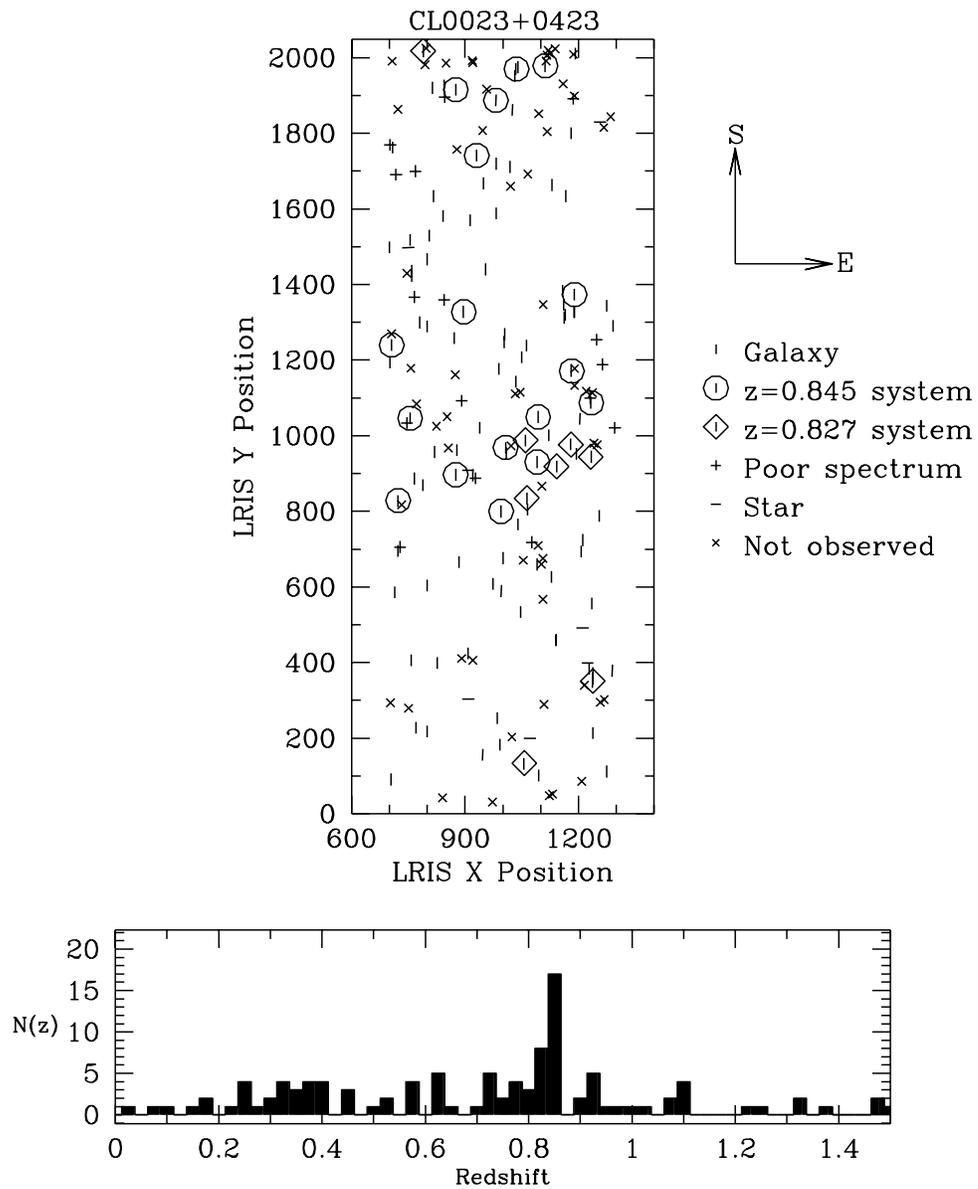}
\caption{The distribution on the sky and in redshift of the spectroscopic
targets in CL0023+0423. One LRIS pixel spans $0.215''$. Stars are not
shown in redshift histogram.}
\label{fig-cl00targ}
\end{figure}

\clearpage
\begin{figure}
\epsscale{1.0}
\plotone{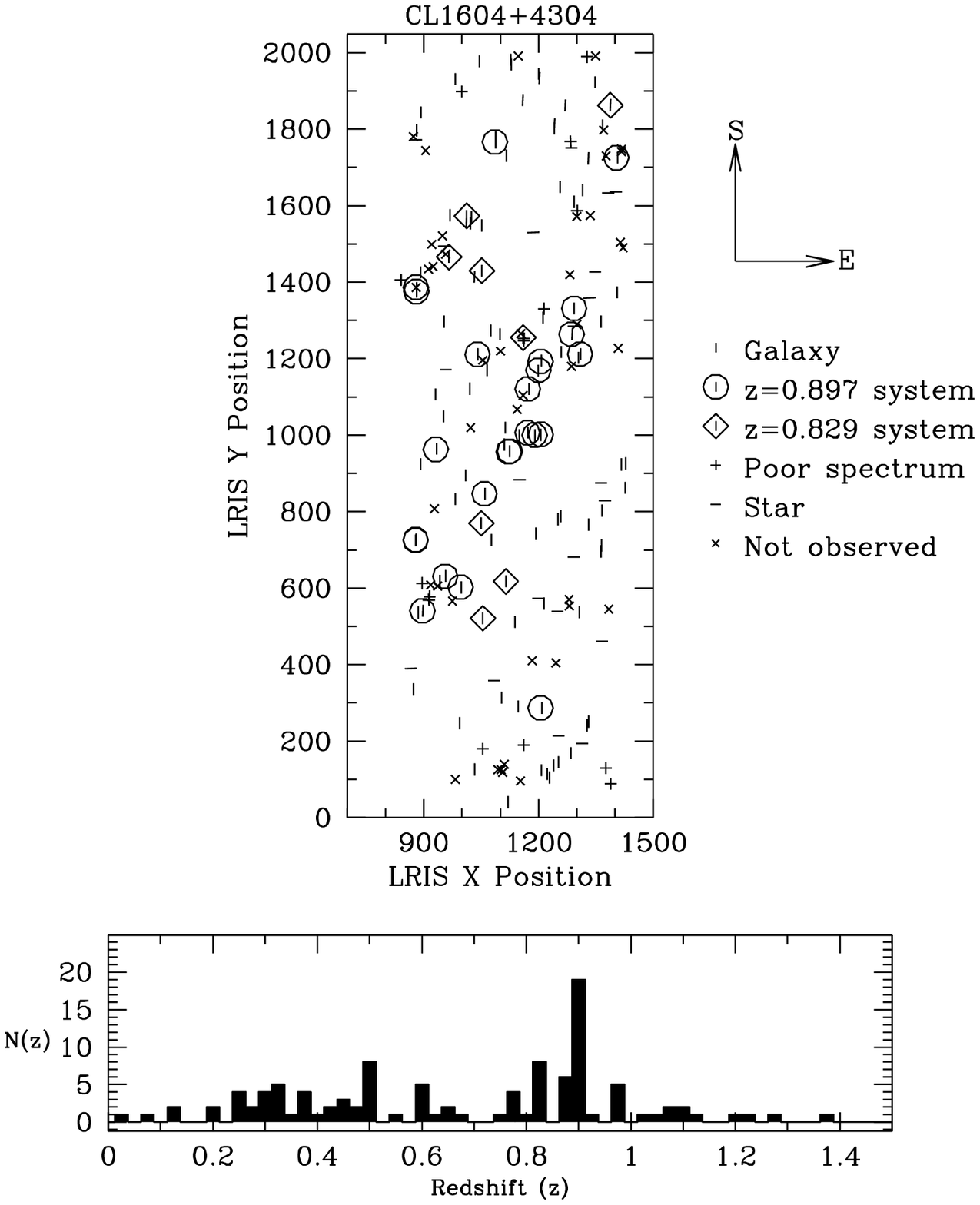}
\caption{Same as Figure~\ref{fig-cl00targ} but for CL1604+4304.}
\label{fig-cl16targ}
\end{figure}

\clearpage

\begin{figure}
\epsscale{0.5}
\plottwo{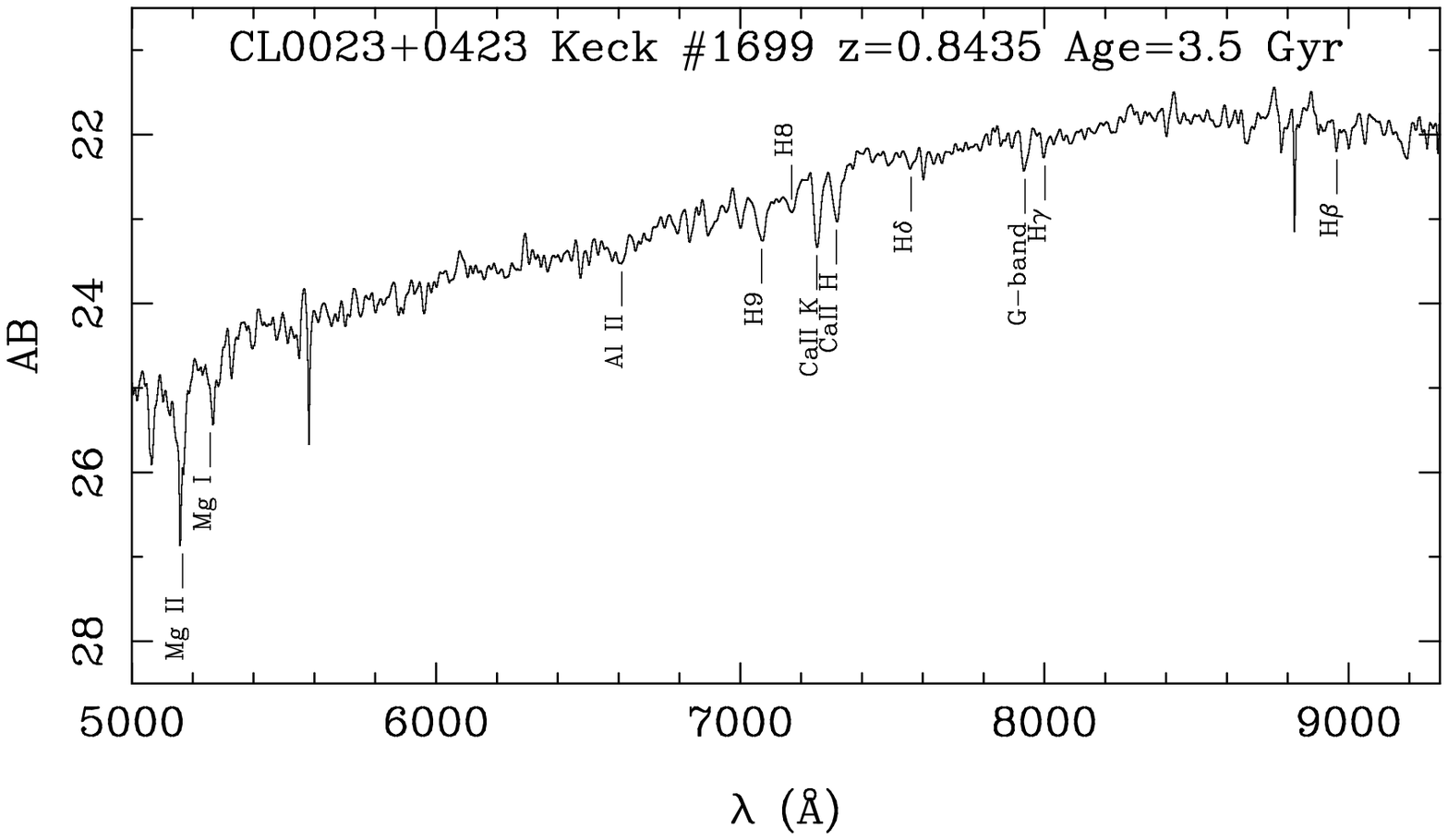}{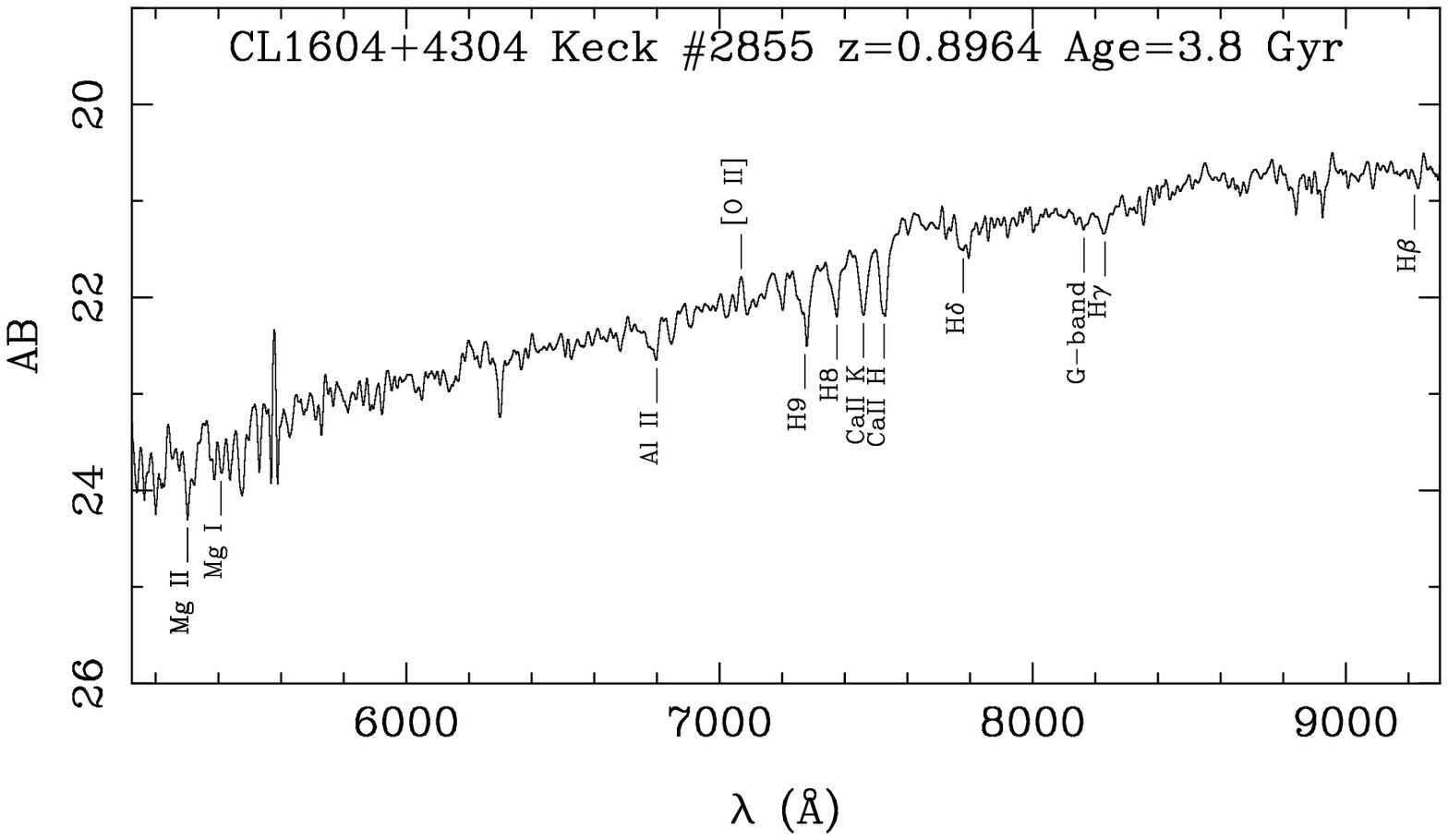}
\epsscale{0.5}
\plottwo{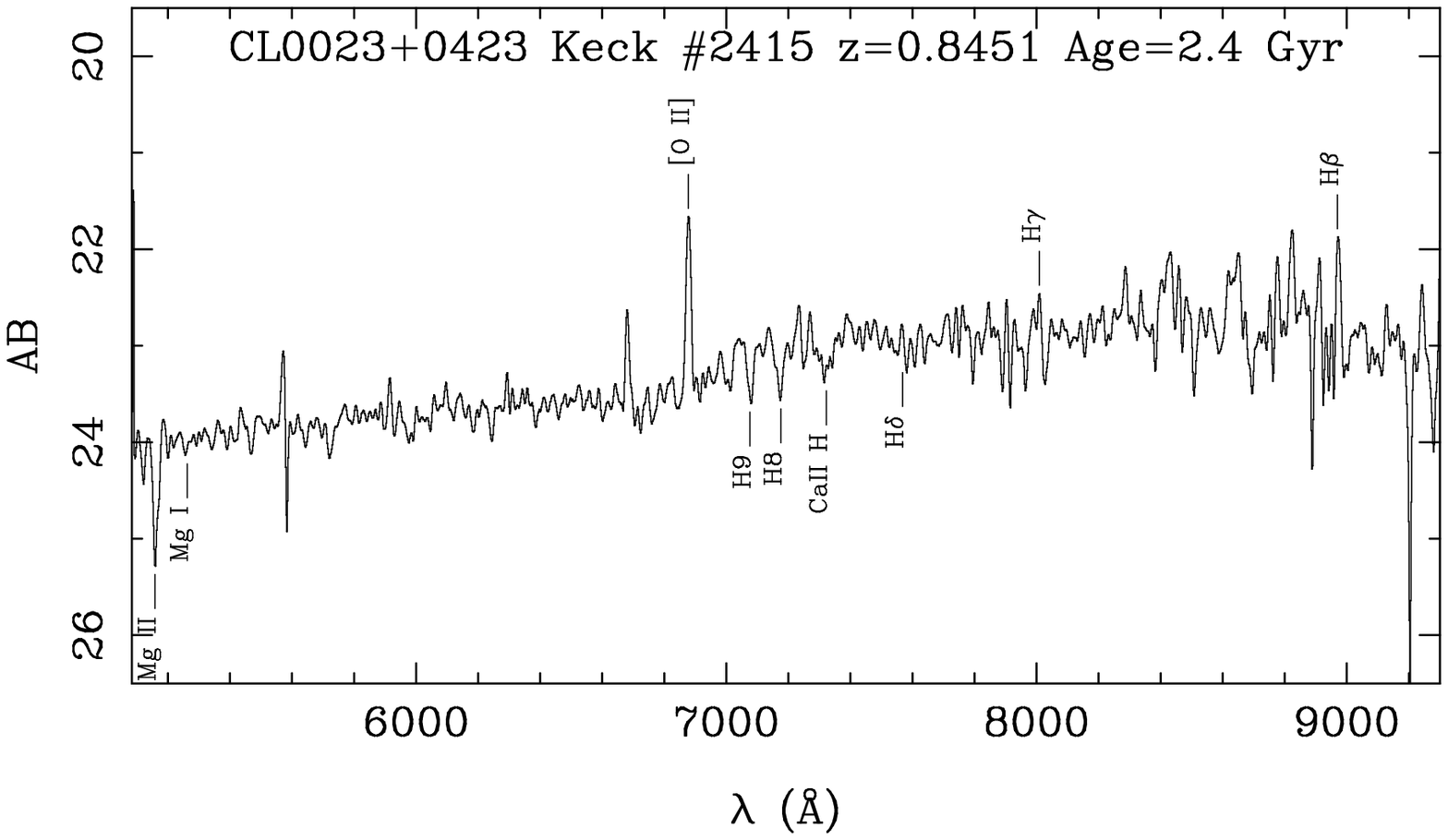}{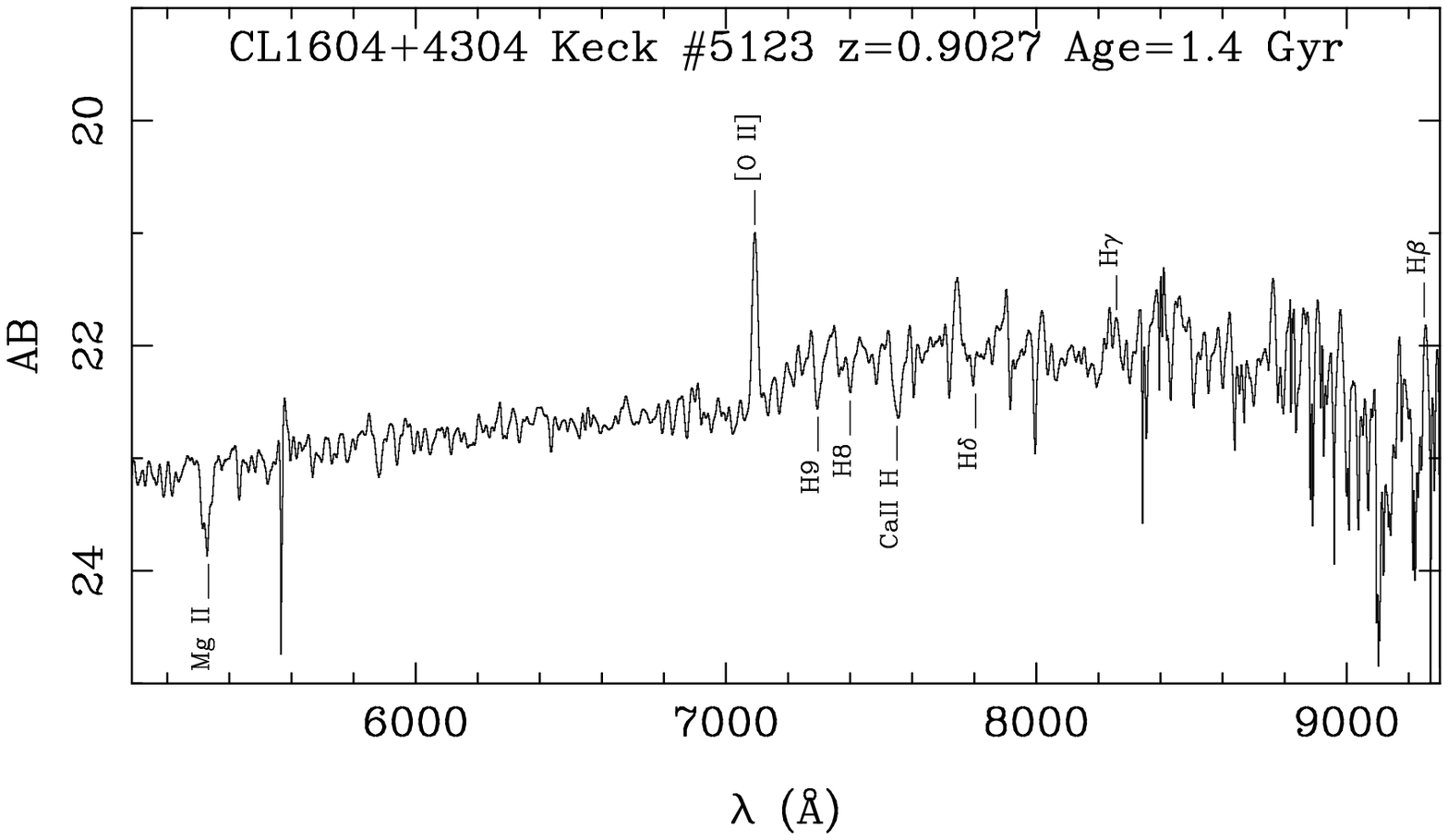}
\epsscale{0.5}
\plottwo{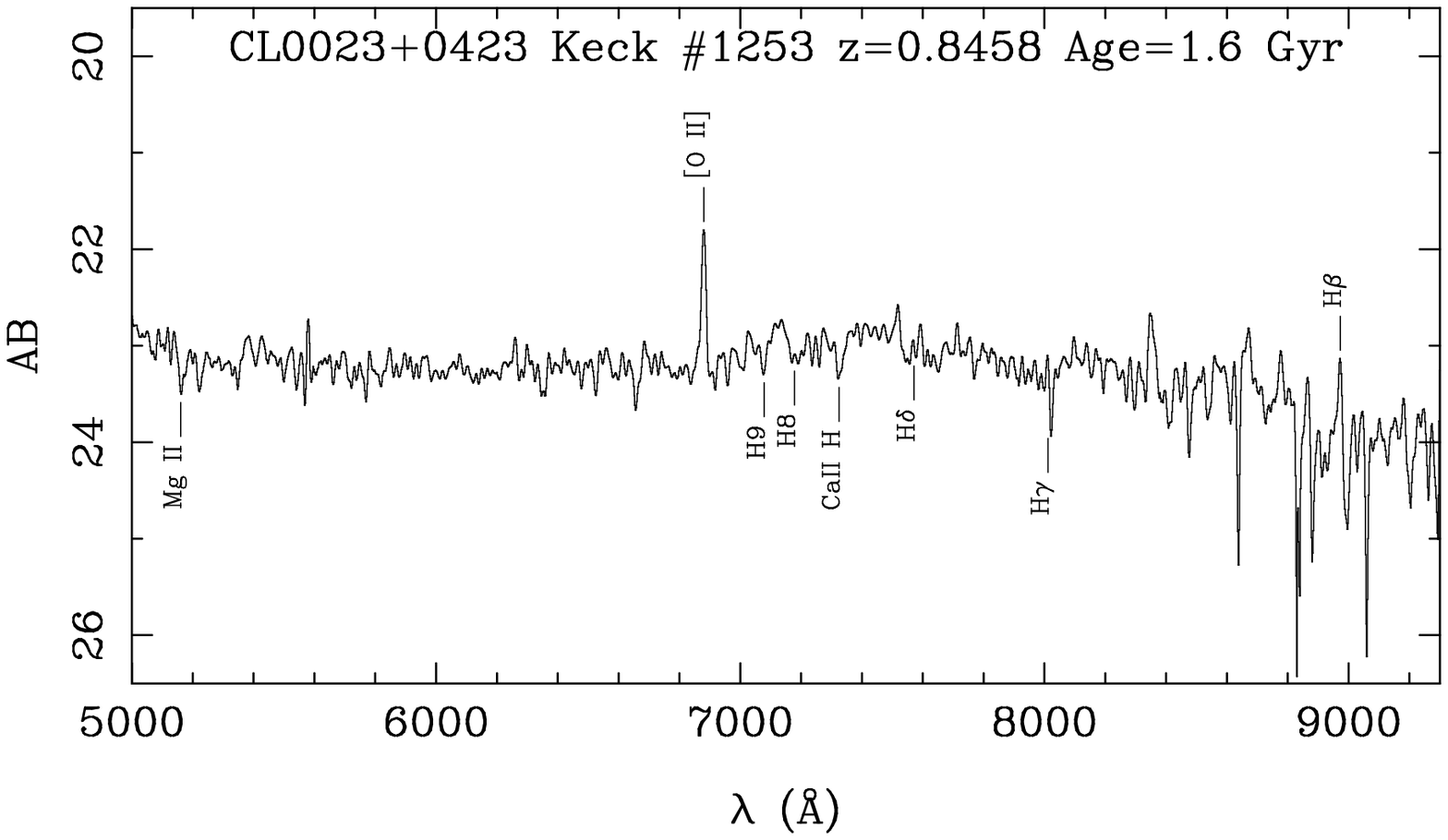}{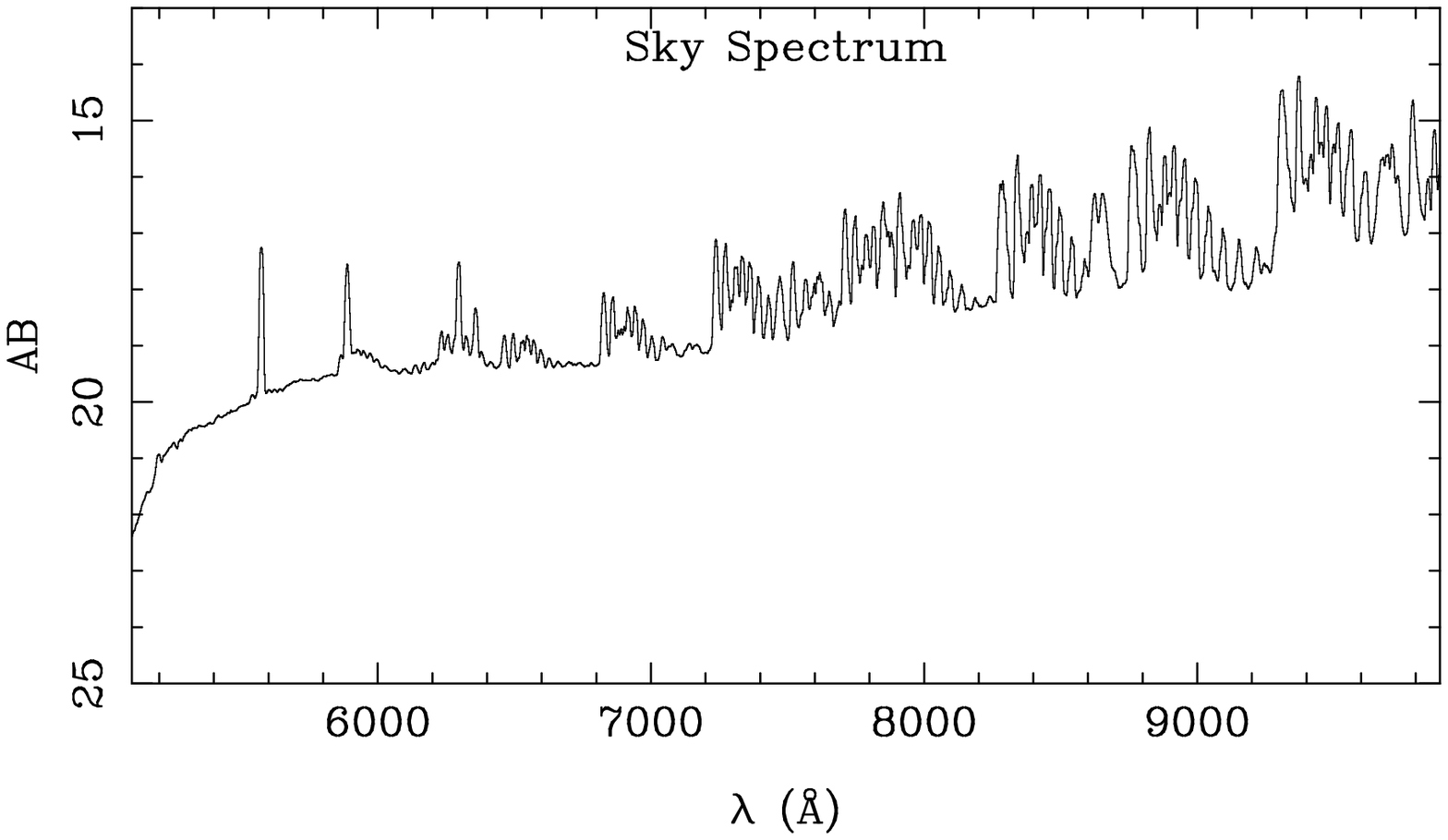}
\caption{The flux, in AB mags, as a function of wavelength, for sample
spectra.}
\label{fig-samplespectra}
\end{figure}

\clearpage

\begin{figure}
\epsscale{1.0}
\plotone{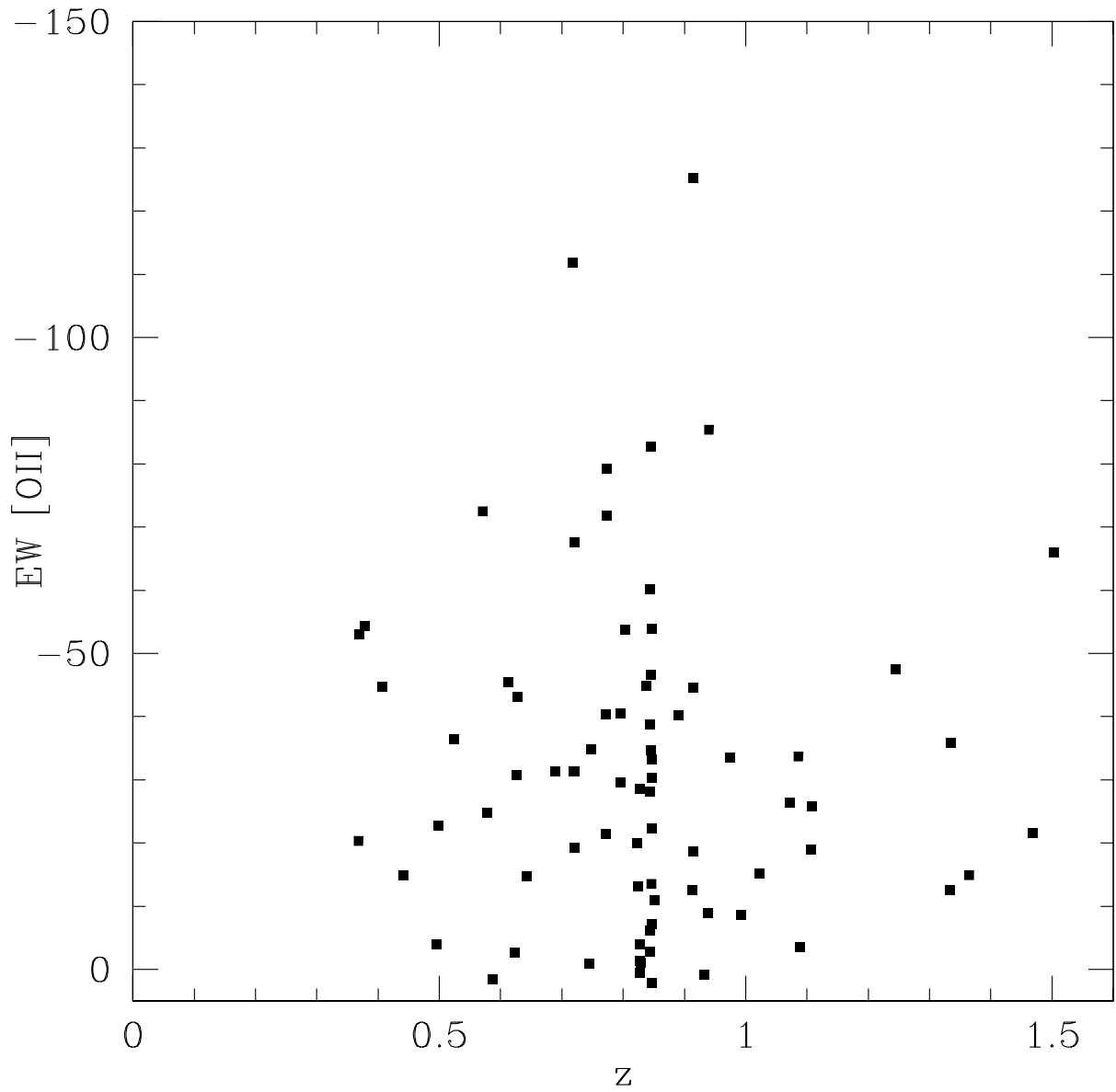}
\caption{The rest equivalent widths of [OII] $\lambda$3727 versus redshift z
for the field CL0023+0423.}
\label{fig-cl00eqwz}
\end{figure}

\clearpage
\begin{figure}
\epsscale{1.0}
\plotone{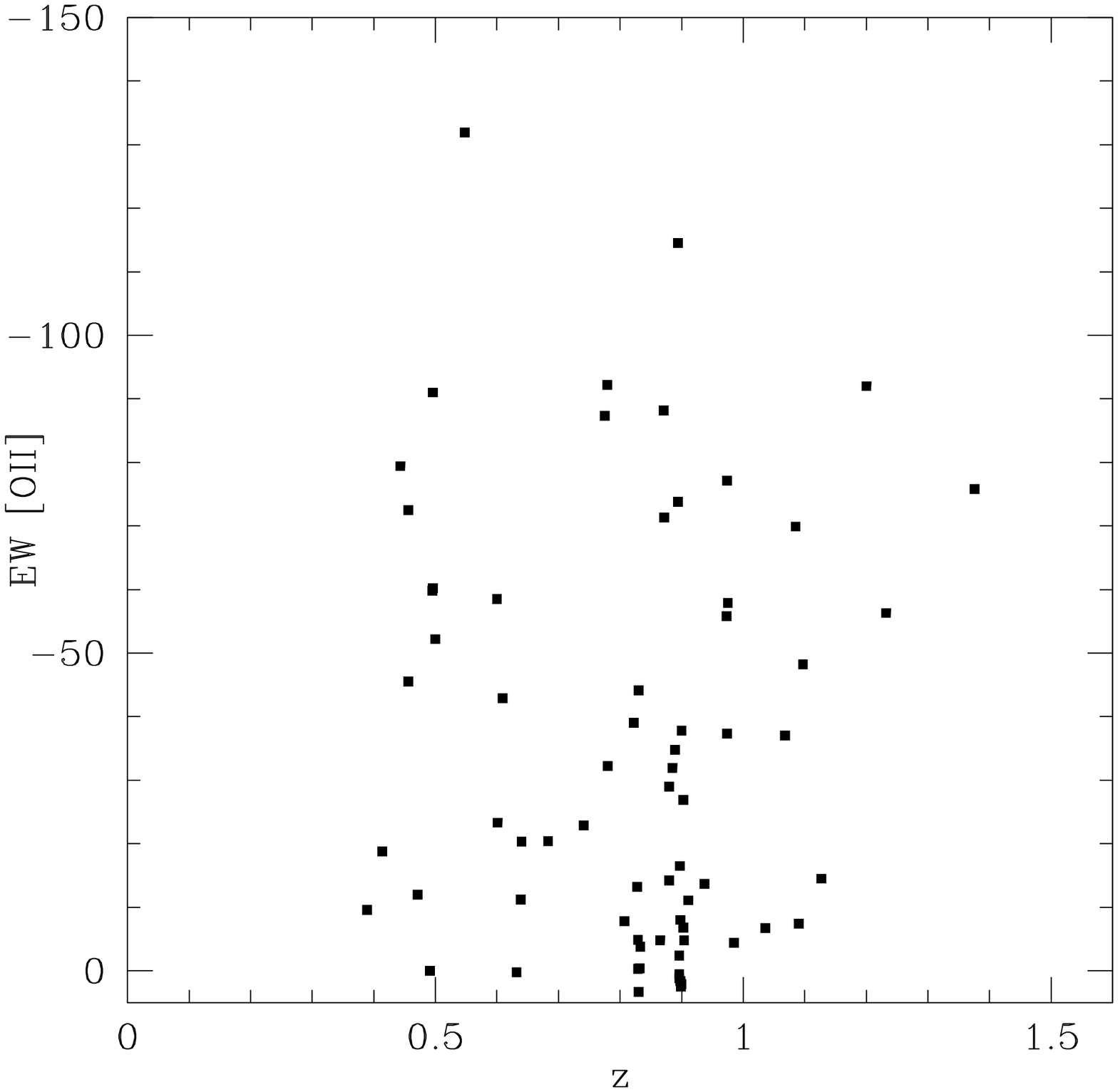}
\caption{Same as Figure~\ref{fig-cl00eqwz} but for CL1604+4304.}
\label{fig-cl16eqwz}
\end{figure}

\clearpage
\begin{figure}
\epsscale{1.0}
\plotone{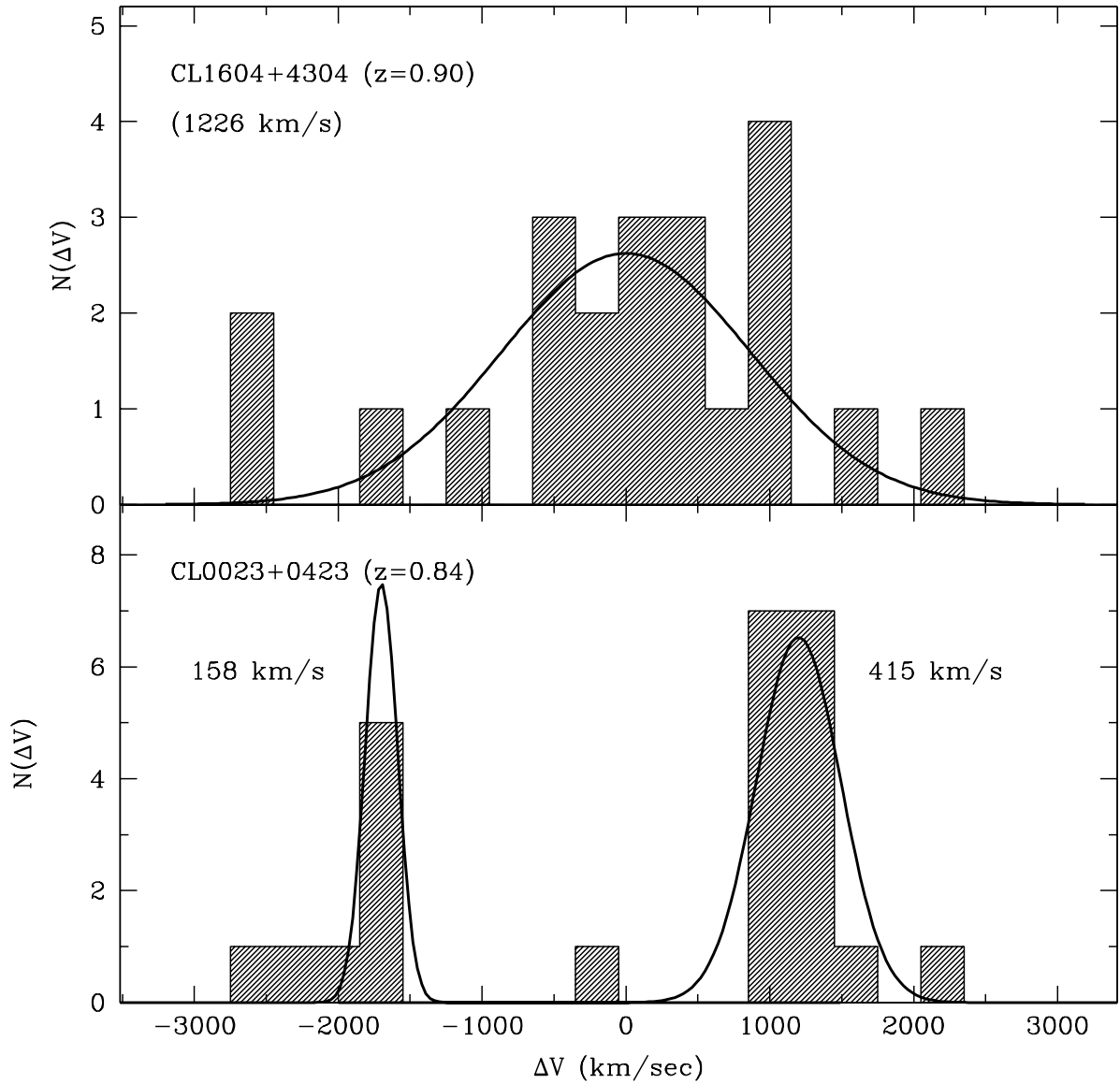}
\caption{Histogram of the relativistically corrected velocity offsets
for CL0023+0423 (bottom) and CL1604+4304 (top). Offsets are relative
to the mean cluster redshift. Best fit Gaussian distributions are shown
for comparison.}
\label{fig-velhist}
\end{figure}

\clearpage
\begin{figure}
\epsscale{1.0}
\plotone{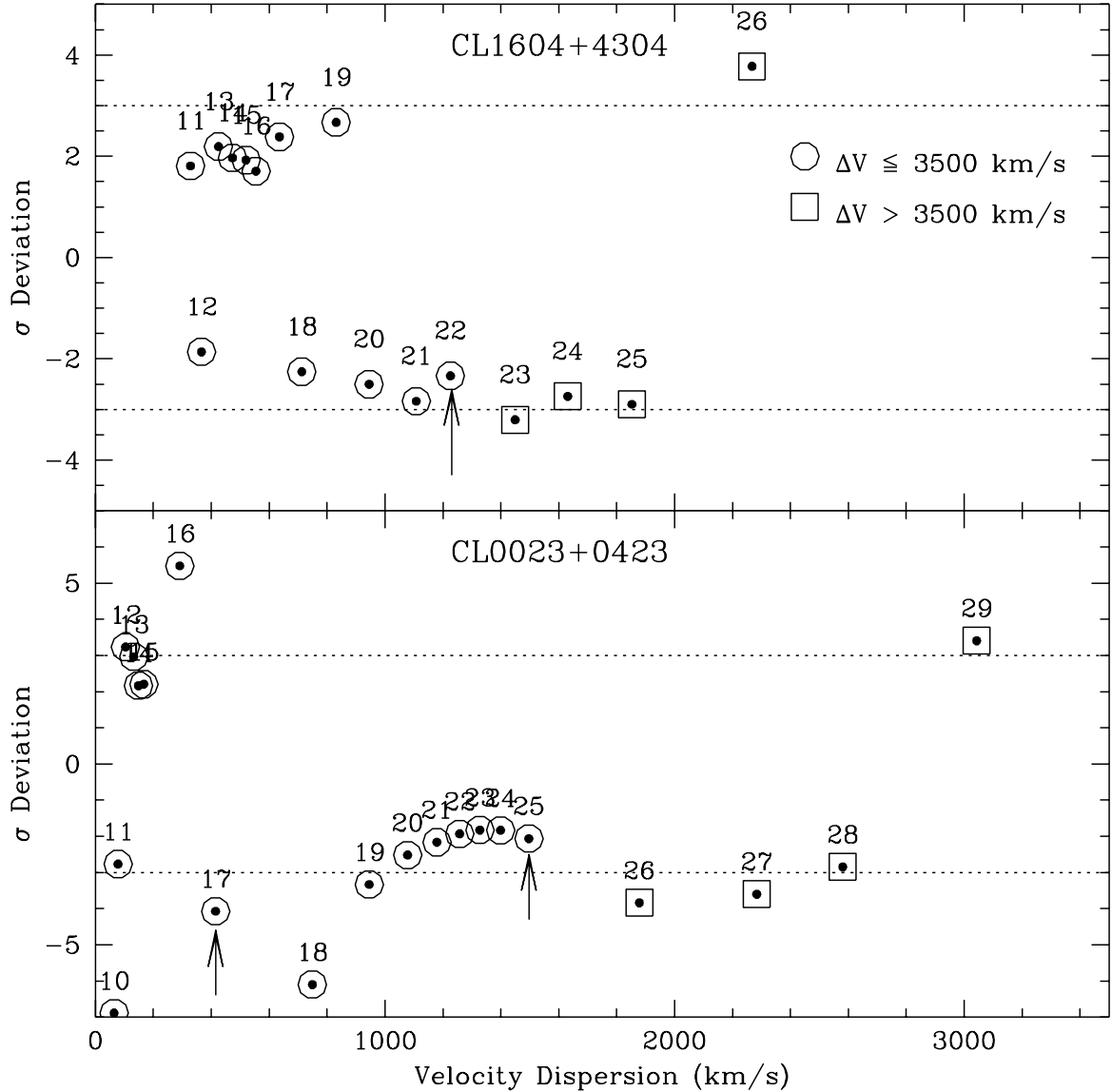}
\caption{The derived velocity dispersion as a function of the
significance of the velocity difference between the cluster bi-weight
mean redshift and that of the most deviant galaxy in the sample. This
shows how the dispersion changes as we remove these outliers from the
analysis.  The number of galaxies remaining in each iteration is shown
above each data point. Those outliers which differ by more than 3500
km s$^{-1}$ are denoted by large open squares. Such galaxies are not
likely to be cluster members even if their deviation is slightly less
than 3$\sigma$. In the case of CL0023+0423, the two arrows highlight
the dispersion after the clipping procedure terminates (1497 km
s$^{-1}$) and the adopted dispersion of the $z=0.8453$ component (415
km s$^{-1}$). The arrow in the plot for CL1604+4304 shows the adopted
dispersion (1226 km s$^{-1}$) after the clipping procedure
terminates.}
\label{fig-veldisp}
\end{figure}

\clearpage
\begin{figure}
\epsscale{1.0}
\plotone{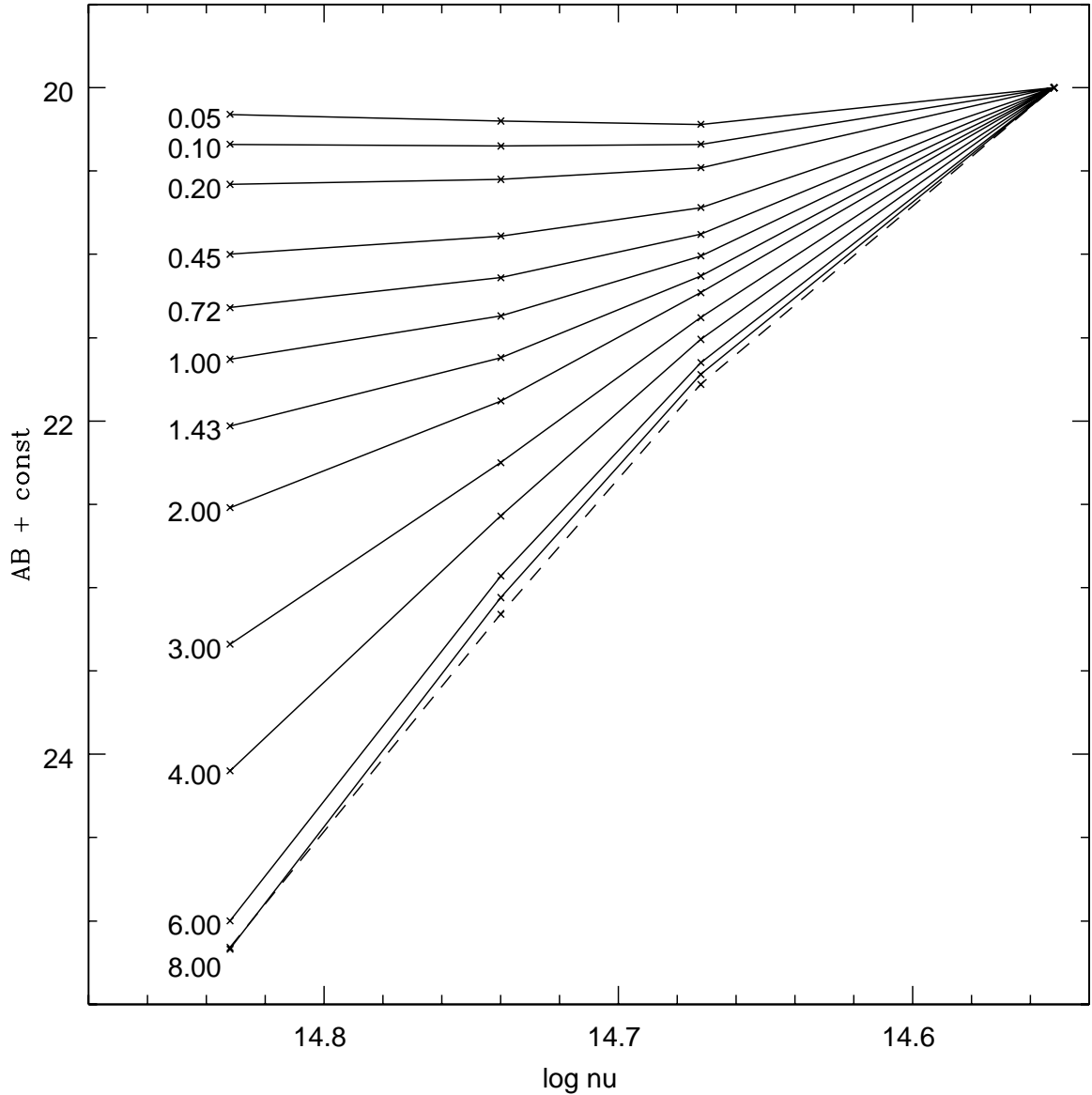}
\caption{Relative values of AB calculated from the tau0.6 models at $z
= 0.8967$ versus log$\nu$.  The four points from left to right
correspond to the B,V,R,I bands.  The ages from top to bottom
correspond to 0.05, 0.10, 0.20, 0.45, 0.72, 1.00, 1.43, 2.00, 3.00,
4.00, 6.00, and 8.00 Gyr.  Note that the energy distributions become
very similar for ages above 6 Gyr.  The energy distribution shown by
the broken curve is for an age of 10 Gyr.}
\label{fig-ABlognu}
\end{figure}

\clearpage
\begin{figure}
\epsscale{1.0}
\plotone{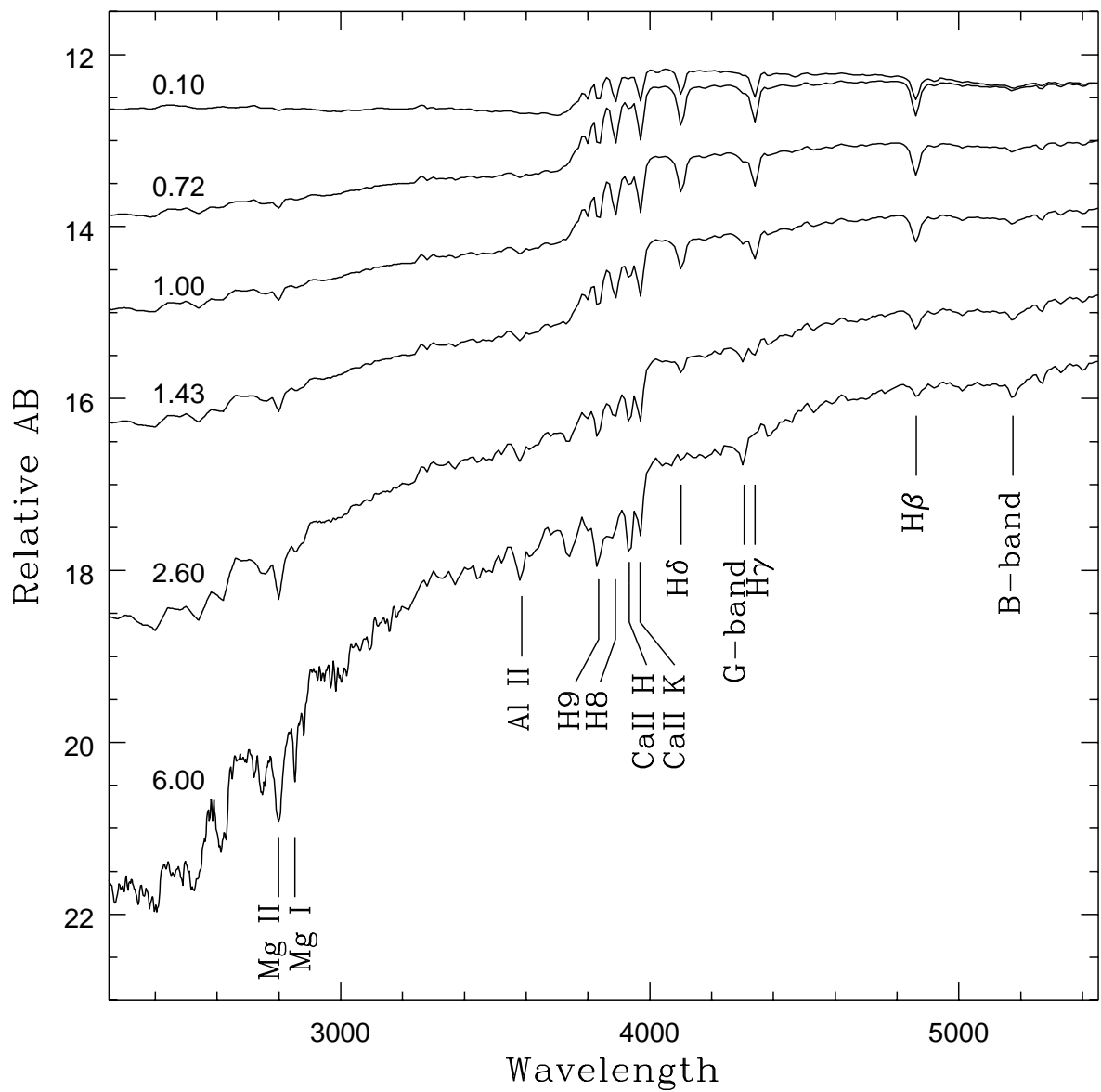}
\caption{A selection of tau 0.6 model spectra.  Relative values of AB
are plotted against the rest wavelength in Angstroms.  The ages
indicated are in Gyr.  The more significant spectral features are
marked.}
\label{fig-modspec}
\end{figure}

\clearpage
\begin{figure}
\epsscale{1.0}
\plotone{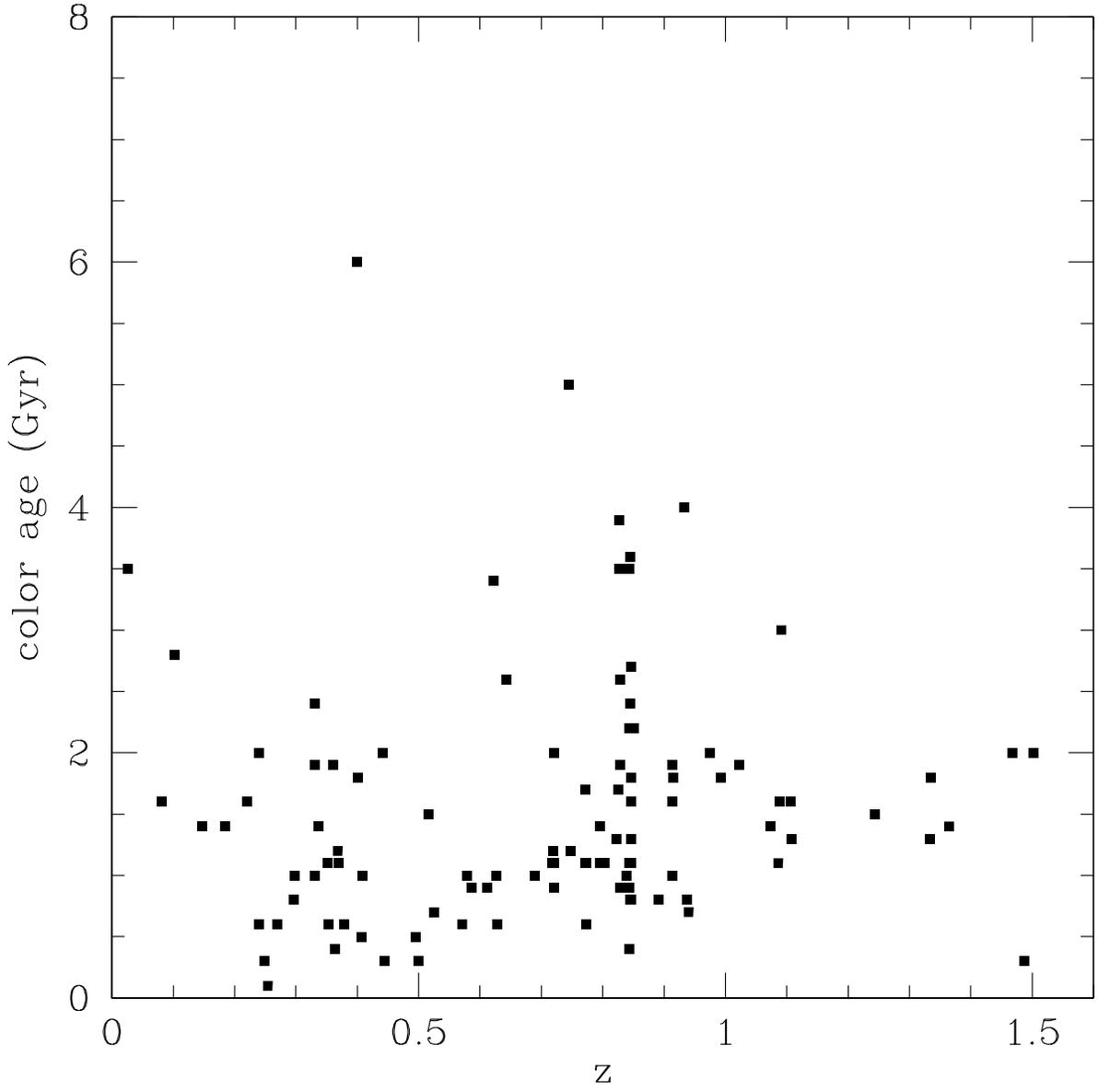}
\caption{The color age based on tau0.6 models versus redshift $z$ for
the CL0023+0423 field.}
\label{fig-cl00agez}
\end{figure}

\clearpage
\begin{figure}
\epsscale{1.0}
\plotone{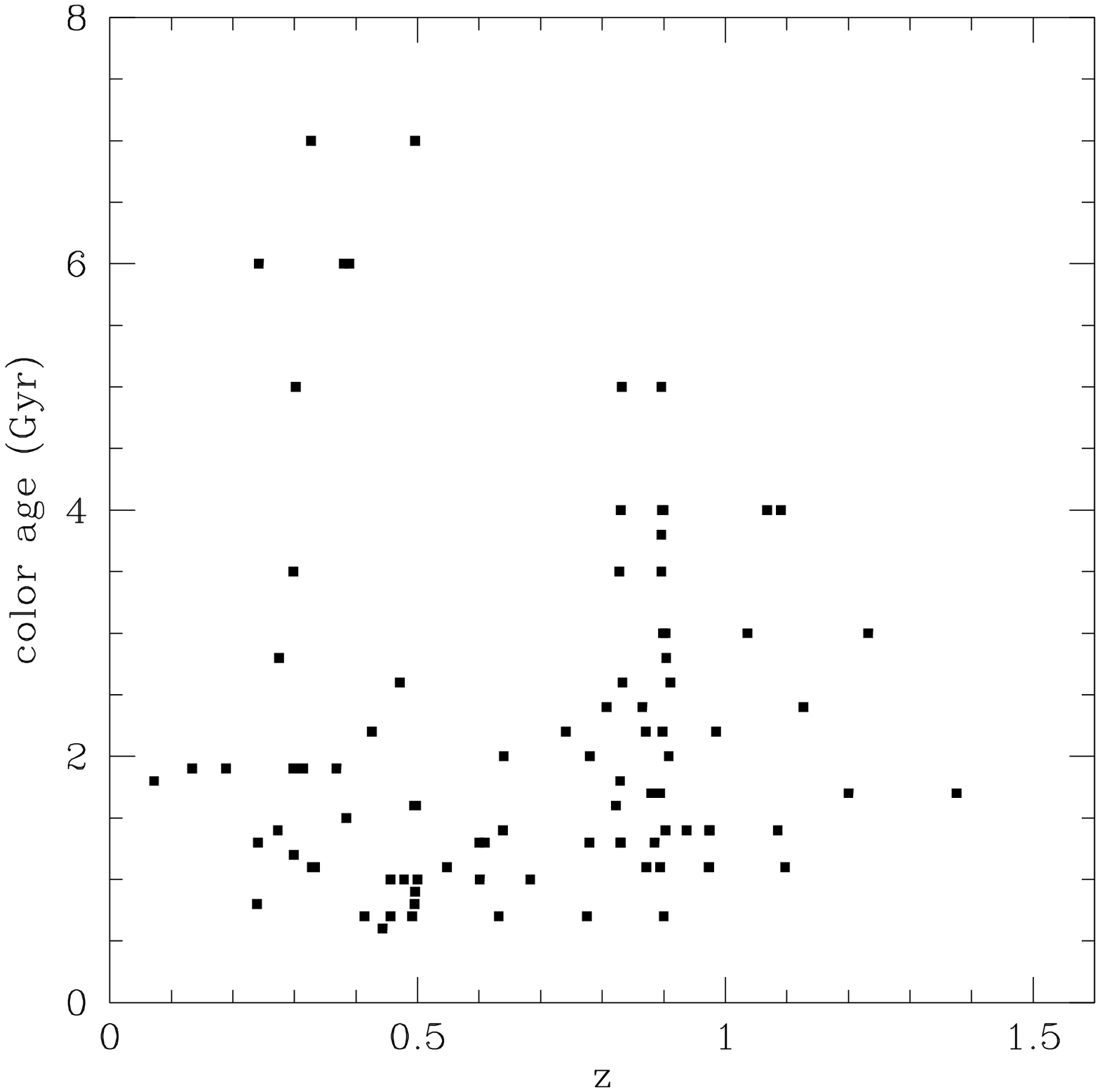}
\caption{Same as Figure~\ref{fig-cl00agez} but for the CL1604+4304 field.}
\label{fig-cl16agez}
\end{figure}

\clearpage
\begin{figure}
\epsscale{1.0}
\plotone{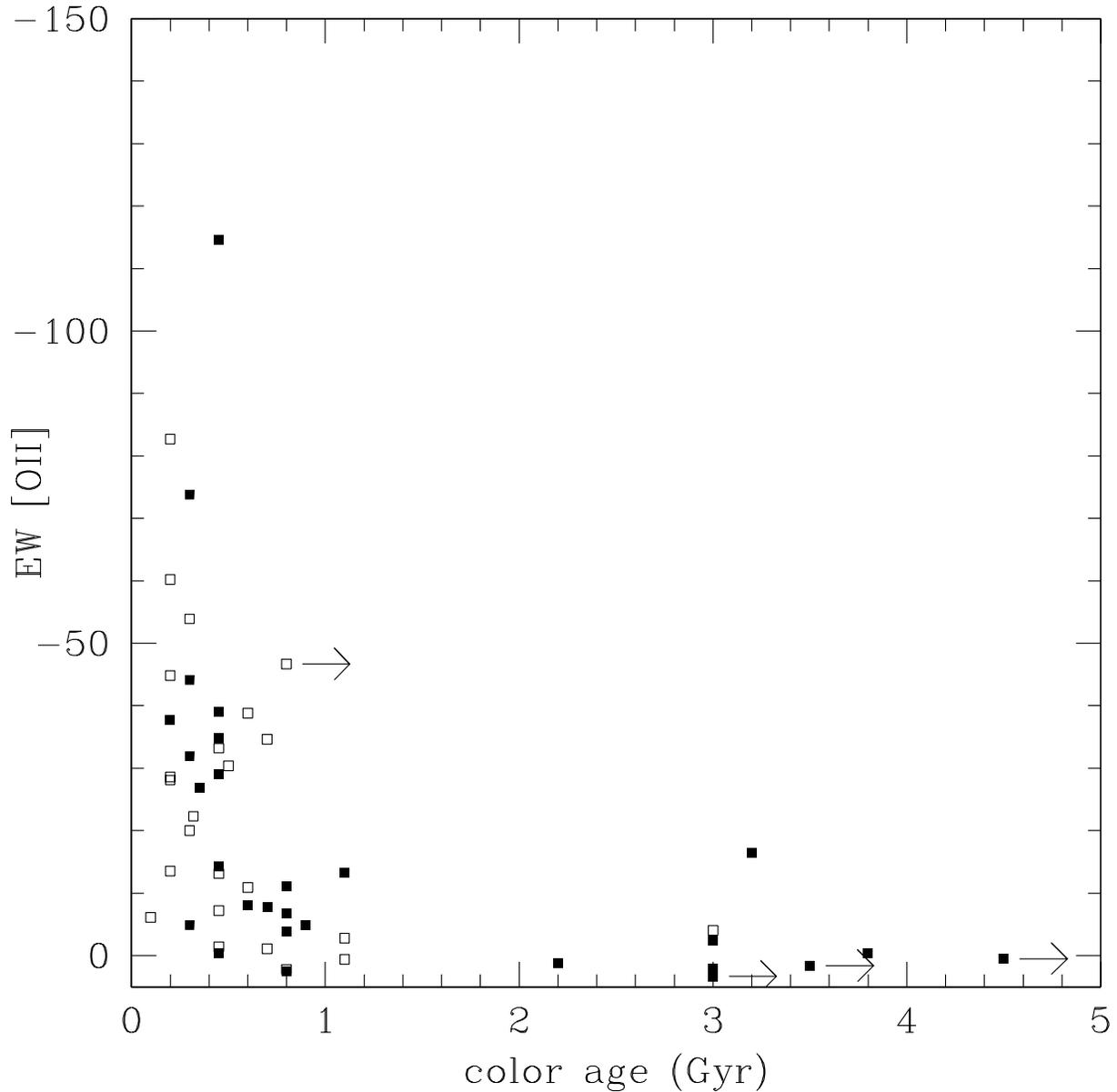}
\caption{The rest [OII] equivalent width versus the color age in Gyr
derived from the ssp models.  The open symbols indicate galaxies
associated with the systems at $z = 0.8274$ and at $z = 0.8453$ in the
CL0023+0423 field, while the closed symbols indicate galaxies
associated with the structure at $z = 0.8290$ and the cluster at $z =
0.8967$ in the CL1604+4304 field. The arrows indicate those values
which are lower limits. The ssp models predict no significant emission
lines after an age of 0.02 Gyr.}
\label{fig-eqwcolssp}
\end{figure}

\clearpage
\begin{figure}
\epsscale{1.0}
\plotone{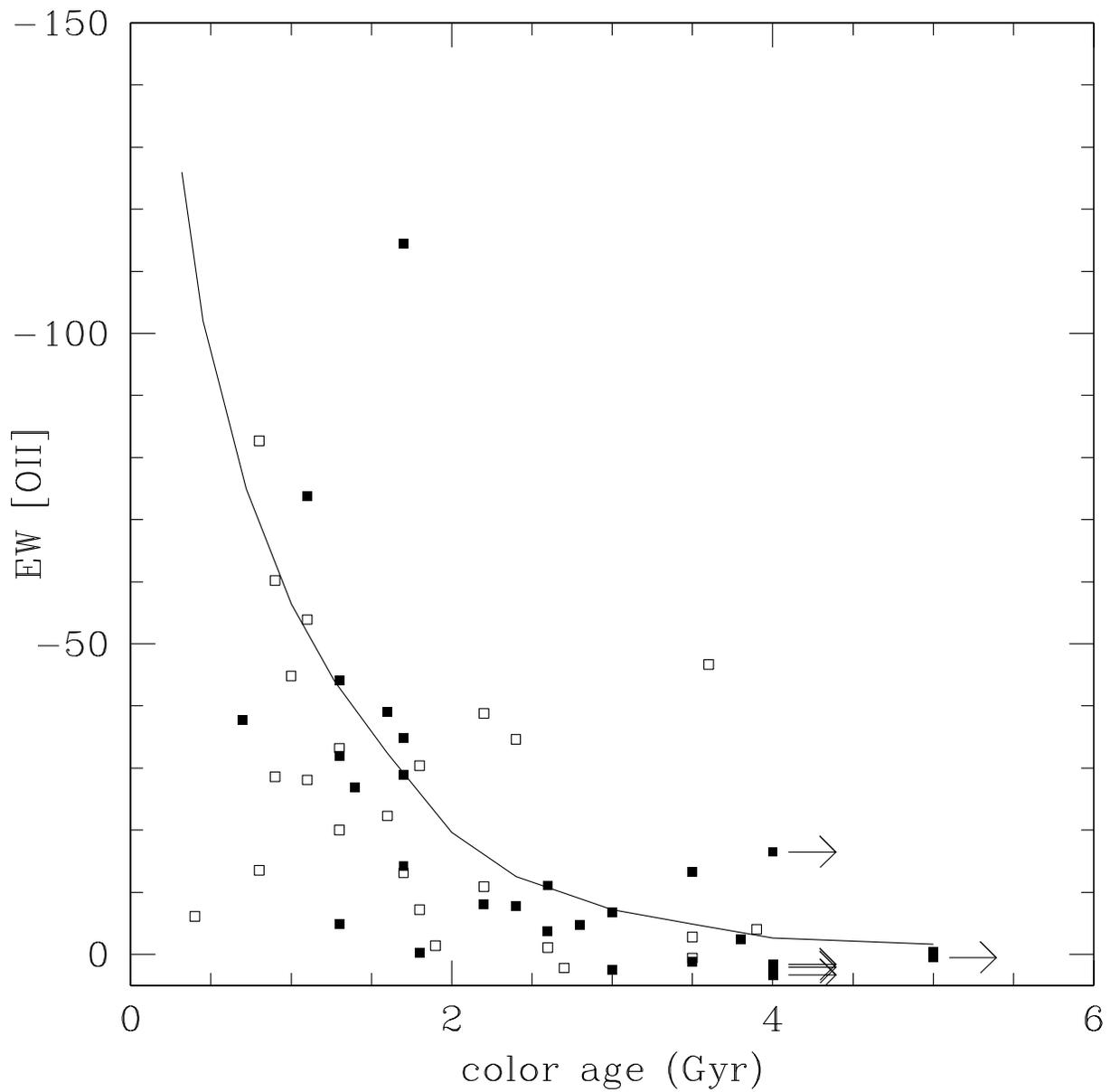}
\caption{Same as Figure~\ref{fig-eqwcolssp} but using ages from the
tau0.6 models.  The arrows indicate those values which are lower
limits.  The solid curve is the predicted equivalent width of [OII]
given in Table 9.}
\label{fig-eqwcoltau0.6}
\end{figure}

\clearpage
\begin{figure}
\epsscale{1.0}
\plotone{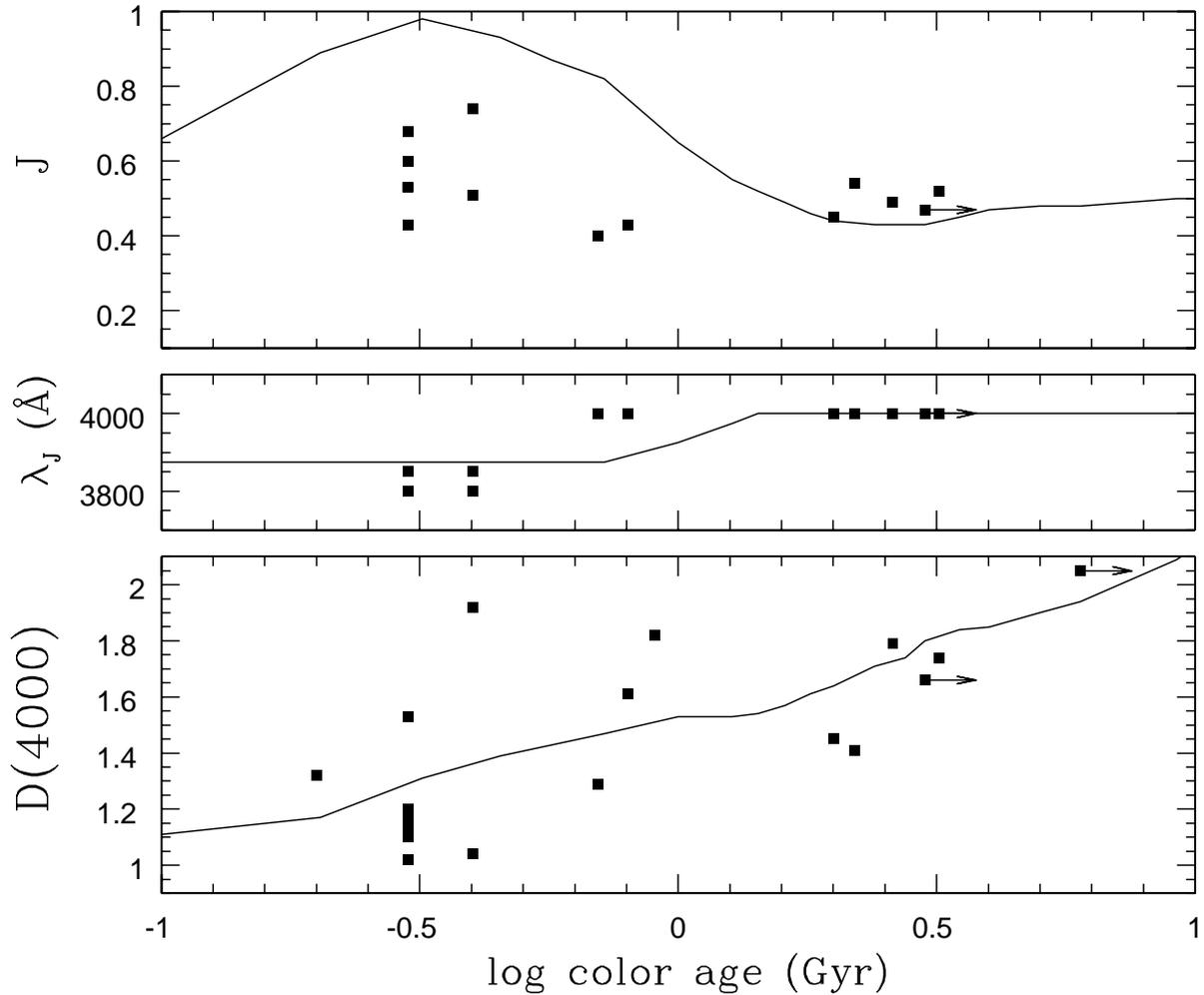}
\caption{Spectral parameters [$J$, $\lambda_J$, and D(4000)] as a
function of the color age derived from the ssp models. The data points
are the values for the groups and clusters listed in Table 10. Color
ages which are lower limits are indicated with arrows (see Sect.\
4.1).  The solid curves in each panel represent the theoretical
relation between the spectral parameter and the color age as derived
from the ssp model energy distributions. {\it Upper Panel} : The
Balmer--4000 \AA\ jump, $J$. {\it Middle Panel} : The wavelength (in
\AA) where the jump occurs, $\lambda_J$.  The wavelength goes from
3875 \AA\ (the Balmer jump) on the left to 4000 \AA\ (the traditional
4000 \AA\ break) on the right. {\it Bottom Panel} : The traditional
4000 \AA\ break measure, D(4000).}
\label{fig-sspjump}
\end{figure}

\clearpage
\begin{figure}
\epsscale{1.0}
\plotone{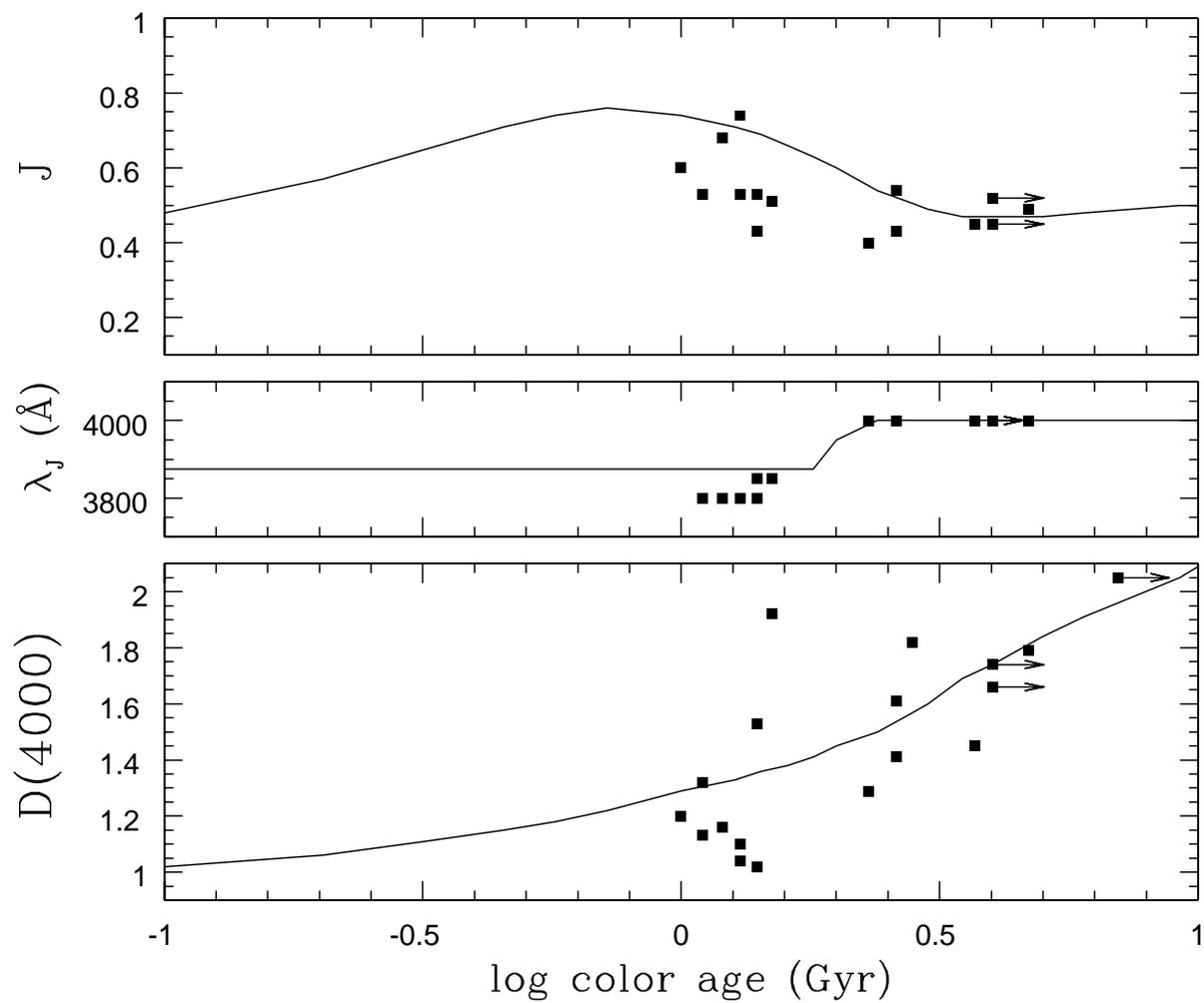}
\caption{Same as Figure~\ref{fig-sspjump} except that tau0.6 models
are used.}
\label{fig-taujump}
\end{figure}

\clearpage
\begin{figure}
\epsscale{1.0}
\plotone{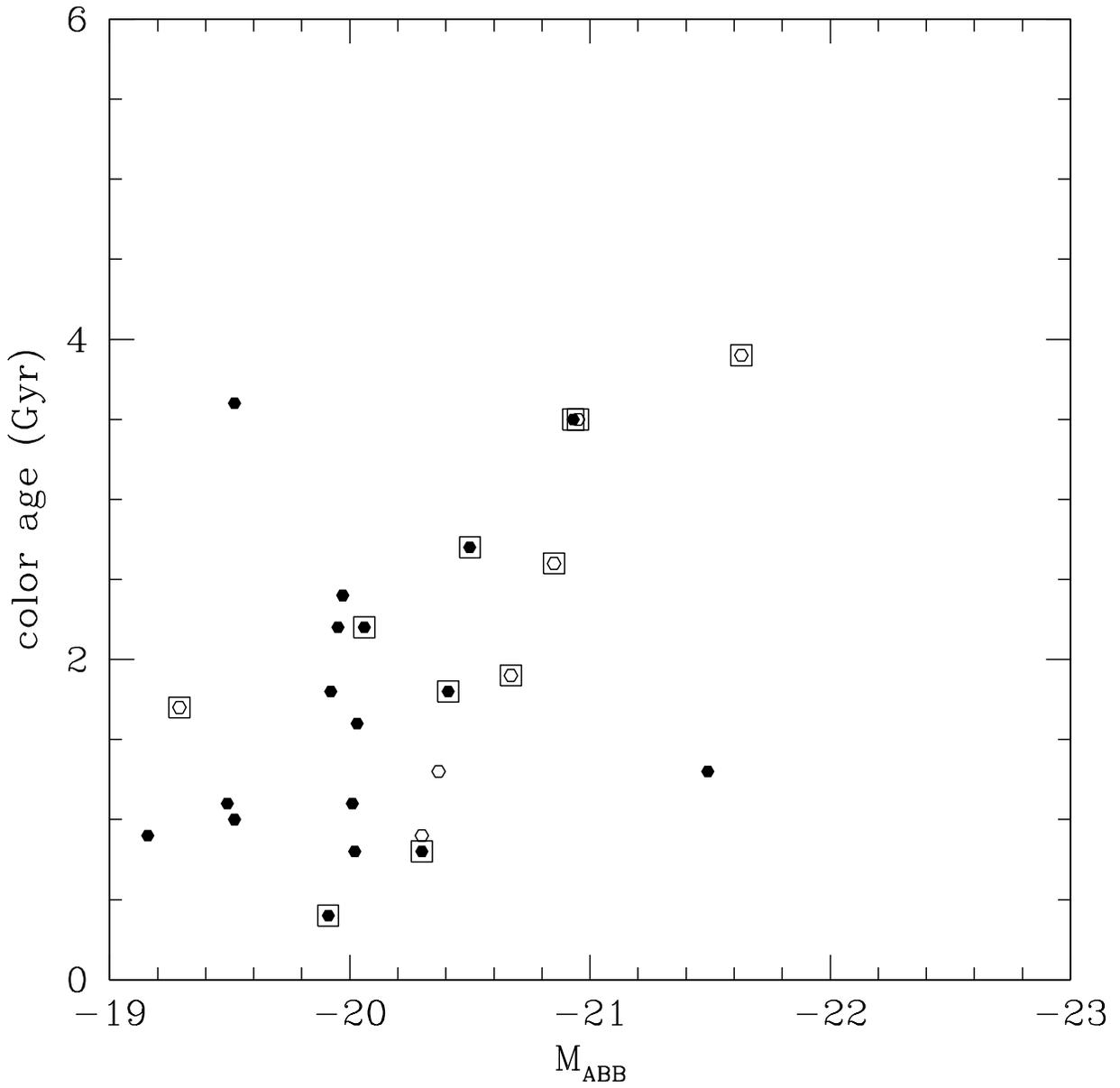}
\caption{Color age based on tau0.6 models versus $M_{ABB}$ for
CL0023+0423 $z = 0.8453$ galaxies (solid dots) and $z = 0.8274$
galaxies (open circles).  Square boxes indicate galaxies where the EW
of [OII] is less than 15 \AA.}
\label{fig-colabs0023}
\end{figure}

\clearpage
\begin{figure}
\epsscale{1.0}
\plotone{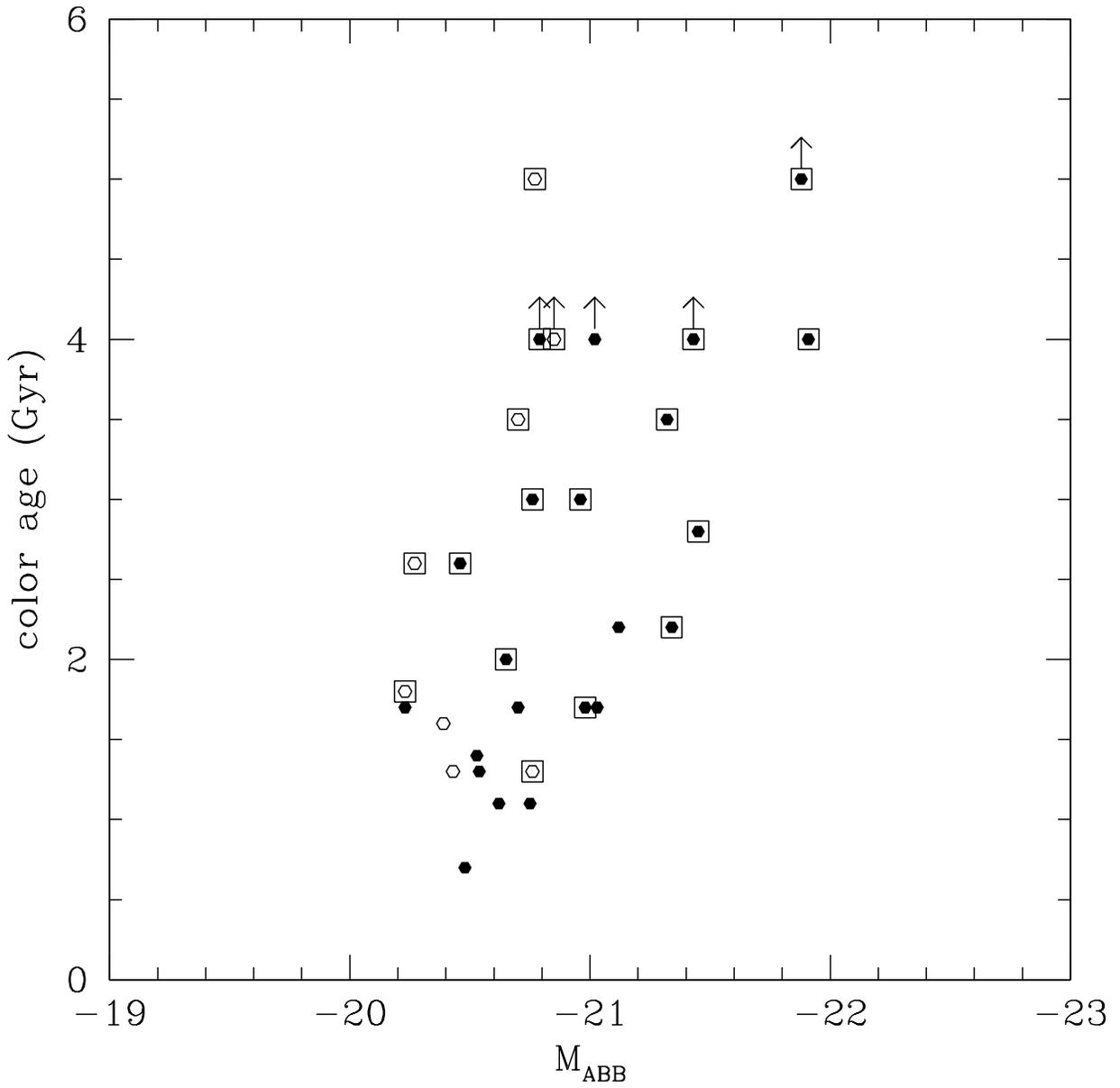}
\caption{Color age based on tau0.6 models versus $M_{ABB}$ for
CL1604+4304 $z = 0.8967$ galaxies (solid dots) and $z = 0.8290$
galaxies (open circles). Square boxes indicate galaxies where the EW
of [OII] is less than 15 \AA.}
\label{fig-colabs1604}
\end{figure}

\clearpage
\begin{figure}
\epsscale{1.0}
\plotone{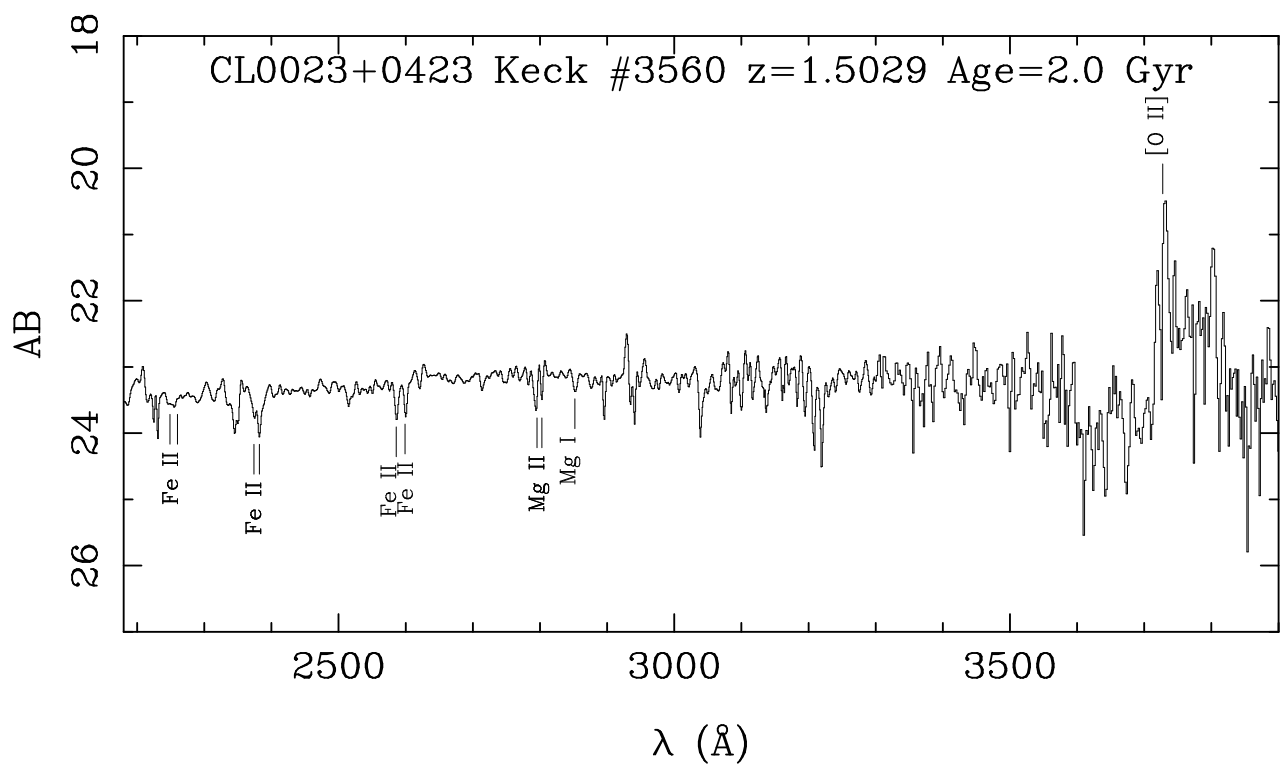}
\caption{The spectrum of Keck \#3560 in CL0023+0423.  Relative AB is
plotted against the rest wavelength.  Lines of FeII, MgII, and MgI are
marked.}
\label{fig-obj3560}
\end{figure}

\clearpage
\begin{figure}
\epsscale{1.0}
\plotone{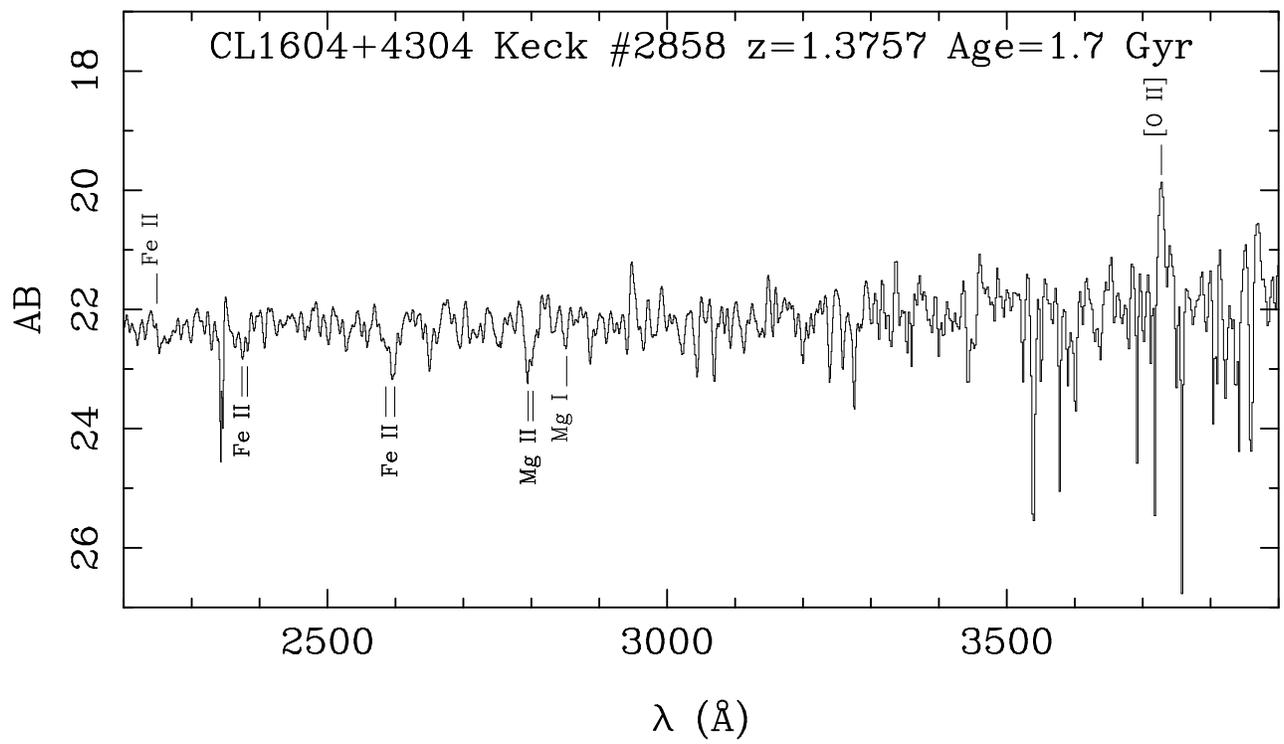}
\caption{Same as Figure~\ref{fig-obj3560} but for Keck \#2858 in CL1604+4304.}
\label{fig-obj2858}
\end{figure}

\end{document}